\documentclass[nofootinbib]{revtex4}
\usepackage{graphicx}
\usepackage{dcolumn}
\usepackage{amsmath,amssymb,epsfig}
\usepackage{paralist}
\usepackage{comment}
\usepackage{graphicx}
\usepackage{multirow}
\usepackage{color,soul}
\usepackage{suffix}
\usepackage{mathtools}

\newcommand{\jgr}{J. Geophys. Res}

\newcommand\VTT[1]{\text{{\color[rgb]{0.75,0,0}~{#1}}}}
\WithSuffix\newcommand\VTT*[1]{\marginpar{\scriptsize{\color[rgb]{0.75,0,0}{#1}}}}
\WithSuffix\newcommand\VTT+[1]{{\bf{\color[rgb]{0.75,0,0}{#1}}}}

\newcommand\SGT[1]{\text{{\color[rgb]{0,0,0.75}~{#1}}}}
\WithSuffix\newcommand\SGT*[1]{\marginpar{\scriptsize{\color[rgb]{0,0,0.75}{#1}}}}

\allowdisplaybreaks

\renewcommand{\vec}[1]{\boldsymbol{\mathrm{#1}}}

\begin{document}


\title{Diffraction of light by the gravitational field of the Sun and the solar corona}

\author{Slava G. Turyshev$^{1}$, Viktor T. Toth$^2$
}%

\affiliation{\vskip 3pt
$^1$Jet Propulsion Laboratory, California Institute of Technology,\\
4800 Oak Grove Drive, Pasadena, CA 91109-0899, USA
}%

\affiliation{\vskip 3pt
$^2$Ottawa, Ontario K1N 9H5, Canada
}%

\date{\today}

\begin{abstract}

We study the optical properties of the solar gravitational lens (SGL) under the combined influence of the static spherically symmetric gravitational field of the Sun---modeled within the first post-Newtonian approximation of the general theory of relativity---and of the solar corona---modeled as a generic, steady-state, spherically symmetric free electron plasma. For this, we consider the propagation of monochromatic electromagnetic (EM) waves near the Sun and develop a Mie theory that accounts for the refractive properties of the gravitational field of the Sun and that of the free electron plasma in the extended solar system. We establish a compact, closed-form solution to the boundary value problem, which extends previously known results into the new regime where gravity and plasma are both present. Relying on the wave-optical approach, we consider three different regions of practical importance for the SGL, including the shadow region directly behind the Sun, the region of geometrical optics and the interference region. We demonstrate that the presence of the solar plasma affects all characteristics of an incident unpolarized light, including the direction of the EM wave propagation, its amplitude and its phase. We show that the presence of the solar plasma leads to a reduction of the light amplification of the SGL and to a broadening of its point spread function.  We also show that the wavelength-dependent plasma effect is
important at radio frequencies, where it drastically reduces both the  amplification factor of the SGL and also its angular resolution. However, for optical and shorter wavelengths, the plasma's contribution to the EM wave is negligibly small, leaving the plasma-free optical properties of the SGL practically unaffected.

\end{abstract}


\maketitle

\section{Introduction}
\label{sec:intro}

When an electromagnetic (EM) wave propagates through a nonmagnetized free electron plasma occupying a region that is much larger than the wavelength, there is a complex interaction between the wave and the medium. As a result, depending on the frequency of the EM wave, the electron plasma frequency and the electron elastic collision frequency, the wave is transmitted, reflected or absorbed by the plasma medium \cite{Ginzburg-book-1964,Landau-Lifshitz:1979}. Understanding this interaction became important with the advent of solar system exploration where EM waves are used for tracking and communicating with deep space probes. This is why, in part, the effect of the solar plasma on the propagation of radio waves was explored extensively \cite{Allen:1947,Ginzburg-Zheleznyakov:1959,Muhleman-etal:1977,Tyler-etal:1977,Muhleman-Anderson:1981}. It is now routinely accounted for in any radio link analysis used either for communication or navigation \cite{DSN-handbook-2017} and, especially, for precision radio science experiments \cite{Giampieri:1994kj,Bertotti-Giampieri:1998,Verma-etal:2013}.

Plasma acts as a dispersive medium. Light rays passing through plasma deviate from light-like geodesics in a way that depends on the frequency \cite{Synge-book-1960,Landau-Lifshitz:1988}. This effect plays a significant role in geometric optics models of gravitational microlensing \cite{Clegg:1997ya,Deguchi-Watson:1987}. Refraction of EM waves from a distant background radio source by an interstellar plasma lens with a Gaussian profile of free-electron column density could lead to observable effects \cite{Clegg:1997ya}. The relative motion of the observer, the lens and the source may modulate the intensity of the background source. There are other effects, including the formation of caustic surfaces, the possible creation of multiple images of the background source and changes in its apparent sky position. The properties of geodesics on a plasma background were investigated extensively. Significant literature on general relativistic ray optics in refractive media is available (for review, \cite{Perlick-book-2000}).

In the context of the optical properties of the solar gravitational lens (SGL) \cite{vonEshleman:1979,Turyshev:2017}, the effects of the solar corona were investigated using a geometric optics approach \cite{Turyshev-Andersson:2002}. It was shown that in the immediate vicinity of the Sun, the propagation of radio waves is significantly affected by the solar plasma, which effectively pushes the focal area of the SGL to larger heliocentric distances. At the same time, one anticipates that the propagation of EM waves at optical frequencies is not significantly affected by the solar plasma. In \cite{Turyshev-Toth:2018-plasma} we show at the required level of accuracy that the direction of travel of visible or near-IR light is indeed unaffected by the plasma. However, the plasma results in a phase shift that depends on the solar impact parameter of an affected ray of light.

In the present paper, we continue to investigate the optical properties of the SGL using a wave theoretical treatment initiated in \cite{Turyshev:2017,Turyshev-Toth:2017}. Specifically, we study light propagation on the background of the solar gravitational monopole and also introduce effects of light refraction in the solar corona. We consider the first post-Newtonian approximation of the general theory of relativity, presented in a harmonic gauge \cite{Fock-book:1959,Turyshev-Toth:2013}. We use a generic model for the electron number density in the solar corona, used in \cite{Muhleman-etal:1977,Tyler-etal:1977,Bertotti-Giampieri:1998,Verma-etal:2013} (using the geometric optics approximation) and in \cite{Turyshev-Toth:2018-plasma,Turyshev-Toth:2018-grav-shadow} (using a wave-optical treatment), which extended the results of \cite{Turyshev-Toth:2018} to the case of a free electron plasma distribution representing the solar corona and the interplanetary medium in the solar system. Here we take a further step and study light propagation on the combined background of the post-Newtonian monopole gravitational field and the solar plasma distribution, thereby extending the results of our earlier work on the SGL \cite{Turyshev:2017,Turyshev-Toth:2017,Turyshev-Toth:2018,Turyshev-Toth:2018-plasma}.

Our main objective here is to investigate the optical properties of the SGL in the presence of the solar corona. What is the effect of the refractive background in the solar system on the structure of the caustic formed by the solar gravitational mass monopole? Specifically, what is the plasma effect on light amplification,  the point-spread function (PSF), and the resulting angular resolution  of the SGL? Are there plasma-induced optical aberrations? How does the solar plasma affect the ultimate image quality? These questions are important for our ongoing efforts to study the application of the SGL for direct high-resolution imaging and spectroscopy of exoplanets \cite{Turyshev-etal:2018,Turyshev-etal:2018-wp}.

This paper is organized as follows:
In Section \ref{sec:eqs-g+p} we discuss the solar plasma and present the model for the electron number density distribution in the solar system.
Section~\ref{sec:em-waves-gr+pl} presents Maxwell's field equations for the EM field on the background of the solar gravitational monopole and the solar plasma.
In Section~\ref{sec:SGL-imaging} we discuss the optical properties of the SGL in the presence of the solar plasma. We also offer some practical considerations for the use of this improved realistic model of the SGL for exoplanet imaging.
In Section~\ref{sec:disc} we discuss the results and their importance to the exploration of exoplanets.
To make the main results more accessible, we placed some material in the Appendices.
Appendix~\ref{app:Debye} discusses the decomposition of the Maxwell equations and their representation in terms of Debye potentials.
In Appendix~\ref{sec:geodesics-phase} we study light propagation in weak and static gravity and steady-state plasma using the geometric optics approximation.
Appendix~\ref{sec:geodesics} is devoted to a study of light's path in weak and static gravity in the presence of the extended solar plasma.
In Appendix~\ref{sec:geom-optics} we study the phase evolution of a plane wave propagating in the vicinity of a massive body in the presence of plasma.
Appendix~\ref{sec:rad_eq_wkb} discusses an approximate solution for the radial equation that relies on the Wentzel--Kramers--Brillouin (WKB) approximation.

\section{EM waves in a static gravitational field in the presence of plasma}
\label{sec:em-waves-gr+pl}

We consider the propagation of monochromatic light emitted by a distant source and received by a detector at the focal area of the SGL. For the purposes of this paper, this light is assumed to originate at a very large distance from the solar system, $r\gg R_\star$. Thus, by the time it reaches the solar system, this light may be approximated as a plane wave whose phase is logarithmically modified due to the presence of the solar gravity \cite{Turyshev-Toth:2017}. As this light reaches the solar system and before it is detected by an imaging telescope, its propagation is affected by the plasma of the solar corona. Our current objective is to investigate the contribution of this plasma to the optical properties of the SGL.

\subsection{Modeling the solar atmosphere and the interplanetary medium}
\label{sec:eqs-g+p}

For an EM wave of angular frequency $\omega$, propagating through a free electron plasma, the dielectric permittivity of the plasma is defined as \cite{Landau-Lifshitz:1979}
\begin{equation}
 \epsilon (t,{\vec r})=1- \frac{4\pi n_e (t,{\vec r})e^2}{m_e\omega^2}= 1 - \frac{\omega_{\tt p}^{ 2}}{\omega^2}, \qquad {\rm where} \qquad
 \omega^2_p=\frac{4\pi n_e e^2}{m_e},
 \label{eq:eps}
 \end{equation}
where $e$ is the electron charge, $m_e$ is its mass, while $n_e=n_e(t,{\vec r})$ is the electron number density. The quantity $\omega_{\tt p}$ is known as the electron plasma (or Langmuir) frequency. As far as magnetic permeability goes, it is reasonable to assume that the solar plasma is non-magnetic, which is captured by setting $\mu=1$.

The effects of the solar plasma are significant at microwave frequencies, but light propagation at optical and IR wavelengths remains almost unaffected \cite{Huber-etal:2013,Lang-book:2009}. Nonetheless, the level of sensitivity of the SGL, given its extreme resolution and light amplification capabilities, makes it obligatory to account for even such minute effects.

To evaluate the plasma contribution, we need to know the electron number density, as given by (\ref{eq:eps}), along the path of a light ray. Much of our knowledge about the solar plasma comes from spacecraft tracking in the inner solar system \cite{Muhleman-etal:1977,Tyler-etal:1977,Muhleman-Anderson:1981,Giampieri:1994kj,Bertotti-Giampieri:1998,Verma-etal:2013}. In addition, distant spacecraft and astronomical observations provide information about the properties and extent of the interplanetary medium towards interstellar space \cite{Belcher-etal:1993,Stone-etal:1996,Decker-etal:2005}.

\begin{figure}
\includegraphics[width=100mm]{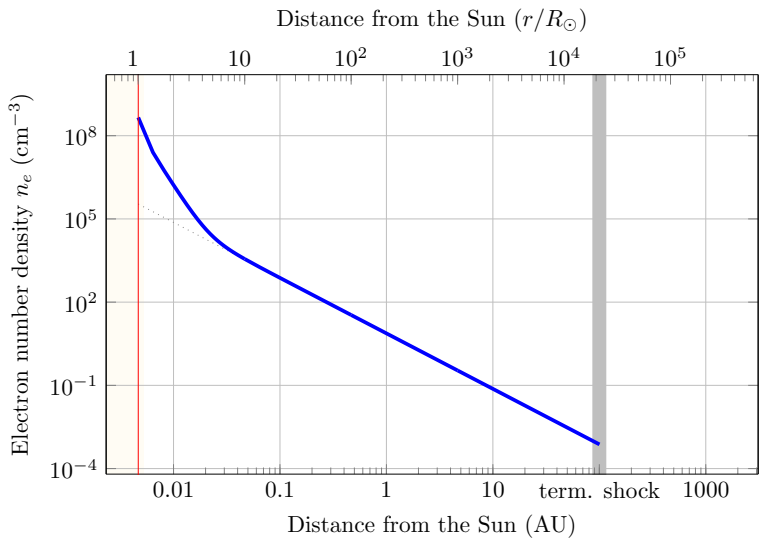}
\caption{The electron number density model (\ref{eq:model}) (thick blue line) given by \cite{Muhleman-etal:1977,Tyler-etal:1977,Muhleman-Anderson:1981}. The leftmost part of the curve, at short heliocentric distances dominated by terms with higher powers of $(R_\odot/r)$ corresponds to the visible solar corona \cite{Lang-book:2009,Lang-ebook:2010}, The thin dotted line shows the contribution of the inverse square term, which dominates beyond a few solar radii. The lightly shaded region on the left represents the solar interior. The approximate location of the termination shock is also marked, beyond which the radial dependence disappears, leaving only an approximately homogeneous interstellar background (not shown). Diagram adapted from \cite{Turyshev-Toth:2018-plasma}, with the horizontal axis extended beyond the termination shock.\label{fig:plasma}}
\end{figure}

In the general case, the electron density shows temporal variability, which we represent by decomposing $n_e$ into a steady-state, spherically symmetric part $\overline{n}_e(r)$, plus a term, $\delta n_e(t,{\bf r})$, describing temporal and spatial fluctuations:
\begin{equation}
n_e(t, {\bf r})= \overline{n}_e(r)+ \delta n_e(t, {\bf r}).
\label{eqelcont}
\end{equation}

The variability of the solar atmosphere, $ \delta n_e(t, {\bf r})$, has no preferred time scale. Variations in the electron number density can be of a magnitude equal to that of the steady-state term \citep{Armstrong-etal:1979}. These variations are carried along by the solar wind, at a typical speed of $\sim 400$~km/s; over integration times measured in the thousands of seconds, the spatial scale of the fluctuations will therefore be comparable to the solar radius.

For the heliocentric regions of interest in the context of the SGL, 650--900 AU \cite{Turyshev-etal:2018}, the corresponding range of the impact parameters is $b\sim (1.1\textup{--}1.3) R_\odot$, where $R_\odot$ is the solar radius. This region, $\sim(0.1\textup{--}0.3)R_\odot$ from the solar surface, is the most violent region of the solar corona, characterized by significant fluctuations of the electron content density.

Consequently, we may reasonably expect that the deflection of a light ray for a given impact parameter $b$ due to spatial and temporal fluctuations will be of the same order as the deflection due to the mean solar atmosphere. This is certainly the case for microwave frequencies \cite{Bertotti-Giampieri:1998,Turyshev-Andersson:2002}.

As these deviations are unpredictable in nature, their contributions must be treated as noise (e.g., as a stochastic component to the convolution matrix that characterizes how the SGL forms an image in the image plane, see \cite{Turyshev-etal:2018} for discussion.) In contrast, the steady-state component of the solar corona is well understood, and the magnitude of its contribution can be estimated. These results can also be used to characterize the noise component due to fluctuations, making it possible to understand the extent to which such contributions will reduce the effective resolution of the SGL, and to devise effective data analysis strategies.

As a result, in the present paper, we focus on the contribution of the steady-state, spherically symmetric component of the electron plasma density and its effect on the SGL. We therefore ignore any dependence on heliographic latitude and any additional spatial and temporal variations. The spherically symmetric, steady-state plasma may be parameterized in the following generic form:
{}
\begin{eqnarray}
n_e({\vec r})=
   \begin{dcases}
~~0,&\phantom{R_\odot}0\leq r<R_\odot,\\
\sum_i \alpha_i \Big(\frac{R_\odot}{r}\Big)^{\beta_i},\quad&
\phantom{0}R_\odot\leq r\leq R_\star,\\
~n_0,&\phantom{R_\odot 0\leq{}}r>R_\star,
  \end{dcases}
 \label{eq:n-eps_n-ism}
 \end{eqnarray}
where $\beta_i>1$ (to match the properties of the solar wind at large heliocentric distances that behaves as $\propto 1/r^2$) and $R_\star$ is the heliocentric distance to the termination shock, which we take to be $R_\star\simeq100$~AU \cite{Belcher-etal:1993,Stone-etal:1996,Decker-etal:2005}. The termination shock is an intermediate border situated before the heliopause, which is the last frontier of the solar wind. It is the boundary at $\sim130$~AU, where the solar wind fades and the interstellar medium begins \cite{Lust:1963}. It is, of course true (as evidenced by, for instance, the findings of Voyager 1 and 2) that the actual distance to the termination shock varies with time and direction. However, as we find below, our main results are not sensitive to the numerical value of $R_\star$ so long as it is of ${\cal O}(100~{\rm AU})$; contributions from the plasma to the propagation of EM waves comes mostly from the region within a few solar radii from the solar surface.

Finally, $n_0$ is the electron number density in the interstellar medium, which is assumed to be homogeneous. The presence of this term is for completeness only. As it does not influence the scattering of light, it may be safely assumed to be that of a vacuum, namely $n_0=0$. Note that the model (\ref{eq:n-eps_n-ism}) neglects the variability in the electron number density within the heliosheath. Any variability, if it exists, does not contribute an observable effect to scattering of light by the SGL.

The steady-state behavior is reasonably well known, and we can use one of
the several plasma models found in the literature \citep{Tyler-etal:1977,Muhleman-etal:1977,Muhleman-Anderson:1981,Verma-etal:2013}.  To be more specific, we make use of the following steady-state, spherically symmetric model of electron distribution (see \cite{Turyshev-Andersson:2002,Turyshev-Toth:2010LLR} and references therein):
{}
\begin{equation}
\overline{n}_e(r)= \Big[\Big(2.99\times10^8 \Big(\frac{R_\odot}{r}\Big)^{16}+1.55 \times10^8 \Big(\frac{R_\odot}{r}\Big)^{6}+ 3.44\times 10^5 \Big(\frac{R_\odot}{r}\Big)^{2}\Big]  ~~   {\rm
cm}^{-3}.
\label{eq:model}
\end{equation}
At a large distance from the source, the model replicates the expected $1/r^2$ behavior of the solar wind. Other existing models are somewhat different from (\ref{eq:model}). Such models may account for the non-sphericity of the electron plasma density and offer a slightly different distance power law (for discussion, \cite{Verma-etal:2013}). These additional features of these plasma models are not important for our purposes, as their effects are below the detection accuracy.  Also, any inhomogeneities of the plasma distribution in the interplanetary medium are small and, thus, they are not expected to yield a significant mechanism of refraction for light propagating through the solar system.

We emphasize that the model (\ref{eq:model}) was developed using the tracking data for interplanetary spacecraft, which was conducted at multiple radio frequencies \cite{Muhleman-etal:1977,Tyler-etal:1977,Muhleman-Anderson:1981}. Astronomical observations conducted on the solar background at optical wavelengths also support this model  \cite{Allen:1947,Lang-ebook:2010}. When studying light propagation in the immediate vicinity of the solar photosphere, the model (\ref{eq:model}) may have to be augmented by terms containing higher powers of $(R_\odot/r)$. However, even in extreme proximity to the Sun, the electron number density would be at most $\overline{n}_e(r)\lesssim 6\times 10^{8}~{\rm cm}^{-3}$ \cite{Verma-etal:2013,Mercier-Chambe:2015}.

\begin{figure}
\includegraphics[width=110mm]{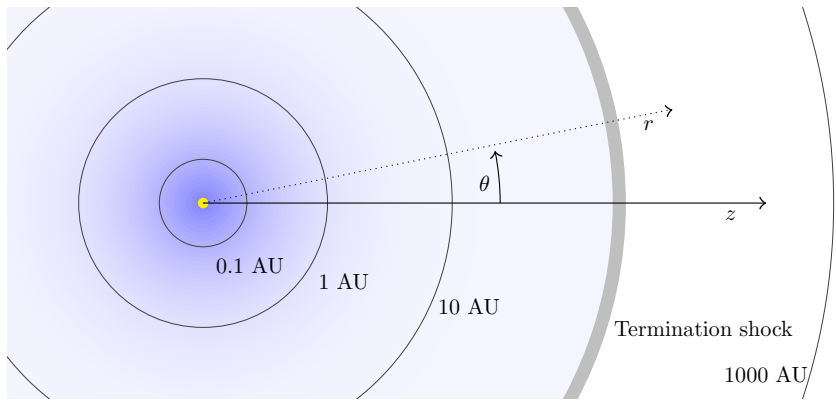}
\caption{Schematic of the solar system using an approximate log-square scale. Shading indicates the solar plasma density that is traversed by an incident plane wave. The termination shock at $\sim 100$~AU is where the solar plasma collides with the interstellar medium, the density of which is constant and does not contribute to the scattering of light. The spherical ($r,\theta,\phi$) coordinate system (with $\phi$ suppressed) and the cylindrical $z$ coordinate used in this paper are indicated. Diagram adapted from \cite{Turyshev-Toth:2018-plasma}, with the heliocentric distance range extended beyond the termination shock.\label{fig:solsys}}
\end{figure}

The plasma frequency $ \omega_{\tt p}^2$ in Eq.~(\ref{eq:eps}), in the case of the spherically symmetric plasma distribution model (\ref{eq:n-eps_n-ism}), in the range of heliocentric distances,  $R_\odot\leq r\leq R_\star$, has the form
 \begin{equation}
 \omega_{\tt p}^2=\frac{4\pi e^2}{m_e}
\sum_i \alpha_i\Big(\frac{R_\odot}{r}\Big)^{\beta_i}.
 \label{eq:n_n-ss}
 \end{equation}
This generic spherically symmetric model for the plasma frequency in the extended solar corona allows us to study the influence of solar plasma on the propagation of EM waves throughout the solar system in the range of heliocentric distances given by $R_\odot\leq r\leq R_\star$. Clearly, the model (\ref{eq:n_n-ss}) may be further extended, for instance, to include known (non-random) effects due to non-sphericity, such as dependence on the solar latitude. If needed, such effects may be treated using the same approach as presented in this paper.

\subsection{Maxwell's equations in three-dimensional form}
\label{sec:maxwell}

We now focus on solving Maxwell's equations on the solar system's background set by gravity and plasma. We rely heavily on \cite{Turyshev:2017,Turyshev-Toth:2017,Turyshev-Toth:2018-grav-shadow} (that were inspired by \cite{Born-Wolf:1999,Herlt-Stephani:1976}), which the reader is advised to consult first.

Following \cite{Turyshev:2017,Turyshev-Toth:2017}, we begin with the generally covariant form of Maxwell's equations:
{}
\begin{eqnarray}
\partial_lF_{ik}+\partial_iF_{kl}+\partial_kF_{li}=0, \qquad
\frac{1}{\sqrt{-g}}\partial_k\Big(\sqrt{-g}F^{ik}\Big)=-\frac{4\pi}{c}j^i,
\label{eq:max-eqs}
\end{eqnarray}
where $g_{mn}$ is the metric tensor and $g=\det g_{mn}$ is its determinant. We use a $(3+1)$ decomposition \cite{Turyshev-Toth:2017} of the  generic interval (see \textsection 84 of  \cite{Landau-Lifshitz:1988}):
\begin{eqnarray}
ds^2=g_{mn}dx^mdx^n&=&
\Big(\sqrt{g_{00}}dx^0+\frac{g_{0\alpha}}{\sqrt{g_{00}}}dx^\alpha\Big)^2-
\kappa_{\alpha\beta}dx^\alpha dx^\beta,
\label{eq:int0}
\end{eqnarray}
where the 3-dimensional symmetric metric tensor $\kappa_{\alpha\beta}$ is given as:
\begin{eqnarray}
 \kappa_{\alpha\beta}=-g_{\alpha\beta}+\frac{g_{0\alpha}g_{0\beta}}{g_{00}}, \qquad \kappa=\det\kappa_{\alpha\beta}.
\label{eq:int10}
\end{eqnarray}

Physical fields are defined as the 3-vectors ${\vec E}, {\vec D}$ and the antisymmetric 3-tensors $B_{\alpha\beta}$ and $H_{\alpha\beta}$ (see the Problem in \textsection 90 of \cite{Landau-Lifshitz:1988}):
{}
\begin{eqnarray}
E_\alpha=F_{0\alpha}, \quad {\cal D}^\alpha=-\epsilon \sqrt{g_{00}}F^{0\alpha}, \quad {\cal B}_{\alpha\beta}=\mu F_{\alpha\beta}, \qquad H^{\alpha\beta}=\sqrt{g_{00}}F^{\alpha\beta},
\label{eq:EDBH}
\end{eqnarray}
where, following \cite{Landau-Lifshitz:1988}, we also introduced the permittivity $\epsilon$ and magnetic permeability $\mu$ of the medium.
The quantities (\ref{eq:EDBH}) are not independent. Introducing the 3-vector $\vec g\equiv -g^{0\alpha}$, we see that the following identities exist:
\begin{eqnarray}
\vec {\cal D}=\epsilon\Big\{\frac{1}{\sqrt{g_{00}}}\vec E+[\vec H\times\vec g]\Big\}, \qquad
\vec {\cal B}=\mu\Big\{\frac{1}{\sqrt{g_{00}}}\vec H+[\vec g\times\vec E]\Big\}.
\label{eq:EDBH*}
\end{eqnarray}

As a result, Eqs.~(\ref{eq:max-eqs}) yield the following 3-dimensional Maxwell's equations:
\begin{eqnarray}
{\rm curl_\kappa\bf E}&=&-\frac{1}{\sqrt{\kappa}}\partial_0\Big(\sqrt{\kappa}{~\vec {\cal B}}\Big), \qquad\qquad ~\,{\rm div_\kappa\vec{\cal B}}=0,
\label{eq:max-set1}\\
{\rm curl_\kappa\bf H}&=&\frac{1}{\sqrt{\kappa}}\partial_0\Big(\sqrt{\kappa}{~\vec {\cal D}}\Big) + \frac{4\pi}{c} {\bf s}, \qquad {\rm div_\kappa\vec{\cal D}}=4\pi \rho,
\label{eq:max-set2}
\end{eqnarray}
where the differential operators ${\rm curl}_\kappa$ and ${\rm div}_\kappa$ are taken with respect to the 3-dimensional metric tensor $\kappa_{\alpha\beta}$ from (\ref{eq:int10}) \cite{Korn-Korn:1968} (also see relevant details in Appendix A of \cite{Turyshev-Toth:2017}).

To describe the optical properties of the SGL in the post-Newtonian approximation, we use a static harmonic metric, for which the line element may be given in spherical coordinates $(r,\theta,\phi)$ as \cite{Fock-book:1959,Turyshev-Toth:2013}:
\begin{eqnarray}
ds^2&=&u^{-2}c^2dt^2-u^2\big(dr^2+r^2(d\theta^2+\sin^2\theta d\phi^2)\big),
\label{eq:metric-gen}
\end{eqnarray}
where, to the accuracy sufficient to describe light propagation in the solar system, the quantity $u$ has the form
\begin{eqnarray}
u=1+c^{-2}U+{\cal O}(c^{-4}),
\qquad
U=G\int\frac{\rho(x') d^3 r'}{|{\vec r}-{\vec r}'|},~~
\label{eq:w-PN}
\end{eqnarray}
with $U$ being the Newtonian gravitational potential.
Using the metric  (\ref{eq:metric-gen}), we may compute Eqs.~(\ref{eq:int10}) and derive Maxwell's equations (\ref{eq:max-set1})--(\ref{eq:max-set2}) in terms of the physical components of the vector ${\vec E}$ and similar components for ${\vec H}, {\vec {\cal D}}$ and ${\vec {\cal B}}$ \cite{Korn-Korn:1968}).
The fact that the chosen metric is static simplifies the expressions for physical fields. Indeed, with $g_{0\alpha}=0$  (thus, $\vec g=0$) and $\partial_0 g_{mn}=0$, expressions (\ref{eq:EDBH*}) take the form:
{}
\begin{eqnarray}
 {\vec {\cal D}}=\frac{1}{\sqrt{g_{00}}}\epsilon {\vec E}= \epsilon u {\vec E}\equiv \epsilon {\vec D}, \qquad\qquad
 {\vec {\cal B}}=\frac{1}{\sqrt{g_{00}}} \mu{\vec H}= \mu u {\vec H}\equiv \mu  {\vec B},
\label{eq:dif-DB}
\end{eqnarray}
where we introduced the quantities ${\vec D}=u{\vec E}$ and ${\vec B}=u{\vec H}$ that describe the EM field in static gravity in the vacuum.

We consider the propagation of an EM wave in the vacuum, where no sources or currents exist, i.e., $j^k\equiv (\rho,{\vec j})=0$. Furthermore, as in  \cite{Turyshev-Toth:2017}, we focus our discussion on the largest contribution to the gravitational scattering of light, which, in the case of the Sun, is due to the gravity field produced by a static monopole. In this case, the Newtonian potential in (\ref{eq:w-PN}) may be given by $c^{-2}U({\vec r})={r_g}/{2r}+{\cal O}(r^{-3},c^{-4}),$ where $r_g=2GM/c^2$ is the Schwarzschild radius of the source. Therefore, the quantity $u$ in (\ref{eq:metric-gen}) and (\ref{eq:w-PN}) has the form
{}
\begin{eqnarray}
u(r)&=&1+\frac{r_g}{2r}+{\cal O}(r^{-3},c^{-4}).
\label{eq:pot_w_1**}
\end{eqnarray}
If needed, one can account for the contribution of higher-order multipoles using the tools developed in \cite{Kopeikin-book-2011,Turyshev-GRACE-FO:2014,Turyshev-Toth:2018-plasma}.

This allows us to present the vacuum form of Maxwell's equations (\ref{eq:max-set1})--(\ref{eq:max-set2}) for the steady-state, spherically symmetric plasma distribution as (see Appendix~\ref{app:Debye} for details)
{}
\begin{eqnarray}
{\rm curl}\,{\vec D}&=&-  \mu u^2\frac{1}{c}\frac{\partial {\vec B}}{\partial t}+{\cal O}(G^2),
\qquad ~{\rm div}\big(\epsilon u^2\,{\vec D}\big)={\cal O}(G^2),
\label{eq:rotE_fl*+*}\\[3pt]
{\rm curl}\,{\vec B}&=& \epsilon u^2\frac{1}{c}\frac{\partial {\vec D}}{\partial t}+{\cal O}(G^2),
\qquad \quad \,
{\rm div }\big(\mu u^2\,{\vec B}\big)={\cal O}(G^2),
\label{eq:rotH_fl*+*}
\end{eqnarray}
where the differential operators ${\rm curl}$  and ${\rm div}$ are now with respect to the 3-dimensional Euclidean flat metric.

Evaluating (\ref{eq:model}) at the shortest relevant heliocentric distance, $r=R_\odot$, we see that the electron density given by this model is of the order of $\overline{n}_e(r)\lesssim 4.54\times 10^{8}~{\rm cm}^{-3}$, which implies that $\omega_{\tt p}=\sqrt{4\pi n_ee^2/m_e}\sim 1.20\times 10^9~{\rm s}^{-1}$ corresponding to a frequency of $\nu_p=\omega_{\rm p}/2\pi=191$~MHz.
For optical frequencies ($\nu=c/\lambda\sim 300$~THz) and for $r=R_\odot$, we see that (\ref{eq:eps}) may contribute at most at the order of $(\omega_{\tt p}/\omega)^2\sim4.08\times10^{-13}$, while for radio frequencies ($\nu\sim 10$~GHz) this ratio is much higher: $(\omega_{\tt p}/\omega)^2\sim3.67\times10^{-4}$. At the same time, the effective contribution of the gravitational monopole to the refraction index from (\ref{eq:pot_w_1**}) is $r_g/r\lesssim 4.25\times 10^{-6}\, (R_\odot/r).$  Therefore, in our discussion below, we need to carry out the necessary analysis up to terms that are linear with respect to gravity and plasma contributions while neglecting higher order terms -- the approach that is certainly justified for optical wavelengths, but may need to be augmented to include higher order contributions, if dealing with radio wavelengths (as was done, for instance, in  \cite{BisnovatyiKogan:2008yg,Bisnovatyi-Kogan:2015dxa}). Nevertheless, our approach remains valid even for lower frequencies, and may be used to provide insight into the physical processes of the EM wave interacting with extended solar corona given by a generic model (\ref{eq:n-eps_n-ism}).

\subsection{Representation of the EM field in terms of Debye potentials}
\label{sec:debye}

In the case of a static, spherically symmetric gravitational field and steady-state, spherically symmetric plasma distributions, solving the  field equations  (\ref{eq:rotE_fl*+*})--(\ref{eq:rotH_fl*+*}) is most straightforward. Following the derivation in \cite{Turyshev-Toth:2017} (see Appendix E therein), we obtain the complete solution of these equations in terms of the electric and magnetic Debye potentials \cite{Born-Wolf:1999}, ${}^e\Pi$ and ${}^m\Pi$. For details, see Appendix~\ref{app:Debye} (see, for instance, derivations leading to (\ref{eq:Dr-em})--(\ref{eq:Bp-em})). The result is a system of equations for the components of the monochromatic EM field with the wavenumber\footnote{When an EM wave is propagating in an electron plasma, its frequency is given by the dispersion relation $\omega^2(k)=k^2 c^2+\omega_{\tt p}^2(k)$ \cite{Landau-Lifshitz:1979}. That is, the plasma modifies the dispersion relation and affects the group and phase velocities. Realizing that the electron number density for the solar plasma is at most $\overline{n}_e(r)\lesssim 6\times 10^{8}~{\rm cm}^{-3}$ \cite{Verma-etal:2013,Mercier-Chambe:2015}, using (\ref{eq:eps}), we compute the largest relevant value of $\omega_{\tt p}^2(k)$ that yeilds $\omega^2(k)=k^2c^2\big(1+ 5.38\times 10^{-13}(\lambda/1~\mu{\rm m})^2\big)$. Therefore, throughout this paper we use $\omega^2=k^2 c^2\big(1+{\cal O}(10^{-12})\big)$, signifying that at the optical and near-IR wavelengths relevant to the SGL, $\lambda\simeq 1~\mu$m, the difference between the group and phase velocities can be neglected.} $k=\omega/c$:
{}
\begin{eqnarray}
{\hat { D}}_r&=&
\frac{1}{\sqrt{\epsilon}u}\Big\{\frac{\partial^2 }{\partial r^2}
\Big[\frac{r\,{}^e{\hskip -1pt}\Pi}{\sqrt{\epsilon}u}\Big]+\Big(\epsilon\mu\,k^2 u^4-\sqrt{\epsilon}u\big(\frac{1}{\sqrt{\epsilon}u}\big)''\Big)\Big[\frac{r\,{}^e{\hskip -1pt}\Pi}{\sqrt{\epsilon}u}\Big]\Big\},
\label{eq:Dr-em0}\\[3pt]
{\hat {  D}}_\theta&=&
\frac{1}{\epsilon u^2r}\frac{\partial^2 \big(r\,{}^e{\hskip -1pt}\Pi\big)}{\partial r\partial \theta}+\frac{ik}{r\sin\theta}
\frac{\partial\big(r\,{}^m{\hskip -1pt}\Pi\big)}{\partial \phi},
\label{eq:Dt-em0}\\[3pt]
{\hat {  D}}_\phi&=&
\frac{1}{\epsilon u^2r\sin\theta}
\frac{\partial^2 \big(r\,{}^e{\hskip -1pt}\Pi\big)}{\partial r\partial \phi}-\frac{ik}{r}
\frac{\partial\big(r\,{}^m{\hskip -1pt}\Pi\big)}{\partial \theta},
\label{eq:Dp-em0}\\[3pt]
{\hat {  B}}_r&=&
\frac{1}{\sqrt{\mu}u}\Big\{\frac{\partial^2}{\partial r^2}\Big[\frac{r\,{}^m{\hskip -1pt}\Pi}{\sqrt{\mu}u}\Big]+\Big(\epsilon\mu\,k^2 u^4-\sqrt{\mu}u\big(\frac{1}{\sqrt{\mu}u}\big)''\Big)\Big[\frac{r\,{}^m{\hskip -1pt}\Pi}{\sqrt{\mu}u}\Big]\Big\},
\label{eq:Br-em0}\\[3pt]
{\hat {  B}}_\theta&=&
-\frac{ik}{r\sin\theta} \frac{\partial\big(r\,{}^e{\hskip -1pt}\Pi\big)}{\partial \phi}+\frac{1}{\mu u^2r}
\frac{\partial^2 \big(r\,{}^m{\hskip -1pt}\Pi\big)}{\partial r\partial \theta},
\label{eq:Bt-em0}\\[3pt]
{\hat { B}}_\phi&=&
\frac{ik}{r}\frac{\partial\big(r\,{}^e{\hskip -1pt}\Pi\big)}{\partial \theta}+\frac{1}{\mu u^2r\sin\theta}
\frac{\partial^2 \big(r\,{}^m{\hskip -1pt}\Pi\big)}{\partial r\partial \phi},
\label{eq:Bp-em0}
\end{eqnarray}
where  the electric and magnetic Debye potentials ${}^e{\hskip -1pt}\Pi$ and  ${}^m{\hskip -1pt}\Pi$ satisfy the wave equations (\ref{eq:Pi-eq+wew1*+-app}), given as
\begin{eqnarray}
\Big(\Delta+\epsilon\mu\,k^2u^4-\sqrt{\epsilon}u\big(\frac{1}{\sqrt{\epsilon}u}\big)''\Big)\Big[\frac{\,{}^e{\hskip -1pt}\Pi}{\sqrt{\epsilon}u}\Big]={\cal O}(r_g^2, r^{-3}),
\label{eq:Pi-eq+wew1*+}\\
\Big(\Delta+\epsilon\mu\,k^2u^4-\sqrt{\mu}u\big(\frac{1}{\sqrt{\mu}u}\big)''\Big)\Big[\frac{\,{}^m{\hskip -1pt}\Pi}{\sqrt{\mu}u}\Big]={\cal O}(r_g^2, r^{-3}).
\label{eq:Pi-eq+wmw1*}
\end{eqnarray}

Expressions (\ref{eq:Dr-em0})--(\ref{eq:Pi-eq+wmw1*}) represent the solution of the Mie problem in terms of Debye potentials \cite{Mie:1908,Born-Wolf:1999}, in the presence of the gravitational field of a mass monopole taken at the first post-Newtonian approximation of the general theory of relativity \cite{Turyshev:2017,Turyshev-Toth:2017} and a steady-state, spherically symmetric distribution of the free electron solar plasma (\ref{eq:n-eps_n-ism}).

For the quantities $\epsilon$ and $u$, given correspondingly by (\ref{eq:eps}) and (\ref{eq:pot_w_1**}), we can rewrite (\ref{eq:Pi-eq+wew1*+}) as the time-independent Schr\"odinger equation that describes the scattering of light by a Coulomb potential and in the presence of plasma:
{}
\begin{eqnarray}
\Big(\Delta +k^2\big(1+\frac{2r_g}{r}-\frac{\omega_{\tt p}^{ 2}(r)}{\omega^2}\big)+\frac{r_g}{r^3}-\frac{(\omega_{\tt p}^2)''}{4\omega^2}\Big)\Big[\frac{\,{}^e{\hskip -1pt}\Pi}{\sqrt{\epsilon}u}\Big]={\cal O}(r_g^2, r^{-3}).
\label{eq:Pi-eq*0+*}
\end{eqnarray}
A similar equation may be obtained for ${}^m{\hskip -1pt}\Pi/\sqrt{\mu}u$  from (\ref{eq:Pi-eq+wmw1*}), which, with $\mu=1$, takes the form:

{}
\begin{eqnarray}
\Big(\Delta +k^2\big(1+\frac{2r_g}{r}-\frac{\omega_{\tt p}^{ 2}(r)}{\omega^2}\big)+\frac{r_g}{r^3}\Big)\Big[\frac{\,{}^m{\hskip -1pt}\Pi}{\sqrt{\mu}u}\Big]={\cal O}(r_g^2, r^{-3}).
\label{eq:Pi-eq+wmw1*+*}
\end{eqnarray}

Eqs.~(\ref{eq:Pi-eq*0+*}) and (\ref{eq:Pi-eq+wmw1*+*}) are almost identical, except for the last term in (\ref{eq:Pi-eq*0+*}), which comes from $\epsilon$ introduced by (\ref{eq:eps}). To demonstrate that this difference is negligible, we note that, as seen from (\ref{eq:eps}) together with (\ref{eq:n-eps_n-ism}), in addition to the purely Newtonian potential of a static relativistic monopole that behaves as $1/r$,  (\ref{eq:Pi-eq*0+*}) has the plasma potential that contains several terms that decay either as $\propto r^{-2}$ or faster. Recognizing that $\omega=kc$ and using the plasma model (\ref{eq:n-eps_n-ism}) in the expression (\ref{eq:eps}) for $\epsilon$, these extra terms may be given as
{}
\begin{eqnarray}
k^2\frac{\omega_{\tt p}^2}{\omega^2}+\frac{(\omega_{\tt p}^2)''}{4\omega^2}\quad \Rightarrow \quad \frac{\omega_{\tt p}^2}{c^2}+\frac{(\omega_{\tt p}^2)''}{4k^2c^2}
=\frac{4\pi e^2}{m_ec^2}
\sum_i \alpha_i \Big(\frac{R_\odot}{r}\Big)^{\beta_i}\Big\{1+
\frac{\beta_i(\beta_i+1)}{4k^2R_\odot^2}
\Big(\frac{R_\odot}{r}\Big)^2\Big\}.
\label{eq:V-sr*+}
\end{eqnarray}

The two terms in the curly brackets of (\ref{eq:V-sr*+}) represent the repulsive potentials due to plasma that, based on the model (\ref{eq:n-eps_n-ism}), vanish as $1/r^2$ or faster.  The second plasma term in this expression is dominated by a factor of $(kR_\odot)^{-2}$, which, given the large value of the solar radius, makes its contribution negligible, especially at optical wavelengths ($\lambda\sim1~\mu$m), for which $(kR_\odot)^2\sim 5.32\times 10^{-32}$ \cite{Turyshev-Toth:2018-plasma}.  Thus, the term $\propto (\omega_{\tt p}^2)''/\omega^2$ may be neglected. Although the remaining terms are small, they may contribute to the phase shifts of the scattered wave and, therefore, they may affect the diffraction of light by the Sun. Thus, we retain these terms for further analysis. As a result, we introduce the steady-state, spherically symmetric plasma potential which,  to ${\cal O}\big((kR_\odot)^{-2}\big)$, is given as
{}
\begin{equation}
V_{\tt p}({ r})=\frac{\omega_{\tt p}^{ 2}({ r})}{c^2}=\frac{4\pi e^2}{m_ec^2}
\sum_i \alpha_i \Big(\frac{R_\odot}{r}\Big)^{\beta_i}+{\cal O}\big((kR_\odot)^{-2}\big).
\label{eq:plasma-mod}
\end{equation}

Also, we note that the last term in (\ref{eq:Pi-eq*0+*}) and (\ref{eq:Pi-eq+wmw1*+*}), representing the $1/r^3$ tail of the gravitational potential, may be discarded as insignificant (see relevant discussion  in Appendix~\ref{sec:rad_eq_wkb} and also in Appendix~F of Ref.~\cite{Turyshev-Toth:2017}).

As a result, and taking into account that magnetic permeability $\mu$ is constant,  (\ref{eq:Pi-eq*0+*}) and (\ref{eq:Pi-eq+wmw1*+*}) take an identical form:
{}
\begin{eqnarray}
\Big(\Delta +k^2\big(1+\frac{2r_g}{r}\big)-V_{\tt p}({ r})\Big)\Big[\frac{\Pi}{u}\Big]={\cal O}(r_g^2, r^{-3}),
\label{eq:Pi-eq*0+*+}
\end{eqnarray}
where the plasma potential $V_{\tt p}$ is given by (\ref{eq:plasma-mod}) and
the quantity $\Pi$ represents either the electric Debye potential, $\,{}^e{\hskip -1pt}\Pi/\sqrt{\epsilon}$, or its magnetic counterpart, $\,{}^m{\hskip -1pt}\Pi/\sqrt{\mu}$,  namely
\begin{equation}
\Pi({\vec r})=\Big(\frac{\,{}^e{\hskip -1pt}\Pi}{\sqrt{\epsilon}}; \frac{\,{}^m{\hskip -1pt}\Pi}{\sqrt{\mu}} \Big).
\label{eq:Pem}
\end{equation}

Therefore, the set of equations (\ref{eq:Dr-em0})--(\ref{eq:Bp-em0}) with  (\ref{eq:Pi-eq*0+*+}) and (\ref{eq:Pem}) is greatly simplified, as now we need to solve only one equation (\ref{eq:Pi-eq*0+*+}), which ultimately determines the Debye potential for the entire problem.

\section{Solution for the EM field}
\label{sec:sol-EM-Deb}

\subsection{Separating variables in the equation for the Debye potential}
\label{sec:sep-var}

Typically \cite{Born-Wolf:1999}, in spherical polar coordinates,
Eq.~(\ref{eq:Pi-eq*0+*+}) is solved by separating variables \cite{Turyshev-Toth:2017,Turyshev-Toth:2018-plasma}:
\begin{eqnarray}
\frac{\Pi}{u}=\frac{1}{r}R(r)\Theta(\theta)\Phi(\phi),
\label{eq:Pi*}
\end{eqnarray}
with integration constants and coefficients that are determined by boundary conditions. Direct substitution into (\ref{eq:Pi-eq*0+*}) reveals that the functions $R, \Theta$ and $\Phi$ must satisfy the following ordinary differential equations:
{}
\begin{eqnarray}
\frac{d^2 R}{d r^2}+\Big(k^2(1+\frac{2r_g}{r}) -\frac{\alpha}{r^2}-V_{\tt p}({r})\Big)R&=&{\cal O}(r_g^2,r_g\frac{\omega_{\tt p}^2}{\omega^2}),
\label{eq:R-bar*}\\
\frac{1}{\sin\theta}\frac{d}{d \theta}\Big(\sin\theta \frac{d \Theta}{d \theta}\Big)+\big(\alpha-\frac{\beta}{\sin^2\theta}\big)\Theta&=&{\cal O}(r_g^2,r_g\frac{\omega_{\tt p}^2}{\omega^2}),
\label{eq:Th*}\\
\frac{d^2 \Phi}{d \phi^2}+\beta\Phi&=&{\cal O}(r_g^2,r_g\frac{\omega_{\tt p}^2}{\omega^2}).
\label{eq:Ph*}
\end{eqnarray}

The solution to (\ref{eq:Ph*}) is given as usual \cite{Born-Wolf:1999}:
{}
\begin{eqnarray}
\Phi_m(\phi)=e^{\pm im\phi}  \quad\rightarrow \quad \Phi_m(\phi)=a_m\cos (m\phi) +b_m\sin (m\phi),
\label{eq:Ph_m}
\end{eqnarray}
where $\beta=m^2$, $m$ is an integer and $a_m$ and $b_m$ are integration constants.

Equation (\ref{eq:Th*}) is well known for spherical harmonics. Single-valued solutions to this equation exist when $\alpha=l(l+1)$ with ($l>|m|,$ integer). With this condition, the solution to (\ref{eq:Th*}) becomes
{}
\begin{eqnarray}
\Theta_{lm}(\theta)&=&P^{(m)}_l(\cos\theta).
\label{eq:Th_lm}
\end{eqnarray}

Now we focus on the equation for the radial function (\ref{eq:R-bar*}), where, because of (\ref{eq:Th*}), we have $\alpha=\ell(\ell+1)$.  As a result, (\ref{eq:R-bar*}) takes the form
{}
\begin{eqnarray}
\frac{d^2 R}{d r^2}+\Big(k^2(1+\frac{2r_g}{r})-\frac{\ell(\ell+1)}{r^2}-V_{\tt p}(r)\Big)R&=&{\cal O}(r^2_g,r_g\frac{\omega_{\tt p}^2}{\omega^2}).
\label{eq:R-bar-k*}
\end{eqnarray}

This equation describes light scattering that is dominated by a spherical relativistic potential due to a gravitational monopole  (which is equivalent to an attractive Coulomb potential discussed in quantum mechanics \cite{Schiff:1968,Landau-Lifshitz:1989,Messiah:1968}).

To determine the solution to (\ref{eq:R-bar-k*}), similarly to \cite{Turyshev-Toth:2018-plasma}, we first separate the terms in plasma potential $V_{\tt p}$, (\ref{eq:plasma-mod}), by isolating the $1/r^2$ term and representing the remaining terms as the short-range potential $V_{\tt sr}$:
{}
\begin{eqnarray}
V_{\tt p}(r)&=& \frac{\mu^2}{r^2}+ V_{\tt sr},
\label{eq:V-sr-m2]}
\end{eqnarray}
where $\mu^2$ and $V_{\tt sr}$ are given by
{}
\begin{eqnarray}
\mu^2&=&\frac{4\pi e^2R^2_\odot}{m_ec^2} \alpha_2, \qquad {} V_{\tt sr}\,=\,\frac{4\pi e^2}{m_ec^2}
\sum_{i>2} \alpha_i \Big(\frac{R_\odot}{r}\Big)^{\beta_i}+{\cal O}\big((kR_\odot)^{-2}\big),
\label{eq:V-sr-m2}
\end{eqnarray}
where $\mu^2$ is\footnote{Note that following \cite{Turyshev-Toth:2018-plasma}, we reuse the symbol $\mu$; do not confuse it with magnetic permeability.} the strength of the $1/r^2$ term in the plasma model at $r=R_\odot$. Using the values from the phenomenological model (\ref{eq:model}),
we can evaluate this term: $\mu^2\simeq 5.89\times 10^{15}$.  Also, we note that the range of $V_{\tt sr}$ is very short. In fact, as we see from Fig.~\ref{fig:plasma}, this potential provides a negligible contribution after $r\simeq 8 R_\odot$. Nevertheless, as it propagates through the solar system, light acquires the largest phase shift as it travels through the range of validity of this potential.

Note that if the model (\ref{eq:plasma-mod}) were to have the term $\propto \alpha_1 (R_\odot/r)$, this would imply the presence in (\ref{eq:R-bar-k*}) of another Coulomb potential of the type $2\mu_1/r$, where  $2\mu_1=({4\pi e^2R_\odot}/{m_ec^2}) \alpha_1$. The presence of such a term may be easily accounted for by modifying the $r_g$ term in (\ref{eq:R-bar-k*}) as $r_g\rightarrow r_g-\mu_1$, with all other calculations unchanged. However, the current observations \cite{Muhleman-etal:1977,Tyler-etal:1977,Muhleman-Anderson:1981,Giampieri:1994kj,Bertotti-Giampieri:1998,DSN-handbook-2017,Verma-etal:2013} suggest that such a term must be  absent in the model, thus $\alpha_1=\mu_1=0$.

The separation of the terms performed in the plasma potential (\ref{eq:V-sr-m2]})--(\ref{eq:V-sr-m2}) allows us to appropriately present the radial equation (\ref{eq:R-bar-k*}) as
{}
\begin{eqnarray}
\frac{d^2 R_L}{d r^2}+\Big(k^2(1+\frac{2r_g}{r})-\frac{L(L+1)}{r^2}- V_{\tt sr}(r)\Big)R_L&=&{\cal O}\big(r^2_g,r_g\frac{\omega_{\tt p}^2}{\omega^2}\big),
\label{eq:R-bar-k*2}
\end{eqnarray}
where the new index $L$ for the plasma-modified centrifugal potential is determined from
\begin{equation}
L(L+1)=\ell(\ell+1)+\mu^2.
\label{eq:L}
\end{equation}
Representing (\ref{eq:L}) equivalently as $(L+{\textstyle\frac{1}{2}})^2=(\ell+{\textstyle\frac{1}{2}})^2+\mu^2$ and requiring that when $\mu\rightarrow0$, the index $L$ must behave as $L\rightarrow \ell$, we find the solution to (\ref{eq:L}):
{}
\begin{equation}
L=\ell+\frac{\mu^2}{\sqrt{(\ell+{\textstyle\frac{1}{2}})^2+\mu^2}+\ell+{\textstyle\frac{1}{2}}}.
\label{eq:L2}
\end{equation}
When $\mu/\ell\ll1$, this solution behaves as
{}
\begin{equation}
L=\ell+\frac{\mu^2}{2\ell+1}-\frac{\mu^4}{(2\ell+1)^3}+\frac{2\mu^{6}}{(2\ell+1)^5}+{\cal O}(\mu^{8}/\ell^7)\approx \ell+\frac{\mu^2}{2\ell+1} +{\cal O}(\mu^{4}/\ell^3).
\label{eq:L2-apr}
\end{equation}

For a typical region where the plasma potential (\ref{eq:n-eps_n-ism}) is present,  the value of $\ell$ may be estimated using its relation to the classical impact parameter, namely $\ell=kb\geq kR_\odot=4.37\times 10^{15}$ at near optical wavelengths. Therefore, using the result for $\mu^2$ given above, we see that the ratio $\mu/\ell\leq 1.75\times 10^{-8}$ is indeed small while $\mu^2/\ell={\cal O}(1)$, justifying the approximation (\ref{eq:L2-apr}), as the order term is $\ell^2$ times smaller than the leading term.

\subsection{Eikonal solution for Debye potential}
\label{sec:eik}

Equations similar to  (\ref{eq:Pi-eq*0+*+}) are typical for many problems of nuclear scattering. However, no exact solution is known for an arbitrary short-range potential $V_{\tt p}$ (\ref{eq:V-sr-m2]})--(\ref{eq:V-sr-m2}). This motivated the development of various approximation tools \cite{Friedrich-book-2006,Friedrich-book-2013}.  One such approximation is known for the case of short-range potentials that decay faster than $1/r^2$, where a small parameter is introduced and the total solution to (\ref{eq:Pi-eq*0+*+}) is presented as a series expansion with respect to this parameter.  This method is called the Born approximation (BA)  \cite{Newton-book-2013}. The method uses the radial Green's function solution to (\ref{eq:R-bar-k*}) (obtained while setting $V_{\tt p}=0$) to determine each successive term in the expansion.
The final solution determines the cumulative phase shift for the EM wave as it traverses the area where the short-range scattering potential is present. Since this is a Born-type approximation for the phase shifts relative to the plasma-free wave, the relevant approximation is referred to as the distorted-wave Born approximation (DWBA). It determines the additional phase shift due to the short-range potential $V_{\tt sr}$ \cite{Klarsfeld:1966,Semon-Taylor:1975}.
However, it is known that for potentials that behave as $1/r^2$, this approximation leads to divergent results, as such potentials do not decay fast enough with distance. This is precisely our situation, where the plasma potential contains the $1/r^2$ terms. Thus, neither the BA nor the DWBA are particularly useful for our purposes.

To solve (\ref{eq:Pi-eq*0+*+}), we follow the approach presented in \cite{Turyshev-Toth:2018-plasma} where we developed a method that relies on the properties of the short-range plasma potential and the eikonal (or high-energy) approximation. The region of scattering of high frequency EM waves on the plasma-induced potential  $V_{\tt p}$ is bounded by the heliocentric distance to the heliopause, $R_\star$ from (\ref{eq:n-eps_n-ism}). We implement the eikonal approximation \cite{Akhiezer-Pomeranchuk:1950,Glauber-Matthiae:1970,Semon-Taylor:1977,Sharma-etal:1988,Sharma-Somerford:1990,Sharma-Sommerford-book:2006}. In this approximation, the short-range plasma potential contributes only a phase shift to the EM wave which can be directly calculated. Here we extend the method introduced in \cite{Turyshev-Toth:2018-plasma} on the curved spacetime induced by the solar gravitational mass monopole.

\subsubsection{Solution with short-range potential $V_{\tt sr}$  absent}
\label{sec:short-range=0}

No analytical solution is known to exist for Eq.~(\ref{eq:Pi-eq*0+*+}) in the general case when $V_{\tt sr}\ne 0$. Therefore, we seek a suitable approximation method. A number of methods were developed to solve equations of this type in scattering problems in quantum mechanics. At large incident energies, for a wavefront moving in the forward direction, a very useful method is the eikonal approximation \cite{Akhiezer-Pomeranchuk:1950,Glauber-Matthiae:1970,Semon-Taylor:1977,Sharma-etal:1988,Sharma-Somerford:1990,Sharma-Sommerford-book:2006}. The eikonal approximation is valid when the following two criteria are satisfied \cite{Sharma-Sommerford-book:2006}:
$kb\gg 1$ and $V_{\tt sr}(r)/k^2\ll 1$, where $k$ is the wave number and $b$ is the impact parameter. In our case, both of these conditions are fully satisfied. The first condition yields
$kb=4.37\times 10^{15}\,(\lambda/1\,\mu{\rm m})(b/R_\odot) \gg 1$. Taking the short-range plasma potential $V_{\tt sr}$ from (\ref{eq:V-sr-m2}), we evaluate  the second condition as $V_{\tt sr}(r)/k^2\leq V_{\tt sr}(R_\odot)/k^2\approx 4.07\times 10^{-13} \,(\lambda/1\,\mu{\rm m})^2\ll1$. Therefore, we may proceed.

To develop a solution to (\ref{eq:Pi-eq*0+*+}) using the eikonal approximation, we first note that when the short-range potential $V_{\tt sr}$ is absent, (\ref{eq:R-bar-k*2})  takes the form
{}
\begin{eqnarray}
\frac{d^2 R_L}{d r^2}+\Big(k^2(1+\frac{2r_g}{r})-\frac{L(L+1)}{r^2}\Big)R_L&=&0.
\label{eq:R-bar-k*20}
\end{eqnarray}
The solution to this equation is well known and is given in terms of the Coulomb functions $F_L(kr_g,kr)$ and $G_L(kr_g,kr)$ \cite{Abramovitz-Stegun:1965,Schiff:1968,Landau-Lifshitz:1989,Messiah:1968,Turyshev-Toth:2017} (the presence of these functions is the main difference from the situation encountered in \cite{Turyshev-Toth:2018-plasma}, where a similar equation, but without the Coulomb $r_g/r$ term,  is solved in terms of the Riccati--Bessel functions):
{}
\begin{eqnarray}
R^{(2)}_L=c_LF_L(kr_g,kr)+d_LG_L(kr_g,kr),
\end{eqnarray}
where we use the superscript ${(2)}$ to indicate that the solution to (\ref{eq:R-bar-k*20}) includes the inverse square term, $1/r^2$, from the plasma potential, which is represented by index $L$ from (\ref{eq:L2-apr}) \cite{Turyshev-Toth:2018-plasma}. When the solution for $R^{(2)}_L$ is known, we combine results for $\Phi(\phi)$, $\Theta(\theta)$, given by (\ref{eq:Ph_m}) and (\ref{eq:Th_lm}), to obtain the corresponding Debye potential, $\Pi^{(2)}({\vec r})$, in the form
{}
\begin{eqnarray}
\Pi^{(2)}({\vec r})&=&\frac{1}{r} \sum_{\ell=0}^\infty\sum_{m=-\ell}^\ell
\mu_\ell R^{(2)}_L(r)\big[ P^{(m)}_l(\cos\theta)\big]\big[a_m\cos (m\phi) +b_m\sin (m\phi)\big],
\label{eq:Pi-degn-sol-00}
\end{eqnarray}
where $L=L(\ell)$ is given by (\ref{eq:L2-apr}) and $\mu_\ell, a_m, b_m$ are arbitrary and as yet unknown constants to be determined later. This solution is well estublished and can be studied with available analytical tools (e.g., \cite{Born-Wolf:1999}).

Examining (\ref{eq:Pi-eq*0+*+}), we see that $\Pi^{(2)}({\vec r})$ is a solution to the following wave equation:
{}
\begin{eqnarray}
\Big(\Delta +k^2(1+\frac{2r_g}{r})-\frac{\mu^2}{r^2}\Big)\Pi^{(2)}({\vec r})=0,
\label{eq:Pi-eq*0+*+0}
\end{eqnarray}
which is the equation for the ``free'' Debye potential in the presence of gravity and $1/r^2$ plasma, $\Pi^{(2)}({\vec r})$, and which is yet ``unperturbed'' by the short-range plasma potential, $V_{\tt sr}$.

\subsubsection{Eikonal wavefunction}
\label{sec:eik-wfr}

We may now proceed with solving  (\ref{eq:Pi-eq*0+*+}), given the relevant form of $V_{\tt sr}$, (\ref{eq:V-sr-m2}), first representing this equation as
{}
\begin{eqnarray}
\Big(\Delta +k^2(1+\frac{2r_g}{r})-\frac{\mu^2}{r^2}-V_{\tt sr}({ r})\Big)\Pi({\vec r})={\cal O}(r_g^2, r^{-3}).
\label{eq:Pi-eq*0+*+1*}
\end{eqnarray}

To apply the eikonal approximation to solve (\ref{eq:Pi-eq*0+*+1*}), we consider a trial solution in the form
{}
\begin{eqnarray}
\Pi({\vec r})=\Pi^{(2)}({\vec r})\phi(\vec r),
\label{eq:Pi-eq*0+*+1}
\end{eqnarray}
where $\Pi^{(2)}({\vec r})$ is the ``free'' Debye potential (\ref{eq:Pi-degn-sol-00}). In other words, in the eikonal approximation the Debye potential $\Pi^{(2)}(\vec r)$, becomes ``distorted'' in the presence of the potential $V_{\tt sr}$ given in Eq.~(\ref{eq:V-sr-m2}), by $\phi$, a slowly varying function of $r$, such that
\begin{equation}
\left| \nabla^2 \phi \right|\ll k\left|\nabla \phi\right|.
\label{eq:eik2h+}
\end{equation}

When substituted into (\ref{eq:Pi-eq*0+*+1*}), the trial solution (\ref{eq:Pi-eq*0+*+1}) yields
{}
\begin{eqnarray}
\Big\{\Delta \Pi^{(2)}({\vec r})+\Big(k^2(1+\frac{2r_g}{r})-\frac{\mu^2}{r^2}\Big)\Pi^{(2)}({\vec r})\Big\}\phi({\vec r})+\Pi^{(2)}({\vec r})\Delta \phi({\vec r})&+&\nonumber\\
+\,2\big({\vec \nabla}\Pi^{(2)}({\vec r})\cdot{\vec \nabla} \phi({\vec r})\big)
- V_{\tt sr}({\vec r})\Pi^{(2)}({\vec r})\phi(\vec r)&=&{\cal O}(r_g^2, r^{-3}).
\label{eq:eik4h0-nab}
\end{eqnarray}

As $\Pi^{(2)}({\vec r})$ is the solution of the homogeneous equation (\ref{eq:Pi-eq*0+*+0}), the first term in (\ref{eq:eik4h0-nab}) is zero. Then, we neglect the second term, $\Pi^{(2)}({\vec r})\Delta \phi({\vec r})$, because of (\ref{eq:eik2h+}). As a result, from the last two terms we have
{}
\begin{eqnarray}
\big({\vec \nabla}\ln \Pi^{(2)}({\vec r})\cdot {\vec \nabla} \ln \phi(\vec r)\big)={\tfrac{1}{2}} V_{\tt sr}({\vec r})+{\cal O}(r_g^2, r^{-3}).
\label{eq:eik4h}
\end{eqnarray}

As we discussed above, the plasma contribution is rather small and it is sufficient  to keep the terms that are first order in $\omega^2_{\tt p}/\omega^2$. Thus, to formally solve (\ref{eq:eik4h}) we may present the solution for $\Pi^{(2)}({\vec r})$ from (\ref{eq:Pi-degn-sol-00}) in series form in terms of the small parameter $\mu/\ell$, which enters there via the index $L$ as shown by
(\ref{eq:L2-apr}). Then, to solve (\ref{eq:eik4h})  it is sufficient to take only the zeroth order term (i.e., with $\mu=0$) in $\Pi^{(2)}({\vec r})$. It is easier, however, to obtain such a solution directly from  (\ref{eq:Pi-eq*0+*+0}) by setting $\mu=0$, which yields the well-known solution for the incident wave
in the presence of a gravitational monopole (see Eq.~(23) in \cite{Turyshev-Toth:2017}).
{}
\begin{eqnarray}
\Pi^{(2)}({\vec r})=e^{\pm ik(z-r_g\ln k(r-z))} +{\cal O}(r_g^2, \omega^2_{\tt p}/\omega^2).
\label{eq:eik4mu*}
\end{eqnarray}

To compute the gradient of $\Pi^{(2)}({\vec r})$, following \cite{Turyshev-Toth:2017}, we represent the unperturbed trajectory of a ray of light as
{}
\begin{eqnarray}
\vec{r}(t)&=&\vec{r}_{0}+\vec{k}c(t-t_0)+{\cal O}(r_g,\omega_{\tt p}^2/\omega^2),
\label{eq:x-Newt0}
\end{eqnarray}
where $\vec k$ is the unit vector on the incident direction of the light ray's propagation path and $\vec r_0$ represents the starting point (see Fig.~\ref{fig:solsys}). Following \cite{Kopeikin:1997,Kopeikin-book-2011,Turyshev-Toth:2017}, we define ${\vec b}=[[{\vec k}\times{\vec r}_0]\times{\vec k}]$ to be the impact parameter of the unperturbed trajectory of the light ray. The vector ${\vec b}$ is directed from the origin of the coordinate system toward the point of the closest approach of the unperturbed path of light ray to that origin.

With (\ref{eq:x-Newt0}), we introduce the parameter $\tau=\tau(t)$ along the path of the light ray (see details in Appendix~\ref{sec:geodesics}):
{}
\begin{eqnarray}
\tau &=&({\vec k}\cdot {\vec r})=({\vec k}\cdot {\vec r}_{0})+c(t-t_0),
\label{eq:x-Newt*=0}
\end{eqnarray}
which may be positive or negative. Note that $\tau=z\cos\alpha$ where $\alpha$ being the angle between ${\vec e}_z$ and ${\vec k}$; $\tau=z$ when the $z$-axis of the chosen Cartesian coordinate system is oriented along the incident direction of the light ray. The new parameter $\tau$ allows us to rewrite (\ref{eq:x-Newt0}) as
{}
\begin{eqnarray}
{\vec r}(\tau)&=&{\vec b}+{\vec k} \tau+{\cal O}(r_g,\omega_{\tt p}^2/\omega^2),
\qquad {\rm with} \qquad ||{\vec r}(\tau)|| \equiv r(\tau) =\sqrt{b^2+\tau^2}+{\cal O}(r_g,\omega_{\tt p}^2/\omega^2).
\label{eq:b0}
\end{eqnarray}

Using  (\ref{eq:b0}), the gradient of $\Pi^{(2)}({\vec r})$ from (\ref{eq:eik4mu*}) may be computed as
{}
\begin{eqnarray}
{\vec \nabla}\ln \Pi^{(2)}({\vec r})=\pm ik\Big({\vec k}(1+\frac{r_g}{r})-\frac{r_g}{b^2}{\vec b}\big(1+\frac{\tau}{r}\big)\Big)+{\cal O}(r_g^2, \omega^2_{\tt p}/\omega^2).
\label{eq:eik4h+}
\end{eqnarray}

As a result, (\ref{eq:eik4h}) takes the form
{}
\begin{eqnarray}
\pm ik\Big(\big({\vec k}(1+\frac{r_g}{r})-\frac{r_g}{b^2}{\vec b}\big(1+\frac{\tau}{r}\big)\big)\cdot {\vec \nabla} \ln \phi(\vec r)\Big)={\tfrac{1}{2}} V_{\tt sr}({\vec r})+{\cal O}(r_g\omega^2_{\tt p}/\omega^2, r_g^2, \omega^4_{\tt p}/\omega^4, r^{-3}).
\label{eq:eik4h**0}
\end{eqnarray}

As we want to identify the largest plasma contribution to light propagation we keep only linear terms with respect to gravity and plasma. As a result, neglecting the $r_g$-dependent terms in (\ref{eq:eik4h**0}), we may present (\ref{eq:eik4h}) as
{}
\begin{eqnarray}
\pm ik({\vec k} \cdot {\vec \nabla}) \ln \phi={\tfrac{1}{2}} V_{\tt sr}+{\cal O}(r_g,\omega^4_{\tt p}/\omega^4).
\label{eq:eik4h**}
\end{eqnarray}

We may now compute the eikonal phase due to the short-range plasma potential $V_{\tt sr}$. Using the representation of the light ray's path as ${\vec r}=({\vec b},\tau)$ given by   (\ref{eq:b0}), we observe that (as was also shown in \cite{Turyshev-Toth:2017}) the gradient ${\vec \nabla}$ may be expressed in terms of the variables along the path as ${\vec \nabla}={\nabla}_b+{\vec k}\,d/d\tau +{\cal O}(r_g,\omega_{\tt p}^2/\omega^2)$, where ${\nabla}_b$ is the gradient along the direction of the impact parameter ${\vec b}$ and $\tau$ being the parameter taken along the path. Thus, the differential operator on the left side of (\ref{eq:eik4h**})  is the derivative along the light ray's path, namely  $({\vec k} \cdot {\vec \nabla})=d/d\tau$.

As a result, for (\ref{eq:eik4h**}) we have
{}
\begin{equation}
\frac{d\ln\phi^\pm}{d\tau} =\pm\frac{1}{2ik} V_{\tt sr}+{\cal O}(r_g,\omega^2_{\tt p}/\omega^2),
\label{eq:eik4h*}
\end{equation}
the solutions of which are
 \begin{eqnarray}
\phi^\pm({\vec b}, \tau)&=&\exp\Big\{\mp\frac{i}{2k}\int^{\tau}_{\tau_0}  V_{\tt sr}({\vec b},\tau') d\tau' \Big\}.
\label{eq:eik5h}
\end{eqnarray}
We therefore have the following two particular eikonal solutions of (\ref{eq:Pi-eq*0+*+1*}) for $\Pi(\vec r)$:
\begin{equation}
\Pi(\vec r)=\Pi^{(2)}(\vec r)\exp\Big\{\pm i \xi_b(\tau) \Big\}+{\cal O}(\omega^4_{\tt p}/\omega^4),
\label{eq:eik6h}
\end{equation}
where we introduced the eikonal phase
\begin{equation}
\xi_b(\tau) =-\frac{1}{2k}\int^{\tau}_{\tau_0}  V_{\tt sr}({\vec b},\tau') d\tau'.
\label{eq:eik7h}
\end{equation}

Given $V_{\tt sr}({\vec r})$ from (\ref{eq:V-sr-m2}), we reduced the problem to evaluating a single integral to determine the Debye potentiual $\Pi({\vec r})$ from (\ref{eq:Pi-eq*0+*+1}), which is a great simplification of the problem. Given the fact that ${\vec b}$ is constant and by taking the short-range plasma potential $V_{\tt sr}({\vec r})$ from (\ref{eq:V-sr-m2}), we evaluate (\ref{eq:eik7h}) as
{}
\begin{eqnarray}
\xi_b(r)
&=& -\frac{2\pi e^2R_\odot}{m_ec^2k}
\sum_{i>2}\alpha_i\Big(\frac{R_\odot}{b}\Big)^{\beta_i-1}\Big\{Q_{\beta_i}(\tau)-Q_{\beta_i}(\tau_0)\Big\},
\label{eq:delta-D*-av0WKB+1*}
\end{eqnarray}
where we introduced the function $Q_{\beta_i}(\tau)$, which, with $\tau=({\vec k}\cdot {\vec r})=\sqrt{r^2-b^2}$, is given as
\begin{eqnarray}
Q_{\beta_i}(\tau)={}_2F_1\Big[{\textstyle\frac{1}{2}},{\textstyle\frac{1}{2}}\beta_i,{\textstyle\frac{3}{2}},-\frac{\tau^2}{b^2}\Big]\frac{\tau}{b},
\label{eq:Q}
\end{eqnarray}
with ${}_2F_1[a,b,c,z]$ being the hypergeometric function \cite{Abramovitz-Stegun:1965}. For $r=b$ or, equivalently, for $\tau=0$, the function (\ref{eq:Q}) is well-defined, taking the value of $Q_{\beta_i}(0)=0$, for each $\beta_i$. For large values of $r$ and, thus for large $\tau$, for any given value of $\beta_i$, the function $Q_{\beta_i}(\tau)$ rapidly approaches a limit:
\begin{equation}
\lim\limits_{\tau\to\infty}Q_{\beta_i}\big(\tau\big)
=\lim\limits_{r\to\infty}Q_{\beta_i}\big(\sqrt{r^2-b^2}\big)
=Q^\star_{\beta_i}, \qquad {\rm where}\qquad Q^\star_{\beta_i}\equiv \frac{{\textstyle\frac{1}{2}}\beta_i}{\beta_i-1}B[{\textstyle\frac{1}{2}}\beta_i+{\textstyle\frac{1}{2}},{\textstyle\frac{1}{2}}],
\label{eq:eik1hQ}
\end{equation}
with $B[x,y]$ being Euler's beta function \cite{Turyshev-Toth:2018-plasma}.
For the values of $\beta_i$ used in the model (\ref{eq:model}) for the electron number density in the solar corona, $\beta_i=\{2,6,16\}$, these values are:
\begin{eqnarray}
Q^\star_{2}=\frac{\pi}{2}, \qquad Q^\star_{6}=\frac{3\pi}{16}, \qquad Q^\star_{16}=\frac{429\pi}{4096}.
\label{eq:Q-val}
\end{eqnarray}
Note that the quantities $Q_{\beta_i}(r)$ (\ref{eq:Q}) for $\beta_i>2$ are always small, $0\leq |Q_{\beta_i}|<1$, and as functions of $r$, they reach their asymptotic values $Q^\star_{\beta_i}$ (\ref{eq:eik1hQ}) quite rapidly, typically after $r\simeq3.2 b$ (thus, they may be treated as being constant for all the relevant distance ranges.)

Next, we place the source at a very large distance from the Sun: $|\tau_0|=\sqrt{r_0^2-b^2}\gg R_\star$ (see Fig.~\ref{fig:solsys}). Then, from definition (\ref{eq:Q}) and the asymptotic behavior given by (\ref{eq:eik1hQ}), we have $Q_{\beta_i}(\tau_0)=-Q^\star_{\beta_i}$. As a result, we express the total eikonal phase shift acquired by the wave along its path through the solar system (\ref{eq:delta-D*-av0WKB+1*}) as
{}
\begin{eqnarray}
\xi^{\rm path}_b(r)
&=& -\frac{2\pi e^2R_\odot}{m_ec^2k}
\sum_{i>2}\alpha_i\Big(\frac{R_\odot}{b}\Big)^{\beta_i-1}\Big\{Q^\star_{\beta_i}+Q_{\beta_i}\big(\sqrt{r^2-b^2}\big)\Big\}.
\label{eq:delta-D*-av0WKB+p}
\end{eqnarray}

Expression (\ref{eq:delta-D*-av0WKB+p}) is the total phase shift induced by the short-range plasma potential along the entire path of the EM wave as it propagates through the solar system. One may see that, as the light propagates from the source to the point of closest approach to the Sun, it acquires the first part of the phase shift, i.e.,  the term proportional to $Q^\star_{\beta_i}$ in (\ref{eq:delta-D*-av0WKB+p}). As it continues to propagate, the second term in (\ref{eq:delta-D*-av0WKB+p}) kicks in, providing an additional contribution.

Substituting  the solution that we obtained for the total eikonal phase shift $\xi_b(r)$ of (\ref{eq:delta-D*-av0WKB+p}) in (\ref{eq:eik6h}) results in the desired solution for the Debye potential $\Pi(\vec r)$. Effectively, this solution demonstrates that the phase of the EM wave is modified by the short-range plasma potential, as expected from the eikonal approximation. Although (\ref{eq:eik6h}) is the solution to (\ref{eq:Pi-eq*0+*+1*}), it still  has arbitrary constants  $\mu_\ell, a_m, b_m$ present in (\ref{eq:Pi-degn-sol-00}). These constants must be chosen to satisfy a particular boundary value problem that we set out to solve: Obtaining the solution for the EM field as it propagates through the solar system with the refractive medium given by (\ref{eq:n-eps_n-ism}).

\subsection{Solution for the radial function $R_L(r)$}

At this point, we already have all the key components  needed to develop the solution for the Debye potentials in the presence of a spherically symmetric gravitational field produced by the solar monopole, and the spherically symmetric solar plasma modeled by (\ref{eq:n-eps_n-ism}). As we observed above, with the short-range plasma potential (\ref{eq:V-sr-m2}), the equation for the radial function (\ref{eq:R-bar-k*}) takes the form (\ref{eq:R-bar-k*2}). Solving this equation leads to a solution for the Debye potential (\ref{eq:Pi*}). Following \cite{Turyshev-Toth:2017}, with the help of (\ref{eq:Pi*}),  a particular solution for the Debye potential, ${\Pi}$, is obtained by multiplying together the functions given by (\ref{eq:Ph_m}), (\ref{eq:Th_lm}) and $R_L$ from (\ref{eq:R-bar-k*2});
we then obtain a general solution to (\ref{eq:Pi-eq*0+*}). Specifically, by combining results for $\Phi(\phi)$ and $\Theta(\theta)$, given by (\ref{eq:Ph_m}) and (\ref{eq:Th_lm}), the solution for the Debye potential takes the form
{}
\begin{eqnarray}
\frac{\Pi}{u}&=&\frac{1}{r} \sum_{\ell=0}^\infty\sum_{m=-\ell}^\ell
\mu_\ell R_L(r)\big[ P^{(m)}_l(\cos\theta)\big]\big[a_m\cos (m\phi) +b_m\sin (m\phi)\big]+{\cal O}\big(r_g^2,k^{-2},r_g\frac{\omega_{\tt p}^2}{\omega^2}\big),
\label{eq:Pi-degn-sol-0}
\end{eqnarray}
where $L=L(\ell)$ is given by (\ref{eq:L2-apr}) and $\mu_\ell, a_m, b_m$ are arbitrary and as yet unknown constants.

As we discussed earlier, no analytic solution to (\ref{eq:R-bar-k*2}) for $R_L$ in the case of an arbitrary form of the short-range plasma potential  $V_{\tt sr}$ is known.  However, we may proceed with solving  (\ref{eq:Pi-degn-sol-0}) by relying on the eikonal approximation discussed in Sec.~\ref{sec:eik-wfr}. For this, we notice that in the plasma-free case (at the great heliocentric distances beyond the termination shock), the entire plasma potential $V_{\tt p}$ is absent, thus $L= \ell$.  The solution to the Maxwell field equations in this case is known and describes the scattering of the  EM waves by a gravitational monopole,  given in \cite{Turyshev-Toth:2017}. In that plasma-free case, to determine the coefficients $\mu_\ell$ in (\ref{eq:Pi-degn-sol-0}), we chose $R_\ell(r)$ to be the regular Coulomb wave function $F_\ell(kr_g, kr)$, and require that the resulting EM field match the incident Coulomb-modified plane EM wave.

As a result, in the vacuum, the solutions for the electric and magnetic potentials of the incident wave, ${}^e{\hskip -1pt}\Pi_0$ and ${}^m{\hskip -1pt}\Pi_0$, were found to be given in terms of a single potential $\Pi_0(r, \theta)$ (see  \cite{Turyshev-Toth:2017} for details):
{}
\begin{align}
  \left( \begin{aligned}
{}^e{\hskip -1pt}\Pi_0/\sqrt{\epsilon}& \\
{}^m{\hskip -1pt}\Pi_0/\sqrt{\mu}& \\
  \end{aligned} \right) =&  \left( \begin{aligned}
\cos\phi \\
\sin\phi  \\
  \end{aligned} \right) \,\Pi_0(r, \theta), &
\hskip 2pt {\rm where} \quad
\Pi_0 (r, \theta)=
\frac{E_0}{k^2}\frac{u}{r}\sum_{\ell=1}^\infty i^{\ell-1}\frac{2\ell+1}{\ell(\ell+1)}e^{i\sigma_\ell}
F_\ell(kr_g,kr) P^{(1)}_\ell(\cos\theta)+{\cal O}(r_g^2).
  \label{eq:Pi_ie*+*=}
\end{align}
In other words, the incident EM wave is not affected by the solar plasma, thus its form is identical to that of the free wave propagating in gravity, discussed in \cite{Turyshev-Toth:2017}.

Now, considering the plasma, we notice that,  for large $r$, the potential $ V_{\tt sr}(r)$ in (\ref{eq:R-bar-k*2}) can be neglected in comparison to the Coulomb potential $U_{\tt c}(r )=2k^2r_g/r$ and this equation reduces to the Coulomb equation discussed in \cite{Turyshev-Toth:2017} with the solution given by  (\ref{eq:Pi_ie*+*=}). The solution of (\ref{eq:R-bar-k*2}) that is regular at the origin can thus be written asymptotically as a linear combination of the regular and irregular Coulomb wave functions $F_L(kr_g, kr)$ and $G_L(kr_g, kr)$, respectively  \cite{Hull-Breit:1959,Friedrich-book-2006,Friedrich-book-2013,Burke-book-2011}, which are solutions of (\ref{eq:R-bar-k*2})
in the absence of the potential $V_{\tt sr}(r )$. Asymptotically, at large values of the argument $(kr)$, these functions behave as \cite{Turyshev-Toth:2017}
{}
\begin{eqnarray}
 F_L(kr_g, kr) &\sim& \sin\Big(k(r+r_g\ln2kr) +\frac{L(L+1)}{2kr}-\frac{\pi L}{2}+\sigma_L\Big),
 \label{eq:F-beh}\\
 G_L(kr_g, kr) &\sim& \cos\Big(k(r+r_g\ln2kr)+\frac{L(L+1)}{2kr}-\frac{\pi L}{2}+\sigma_L \Big).
\label{eq:G-beh}
\end{eqnarray}

Since the Coulomb potential falls off slower than the centrifugal potential (i.e., the $L(L+1)/r^2$ term in (\ref{eq:R-bar-k*2})) at large distances, it dominates the asymptotic behavior of the effective potential in every partial wave. Hence, we look for a solution satisfying the following boundary conditions \cite{Burke-book-2011}
{}
\begin{eqnarray}
R_L (r) &\underset{r\rightarrow 0}{\sim}& nr^{L+1},
\label{eq:R-anz0}\\
R_L (r) &\underset{r\rightarrow \infty}{\sim}&  F_L(kr_g, kr) + \tan \delta_\ell \,G_L(kr_g, kr)
\underset{kr\rightarrow \infty}{\propto}
\sin\Big(k(r+r_g\ln2kr)+\frac{L(L+1)}{2kr}-\frac{\pi L}{2}+\sigma_L +\delta_\ell\Big),
\label{eq:R-anz}
\end{eqnarray}
where $n$ is a normalization factor and $F_L(kr_g, kr)$ and $G_L(kr_g, kr)$ are solutions of (\ref{eq:R-bar-k*2}) in the absence of the potential $V_{\tt sr}(r)$, which, as we discussed above, are, respectively, regular and irregular at the origin. The real quantities $\delta_\ell(k)$ introduced by these equations are the phase shifts for spherically symmetric scattering \cite{Grandy-book-2005}  due to the short-range potential $V_{\tt sr}(r)$ (\ref{eq:V-sr-m2}) in the presence of the Coulomb potential $U_{\tt c}(r )=2k^2r_g/r$ in (\ref{eq:R-bar-k*2}). We note that $\delta_\ell(k)$ fully describes the non-Coulombic part of the scattering and vanishes when this short-range potential is not present. We generalized these expressions to the case where the additional plasma potential has an $1/r^2$ term, which creates an additional centrifugal potential in (\ref{eq:R-bar-k*2}) that was absorbed by the substitution $\ell\rightarrow L$.

We can satisfy the conditions (\ref{eq:R-anz0})--(\ref{eq:R-anz}) by choosing the  function $R_L(r)$ as a linear combination of the two solutions (\ref{eq:eik6h}). One way to do that is by relying on the two solutions to (\ref{eq:eik6h}) taken in the form of the incoming and outgoing waves \cite{Thomson-Nunes-book:2009}, which are given by the functions $H^{-}_L(kr_g,kr)$ and  $H^{+}_L(kr_g,kr)$, correspondingly, and to show explicit dependence on the eikonal phase shift, $\xi_b(r)$, which can be captured in the following form:
{}
\begin{eqnarray}
R_L (r) = \frac{1}{2i}\Big(H^+_L(kr_g,kr)e^{i \xi_b (r)} -H^-_L(kr_g,kr)e^{-i \xi_b(r)}\Big),
\label{eq:R-L}
\end{eqnarray}
where the Coulomb--Hankel functions $H_L^{(\pm)}$ are related to the Coulomb functions by $H_L^{\pm}(kr_g,kr)=G_L(kr_g,kr)\pm iF_L(kr_g,kr)$ (for discussion, see Appendix~A of \cite{Turyshev-Toth:2017}) and their asymptotic behavior given by (see Appendix F of \cite{Turyshev-Toth:2017}):
{}
\begin{equation}
H^\pm_L(kr_g,kr)\underset{kr\to\infty}{\sim} \exp\Big\{\pm i\Big(k(r+r_g\ln 2kr)+\frac{L(L+1)}{2kr}-\frac{\pi L}{2}+\sigma_L\Big)\Big\}.
\label{eq:free_u+}
\end{equation}
Clearly, using the approach demonstrated in Appendix~\ref{sec:rad_eq_wkb} and especially (\ref{eq:R_solWKB+=_bar-imp}), this expression may be extended to include terms with higher powers of $1/kr$. In addition, $\xi_b(r)$ in (\ref{eq:R-L}) is the eikonal phase shift that is accumulated by the EM wave starting from the point of closest approach, $r=b$. The expression for the quantity is obtained directly from  (\ref{eq:delta-D*-av0WKB+1*}) by setting $z_0=0$ (or, equivalently, from (\ref{eq:delta-D*-av0WKB+p}) by dropping the $Q^\star_{\beta_i}$ term),   which results in
{}
\begin{eqnarray}
\xi_b(r)
&=& -\frac{2\pi e^2R_\odot}{m_ec^2k}
\sum_{i>2}\alpha_i\Big(\frac{R_\odot}{b}\Big)^{\beta_i-1}Q_{\beta_i}\big(\sqrt{r^2-b^2}\big).
\label{eq:delta-D*-av0WKB+}
\end{eqnarray}

The form of the radial function $R_L$ from (\ref{eq:R-L}) captures our expectation that, in the presence of a potential $V_{\tt sr}(r)$ from (\ref{eq:V-sr-m2}),  the Coulomb--Hankel functions (which represent the radial free-particle wavefunction solutions of the homogeneous equation (\ref{eq:R-bar-k*20})), become ``distorted'' by this short-range potential.  Clearly, (\ref{eq:R-L})  satisfies the radial  equation (\ref{eq:R-bar-k*2}). We can verify that $R_L$ in the form of (\ref{eq:R-L}) also satisfies the asymptotic boundary conditions (\ref{eq:R-anz0})--(\ref{eq:R-anz}). Indeed, as the plasma potential exists only for $R_\odot \leq r\leq R_\star$ (which is evident from (\ref{eq:n-eps_n-ism})), the eikonal phase $\xi_b$ is zero for $r<R_\odot$. Therefore, as $r\rightarrow 0$, the index $L\rightarrow \ell$ and the radial function  (\ref{eq:R-L}) becomes $R_L (r) \rightarrow F_\ell(kr_g,kr)$, where the function  $F_\ell(kr_g,kr)$ obeys the condition (\ref{eq:R-anz0}). Next, we consider another limit, when $r\rightarrow \infty$. Using the asymptotic behavior of $H^\pm_L$ from (\ref{eq:free_u+}), we see that, as $r\rightarrow \infty$, the radial function obeys the asymptotic condition (\ref{eq:R-anz}) taking the form where the phase shift $\delta_\ell$ is given by the eikonal phase $\xi_{b}$ introduced by (\ref{eq:eik7h}).  As  a result, we established that the radial function (\ref{eq:R-L}) represents a desirable solution to (\ref{eq:R-bar-k*2}) inside the termination shock boundary, $0\leq r\leq R_\star$ and, of course, it is good choice for the radial function for the region outside the solar system, $r> R_\star$.

We may put the result (\ref{eq:R-L})  it in the following equivalent form:
{}
\begin{eqnarray}
R_L (r) &=&  \cos\xi_b(r)F_L(kr_g, kr) + \sin\xi_b(r)\,G_L(kr_g, kr),
\label{eq:R-L3}
\end{eqnarray}
which explicitly shows the phase shift, $\xi_b(r)$, induced by the short-range plasma potential, clearly satisfying the boundary condition (\ref{eq:R-anz})  with the quantity $\xi_b(r)$ from (\ref{eq:delta-D*-av0WKB+}) being the anticipated phase shift $\delta_\ell(k)$.

In conjunction with (\ref{eq:R-L3}), Eq.~(\ref{eq:Pi-degn-sol-0}) describes the potential inside the termination shock, $r<R_\star$. Outside the termination shock, $r>R_\star$, we model the solution for the Debye potential, as usual, as a combination of that of a Coulomb-modified plane wave and a scattered wave. These two solutions must be consistent on the boundary, that is, at $r=R_\star$.

To match the potentials (\ref{eq:Pi-degn-sol-0}) inside the termination shock with those of the incident and scattered waves outside, the latter must be expressed in a similar form but with arbitrary coefficients. Only the function $F_\ell(kr_g,kr)$ may be used in the expression for the potential inside the sphere (i.e., the termination shock boundary)  since $G_\ell(kr_g,kr)$ becomes infinite at the origin. On the other hand, the scattered wave must vanish at infinity. The Coulomb--Hankel functions $H^+_L(kr_g,kr)$ impart precisely this property. These functions are suitable as representations of scattered waves. For large values of the argument $kr$, the result behaves as  $e^{ik(r+r_g\ln2kr)}$ and the Debye potential $\Pi\propto e^{ik(r+r_g\ln2kr)}/r$ for large $r$. Thus, at large distances from the sphere the scattered wave is spherical (with the $\ln$ term in the phase due to the modification by the Coulomb potential), with its center at the origin  $r=0$. Accordingly, we use it in the expression for the scattered wave, i.e., in the trial solution for the Debye potentials of the scattered wave for $r>R_\star$.

Collecting results for the functions $\Phi(\phi)$ and $\Theta(\theta)$, respectively given by (\ref{eq:Ph_m}) and (\ref{eq:Th_lm}), and $R_L(r)=H^+_L(kr_g, kr)e^{i\xi_{b}(r)}$ from (\ref{eq:eik6h}), to the order of ${\cal O}(r_g^2,r_g{\omega_{\tt p}^2}/{\omega^2})$, we obtain the Debye potential for the scattered wave:
{}
\begin{eqnarray}
\Pi_{\tt s}&=&\frac{u}{r} \sum_{\ell=0}^\infty\sum_{m=-\ell}^\ell
a_\ell H^+_L(kr_g, kr)e^{i\xi_b(r)}\big[ P^{(m)}_\ell(\cos\theta)\big]\big[a'_m\cos (m\phi) +b'_m\sin (m\phi)\big],
\label{eq:Pi-degn-sol-s}
\end{eqnarray}
where $a_\ell, a'_m, b'_m$ are arbitrary and as yet unknown constants and the relation between $L$ and $\ell$ is given by (\ref{eq:L2-apr}).

Representing the potential inside the termination shock via $F_\ell(kr_g,kr)$ is appropriate. The trial solution to (\ref{eq:Pi-eq*0+*+}) for the electric and magnetic Debye potentials inside the termination shock boundary (i.e., $0\leq r \leq R_\star$) relies on the radial function $R_L(r)$ given by (\ref{eq:R-L3}) and has the form
{}
\begin{eqnarray}
\Pi_{\tt in}&=&\frac{u}{r} \sum_{\ell=0}^\infty\sum_{m=-\ell}^\ell
b_\ell \Big\{ \cos\xi_{b}(r)F_L(kr_g, kr) + \sin\xi_{b}(r)\,G_L(kr_g, kr)\Big\}\big[ P^{(m)}_\ell(\cos\theta)\big]\big[a_m\cos (m\phi) +b_m\sin (m\phi)\big],
~~~~~
\label{eq:Pi-degn-sol-in}
\end{eqnarray}
where $b_\ell, a_m, b_m$ are arbitrary and yet unknown constants.

The boundary (continuity) conditions mentioned in Appendix~\ref{app:Debye} (see also discussion in \cite{Born-Wolf:1999}), imposed on the quantities (\ref{eq:bound_cond}) at the termination shock boundary $r=R_\star$, using the electron plasma distribution (\ref{eq:eps}) and (\ref{eq:n-eps_n-ism}) with $n_0=0$ and, thus, with $\epsilon(R_\star)=\mu(R_\star)=1$, are written in full as
{}
\begin{eqnarray}
\frac{\partial }{\partial r}\Big[\frac{r\,{}^e{\hskip -1pt}\Pi_0}{u}+\frac{r{}^e{\hskip -1pt}\Pi_{\tt s}}{\sqrt{\epsilon}u}\Big]\Big|^{}_{r=R_\star}&=&\frac{\partial }{\partial r}\Big[\frac{r\,{}^e{\hskip -1pt}\Pi_{\tt in}}{\sqrt{\epsilon}u}\Big]\Big|^{}_{r=R_\star},
\label{eq:bound_cond-expand1+}\\[3pt]
\frac{\partial }{\partial r}\Big[\frac{r\,{}^m{\hskip -1pt}\Pi_0}{u}+\frac{r{}^m{\hskip -1pt}\Pi_{\tt s}}{\sqrt{\mu}u}\Big]\Big|^{}_{r=R_\star}&=&\frac{\partial }{\partial r}\Big[\frac{r\,{}^m{\hskip -1pt}\Pi_{\tt in}}{\sqrt{\mu}u}\Big]\Big|^{}_{r=R_\star},
\label{eq:bound_cond-expand2+}\\[3pt]
\Big[ \frac{r\,{}^e{\hskip -1pt}\Pi_0}{u}+\frac{r{}^e{\hskip -1pt}\Pi_{\tt s}}{\sqrt{\epsilon}u}\Big]\Big|^{}_{r=R_\star}&=&\Big[\frac{r\,{}^e{\hskip -1pt}\Pi_{\tt in}}{\sqrt{\epsilon}u}\Big]\Big|^{}_{r=R_\star},
\label{eq:bound_cond-expand3+}\\[3pt]
\Big[\frac{r\,{}^m{\hskip -1pt}\Pi_0}{u}+\frac{{}^m{\hskip -1pt}\Pi_{\tt s}}{\sqrt{\mu}u}\Big]\Big|^{}_{r=R_\star}&=&\Big[\frac{r\,{}^m{\hskip -1pt}\Pi_{\tt in}}{\sqrt{\mu}u}\Big]\Big|^{}_{r=R_\star}.
\label{eq:bound_cond-expand4+}
\end{eqnarray}

We now make use of the symmetry of the geometry of the problem \cite{Born-Wolf:1999} and by applying the boundary conditions (\ref{eq:bound_cond-expand1+})--(\ref{eq:bound_cond-expand4+}). We recall that we can use a single Debye potential $\Pi$ in (\ref{eq:Pi-degn-sol-s}) and (\ref{eq:Pi-degn-sol-in}) to represent electric and magnetic fields (\ref{eq:Pem}). We find that the constants $a_m$ and $b_m$ for the electric Debye potentials are $a_1=1$, $b_1=0$ and $a_m=b_m=0$ for $m\ge 2$. For the magnetic Debye potentials, we obtain $a_1=0$, $b_1=1$ and $a_m=b_m=0$ for $m\ge 2$. The values are identical for $a'_m$ and $b'_m$.

As a result, the solutions for the electric and magnetic potentials of the scattered wave (for the region $r>R_\star$, where, based on the plasma model (\ref{eq:n-eps_n-ism}),  $\epsilon=\mu\equiv1$), ${}^e{\hskip -1pt}\Pi_{\tt s}$ and ${}^m{\hskip -1pt}\Pi_{\tt s}$, may be given in terms of a single potential $\Pi_{\tt s}(r, \theta)$ (see  \cite{Turyshev-Toth:2017} for details), which, to ${\cal O}(r_g^2)$, is given by
\begin{align}
  \left( \begin{aligned}
{}^e{\hskip -1pt}\Pi_{\tt s}& \\
{}^m{\hskip -1pt}\Pi_{\tt s}& \\
  \end{aligned} \right) =&  \left( \begin{aligned}
\cos\phi\\
\sin\phi  \\
  \end{aligned} \right) \,\Pi_{\tt s}(r, \theta),& {\rm where} \hskip 30pt
 \Pi_{\tt s}(r, \theta)=\frac{u}{r} \sum_{\ell=1}^\infty
a_\ell H^+_L(kr_g, kr)e^{i\xi_{b}(r)} P^{(1)}_\ell(\cos\theta). ~~~~~
  \label{eq:Pi_s+}
\end{align}

In a relevant scattering scenario, the EM wave and the Sun are well separated initially, so the Debye potential for the incident wave can be expected to have the same form as for the pure plasma-free case that includes only the Coulomb potential that is given by (\ref{eq:Pi_ie*+*=}).
Therefore, the Debye potential for the inner region has the form:
\begin{align}
  \left( \begin{aligned}
{}^e{\hskip -1pt}\Pi_{\tt in}/\sqrt{\epsilon}& \\
{}^m{\hskip -1pt}\Pi_{\tt in}/\sqrt{\mu}& \\
  \end{aligned} \right) =&  \left( \begin{aligned}
\cos\phi\\
\sin\phi  \\
  \end{aligned} \right) \,\Pi_{\tt in}(r, \theta),
  \label{eq:Pi_in+}
\end{align}
with the potential $\Pi_{\tt in}$ given, to ${\cal O}(r_g^2,r_g{\omega_{\tt p}^2}/{\omega^2})$, as
{}
\begin{eqnarray}
\Pi_{\tt in}(r, \theta)&=&\frac{u}{r} \sum_{\ell=1}^\infty
b_\ell \Big\{ \cos\xi_{b}(r)F_L(kr_g, kr) + \sin\xi_{b}(r)\,G_L(kr_g, kr)\Big\}P^{(1)}_\ell(\cos\theta).
\label{eq:Pi-in}
\end{eqnarray}

We thus expressed all the potentials in the series (\ref{eq:Pi-degn-sol-0}) and any unknown constants can now be determined easily. If we now substitute the expressions (\ref{eq:Pi_ie*+*=}), (\ref{eq:Pi_s+}) and (\ref{eq:Pi_in+})--(\ref{eq:Pi-in}) into the boundary conditions (\ref{eq:bound_cond-expand1+})--(\ref{eq:bound_cond-expand4+}), we obtain the following linear relationships between the coefficients $a_\ell$ and $b_\ell$:
{}
\begin{eqnarray}
\Big[\frac{E_0}{k^2}i^{\ell-1}\frac{2\ell+1}{\ell(\ell+1)}e^{i\sigma_\ell}
F'_\ell(kr_g,kr)+a_\ell \Big(H^+_L(kr_g,kr)e^{i\xi_{b}(r)}\Big)'\Big]\Big|_{r=R_\star}&=&b_\ell R'_L(r)\Big|_{r=R_\star},
\label{eq:bound-cond*1}\\
\Big[\frac{E_0}{k^2}i^{\ell-1}\frac{2\ell+1}{\ell(\ell+1)}e^{i\sigma_\ell}
F_\ell(kr_g,kr)+a_\ell H^+_L(kr_g,kr)e^{i\xi_{b}(r)}\Big]\Big|_{r=R_\star}&=&b_\ell R_L(r)\Big|_{r=R_\star},
\label{eq:bound-cond*2}
\end{eqnarray}
where $R_L(r)$ is from (\ref{eq:R-L3}) and $'=d/dr$.  From the definition of the eikonal phase (\ref{eq:eik7h}), we see that
\begin{equation}
\xi'_{b}(R_\star)\equiv \xi'_{ b}(r)\big|_{r=R_\star} =-\frac{1}{2k} V_{\tt sr}(r)\Big|_{r=R_\star}=0.
\label{eq:eik7h+}
\end{equation}
For the electron plasma distribution (\ref{eq:eps}) and (\ref{eq:n-eps_n-ism}), especially with the values taken from the phenomenological model (\ref{eq:model}), the value $\xi'_{b}(R_\star)$ is extremely small and may be neglected.  We now define, for convenience, $\alpha_\ell$ and $\beta_\ell$ as
{}
\begin{eqnarray}
a_\ell=\frac{E_0}{k^2}i^{\ell-1}\frac{2\ell+1}{\ell(\ell+1)}e^{i\sigma_\ell} \alpha_\ell \qquad {\rm and}\qquad b_\ell=\frac{E_0}{k^2}i^{\ell-1}\frac{2\ell+1}{\ell(\ell+1)}e^{i\sigma_\ell}\beta_\ell.
\label{eq:a-b}
\end{eqnarray}
From (\ref{eq:bound-cond*1})--(\ref{eq:bound-cond*2}), we have:
{}
\begin{eqnarray}
F'_\ell(R_\star)+\alpha_\ell {H^+_L}'(R_\star)e^{i\xi_{b}(R_\star)}&=&\beta_\ell R'_L(R_\star),
\label{eq:bound-cond*1+}\\[3pt]
F_\ell(R_\star)+\alpha_\ell H^+_L(R_\star)e^{i\xi_{b}(R_\star)}&=&\beta_\ell R_L(R_\star),
\label{eq:bound-cond*2+}
\end{eqnarray}
where $F_\ell(R_\star)=F_\ell(kr_g,kR_\star)$ and $H^+_L(R_\star)=H^+_L(kr_g,kR_\star)$ with similar definitions for the derivatives of these functions. Equations (\ref{eq:bound-cond*1+})--(\ref{eq:bound-cond*2+}) may now be solved to determine the two sets of coefficients $\alpha_\ell$ and $\beta_\ell$:
{}
\begin{eqnarray}
\alpha_\ell&=&e^{-i\xi_{b}(R_\star)}\frac{F_\ell(R_\star)R'_L(R_\star)-F'_\ell(R_\star)R_L(R_\star)}{R_L(R_\star){H^+_L}'(R_\star)-R'_L(R_\star)H^+_L(R_\star)},
\label{eq:a_l*}\\
\beta_\ell&=&\frac{F_l(R_\star){H^+}'_L(R_\star)-F'_\ell(R_\star)H^+_L(R_\star)}{R_L(R_\star){H^+_L}'(R_\star)-R'_L(R_\star)H^+_L(R_\star)}.
\label{eq:b_l*}
\end{eqnarray}

Taking into account the asymptotic behavior of all the functions involved: namely (\ref{eq:free_u+}) for $H^+_L$ and (\ref{eq:F-beh})--(\ref{eq:G-beh}) for $F_L$ and $G_L$, we have the following solution for the coefficients $\alpha_\ell$ and $\beta_\ell$:
 {}
\begin{eqnarray}
\alpha_\ell&=&\sin\delta_\ell^\star, \qquad \beta_\ell=e^{i\delta_\ell^\star},
\qquad {\rm where}\qquad \delta_\ell^\star=-\frac{\pi}{2}(L-\ell)+\sigma_L-\sigma_\ell+\xi^\star_{b},
\label{eq:a_b_del}
\end{eqnarray}
with $\xi^\star_b=\xi_b(R_\star)$ and $\delta^*_\ell$ being the phase shift induced by the plasma to the phase of the EM wave propagating through the solar system, as measured at the termination shock, $\delta^*_\ell=\delta_\ell(R_\star)$.

As expected, when the plasma is absent, $L=\ell$ and $\xi_b=0$, the total plasma phase shift vanishes, resulting in $\delta_\ell=0$. However, in the case of scattering by the plasma, $\xi^*_b=\xi_b(R_\star)\not=0$ and $\delta_\ell$ is important. Also, for large heliocentric distances along the incident direction, for which $r\gg b$, and certainly for the region outside the termination shock, $r>R_\star$, the  eikonal phase shift $\xi_b^\star=\xi_b(R_\star)$, given by  (\ref{eq:delta-D*-av0WKB+}), together with (\ref{eq:eik1hQ}), is
{}
\begin{eqnarray}
\xi_b^\star\approx -\frac{2\pi e^2R_\odot}{m_ec^2k}
\sum_{i>2}\alpha_iQ_{\beta_i}^\star\Big(\frac{R_\odot}{b}\Big)^{\beta_i-1},
\label{eq:delta-D*-av0WKB}
\end{eqnarray}
which, for any given $b$, is a constant value. In the case when $\mu/\ell\ll1$ and (\ref{eq:L2-apr}) is valid, expression (\ref{eq:a_b_del}) for the plasma-induced delay, to ${\cal O}(\mu^4)$,  takes the form (see \cite{Turyshev-Toth:2018-plasma} for a similar discussion):
 {}
\begin{eqnarray}
\delta_\ell^\star=-\frac{\pi}{2}\frac{\mu^2}{2\ell}+\sigma_L-\sigma_\ell+\xi_b^\star.
\label{eq:a_b_del-r_mu0}
\end{eqnarray}

We can evaluate the contribution of the plasma to the phase of the EM wave as the wave traverses the solar system.  In the case of the electron number density model (\ref{eq:model}) and from (\ref{eq:a_b_del}), the plasma phase shift $\delta^*_\ell$ in (\ref{eq:a_b_del-r_mu0}) is given as
{}
\begin{eqnarray}
\delta^*_{\ell}&=&\sigma_L-\sigma_\ell-\eta_2\frac{R_\odot}{b}-\eta_6Q^\star_{6}\Big(\frac{R_\odot}{b}\Big)^5-\eta_{16}Q^\star_{16}\Big(\frac{R_\odot}{b}\Big)^{15}+...,
\label{eq:s-d1}
\end{eqnarray}
with $\eta_2, \eta_6$ and $\eta_{16}$ having the form
{}
\begin{eqnarray}
\eta_2&=&\frac{\pi}{2}\frac{2\pi e^2R_\odot}{m_ec^2k}
{\alpha_2},\qquad
\eta_6=\frac{2\pi e^2R_\odot}{m_ec^2k}{\alpha_6},\qquad
\eta_{16}=\frac{2\pi e^2R_\odot}{m_ec^2k}{\alpha_{16}},
\label{eq:delta-etas}
\end{eqnarray}
where, to derive the expression for $\eta_2$, we used $\mu^2$ from (\ref{eq:V-sr-m2}) and  approximated (\ref{eq:a_b_del}) for the case of $\mu/\ell \ll1$ by using (\ref{eq:L2-apr}) with $Q^\star_{\beta_i}$ in the incident direction as given by (\ref{eq:eik1hQ}).  Note that this approach results in the additional factor of $\pi/2$ (which came from the first term in (\ref{eq:a_b_del})) that is characteristic to the eikonal approximation  (see discussion in \cite{Friedrich-book-2006,Friedrich-book-2013}). To derive  $\eta_6$ and $\eta_{16}$, we used (\ref{eq:delta-D*-av0WKB}). The empirical model for the free electron number density in the solar corona (\ref{eq:model}) results in the following values for the constants $\eta_2, \eta_6$ and $\eta_{16}$  in (\ref{eq:delta-etas}):
{}
\begin{eqnarray}
\eta_2&=&1.06 \,\Big(\frac{\lambda}{1~\mu{\rm m}}\Big),\,\qquad
\eta_6=303.87 \,\Big(\frac{\lambda}{1~\mu{\rm m}}\Big),\qquad
\eta_{16}=586.17 \,\Big(\frac{\lambda}{1~\mu{\rm m}}\Big).
\label{eq:delta-etas-m}
\end{eqnarray}
Beyond $b\simeq3.65 R_\odot$, the contribution from the $\eta_2$ term rapidly becomes dominant. However, for small impact parameters characteristic for imaging with the SGL, the plasma phase shift is driven by the terms with larger powers of $R_\odot/b$ in free electron number density model of the solar corona (\ref{eq:model}).

Therefore, using the value for $a_\ell$ from (\ref{eq:a-b}), together with $\alpha_\ell$ from (\ref{eq:a_b_del}), we determine that the solution for the scattered potential (\ref{eq:Pi_s+}) for $r>R_\star$ takes the form
{}
\begin{eqnarray}
\Pi_{\tt s}(r, \theta)&=& \frac{E_0}{k^2} \frac{u}{r}\sum_{\ell=1}^\infty
i^{\ell-1}\frac{2\ell+1}{\ell(\ell+1)}e^{i\sigma_\ell} \sin\delta_\ell^\star H^+_L(kr_g, kr)e^{i\xi_{b}^\star} P^{(1)}_\ell(\cos\theta).
\label{eq:Pi-s_a*=}
\end{eqnarray}

Next, using the asymptotic behavior of $H^{(+)}_L$ from (\ref{eq:free_u+}) together with the expression for the phase shift $\delta_\ell^\star$ (\ref{eq:a_b_del}), we notice that at large distances from the Sun the following relation exists:  $H^{+}_L(kr_g,kr)e^{i\xi_b^\star}  \approx H^{+}_\ell (kr_g,kr)e^{i\delta^*_\ell}+{\cal O}(\mu^2/2kr)$, which allows us to present (\ref{eq:Pi-s_a*=}) as
{}
\begin{eqnarray}
\Pi_{\tt s}(r, \theta)&=&\frac{E_0}{2ik^2} \frac{u}{r}
\sum_{\ell=1}^\infty
i^{\ell-1}\frac{2\ell+1}{\ell(\ell+1)}e^{i\sigma_\ell} H^+_\ell(kr_g, kr) \Big(e^{2i\delta_\ell^\star} -1\Big)P^{(1)}_\ell(\cos\theta).
\label{eq:Pi-s_a*}
\end{eqnarray}

In the region outside the termination shock, $r>R_\star$, we may take the asymptotic form for the Coulomb--Hankel function and present (\ref{eq:Pi-s_a*}) as
{}
\begin{eqnarray}
\Pi_{\tt s}(r, \theta)&=&- \frac{E_0}{2k^2} \frac{u}{r}e^{ik(r+r_g\ln 2kr)}
\sum_{\ell=1}^\infty
\frac{2\ell+1}{\ell(\ell+1)}e^{i(2\sigma_\ell+\frac{\ell(\ell+1)}{2kr})} \Big(e^{2i\delta_\ell^\star}-1\Big)P^{(1)}_\ell(\cos\theta).
\label{eq:Pi-s_ass}
\end{eqnarray}

As a result, using (\ref{eq:Pi_ie*+*=}) and (\ref{eq:Pi-s_a*}), we present the Debye potential in the region outside the termination shock boundary, $r>R_\star$,  in the following form:
{}
\begin{eqnarray}
\Pi_{\tt out}(r, \theta)=\Pi_0(r,\theta)+\Pi_{\tt s}(r,\theta)&=& \frac{E_0}{k^2}\frac{u}{r}\sum_{\ell=1}^\infty i^{\ell-1}\frac{2\ell+1}{\ell(\ell+1)}e^{i\sigma_\ell}
\Big\{F_\ell(kr_g,kr) + \frac{1}{2i}\Big(e^{2i\delta_\ell^\star}-1\Big) H^+_\ell(kr_g, kr)\Big\}P^{(1)}_\ell(\cos\theta).~~~~~~
\label{eq:Pi-s_a1*0}
\end{eqnarray}

Similarly,  substituting the value for $b_\ell$ from (\ref{eq:a-b}), together with $\beta_\ell$ from (\ref{eq:a_b_del}), we determine the solution for the inner Debye potential (\ref{eq:a_b_del}) in the form
{}
\begin{eqnarray}
\Pi_{\tt in}(r, \theta)&=&\frac{E_0}{k^2} \frac{u}{r} \sum_{\ell=1}^\infty
i^{\ell-1}\frac{2\ell+1}{\ell(\ell+1)}e^{i(\sigma_\ell +\delta_\ell)} \Big\{ \cos\xi_b(r)F_L(kr_g, kr) + \sin\xi_b(r)\,G_L(kr_g, kr)\Big\}P^{(1)}_\ell(\cos\theta).
\label{eq:Pi-in+}
\end{eqnarray}

As solar gravity is rather weak, we may use the asymptotic expressions for $F_L, G_L$ and $H^\pm_L$ for $r\geq R_\odot$.  Therefore, the radial function  $R_L (r) $ from (\ref{eq:R-L}) (or, equivalently, from (\ref{eq:R-L3})) in the region of heliocentric distances $R_\odot \leq r\leq R_\star$, may be given as
{}
\begin{eqnarray}
R_L (r) &=& \frac{1}{2i}\Big(H^+_L(kr_g,kr)e^{i \xi_{b} (r)} -H^-_L(kr_g,kr)e^{-i \xi_{b}(r)}\Big)\simeq e^{-i \delta_\ell(r)}\Big\{F_\ell(kr_g,kr) + \frac{1}{2i}\big(e^{2i \delta_\ell(r)}-1\big)H^+_\ell(kr_g,kr)\Big\},~~~~~~~
\label{eq:R-L3+*}
\end{eqnarray}
where $\delta_\ell(r)$ has the form given by (\ref{eq:a_b_del}) where  the eikonal phase at the termination shock $\xi_{b}=\xi_{b}(R_\star)$ is replaced with its original form (\ref{eq:delta-D*-av0WKB+1*}) that depends on the heliocentric distance, namely $\xi_{b}=\xi_{b}(r)$, thus
 {}
\begin{eqnarray}
 \delta_\ell(r)=-\frac{\pi}{2}(L-\ell)+\sigma_L-\sigma_\ell+\xi_{b}(r).
\label{eq:a_b_del-r}
\end{eqnarray}
Similarly to (\ref{eq:a_b_del-r_mu0}), in the case when $\mu/\ell\ll1$ and (\ref{eq:L2-apr}) is valid, expression (\ref{eq:a_b_del-r}), to order ${\cal O}(\mu^4)$,  takes the form
 {}
\begin{eqnarray}
\delta_\ell(r)=-\frac{\pi}{2}\frac{\mu^2}{2\ell}+\sigma_L-\sigma_\ell+\xi_{b}(r).
\label{eq:a_b_del-r_mu}
\end{eqnarray}
Thus, in the eikonal approximation, distance dependence in the plasma delay comes from the terms in the short-range plasma potential $V_{\tt sr}$ for which $i>2$. The term with $i=2$ (i.e., the first term in (\ref{eq:a_b_del-r_mu0})
 and (\ref{eq:a_b_del-r_mu})) provides no distance dependence. The physical interpretation of this observation follows, in part, from (\ref{eq:model}). It can be seen that near the solar surface, $b\gtrsim R_\odot$, the potential is dominated by terms containing higher powers of $(R_\odot/r)$. The inverse square term contributes an approximately uniform background potential that, at these small heliocentric ranges, is several orders of magnitude smaller compared to the other terms (see Fig.~\ref{fig:plasma}). This $1/r^2$-term becomes dominant only at greater distances from the Sun, where $V_{\tt sr}$ is several orders of magnitude smaller than it is near the solar surface.

As  a result, outside the Sun, we may present (\ref{eq:Pi-in+}) in the following equivalent form:
{}
\begin{eqnarray}
\Pi_{\tt in}(r, \theta)&=&\frac{E_0}{k^2} \frac{u}{r} \sum_{\ell=1}^\infty
i^{\ell-1}\frac{2\ell+1}{\ell(\ell+1)}e^{i\big(\sigma_\ell +\delta_\ell -\delta_\ell(r)\big)} \Big\{F_\ell(kr_g, kr)+ \frac{1}{2i}\Big(e^{2i\delta_\ell(r)}-1\Big)H^+_\ell(kr_g, kr)\Big\}P^{(1)}_\ell(\cos\theta).
\label{eq:Pi-in+sl}
\end{eqnarray}

With the plasma model (\ref{eq:n-eps_n-ism}) the phase shift vanishes inside the Sun, $\delta_\ell=0$, and (\ref{eq:Pi-in+sl}) reduces to the plasma-free solution (\ref{eq:Pi_ie*+*=}). As a result, the solution for the Debye potential, $\Pi (r, \theta)$ from (\ref{eq:Pi-in+sl}), describing the propagation of the EM wave in the solar system on the background of the static gravitational monopole and a steady-state, spherically symmetric plasma distribution takes the form
{}
\begin{eqnarray}
\Pi_{\tt in}(r, \theta)&=&
\frac{E_0}{k^2}\frac{u}{r}\sum_{\ell=1}^\infty i^{\ell-1}\frac{2\ell+1}{\ell(\ell+1)}e^{i\big(\sigma_\ell +\delta_\ell -\delta_\ell(r)\big)} F_\ell(kr_g,kr) P^{(1)}_\ell(\cos\theta)+\nonumber\\
&+&\frac{E_0}{2ik^2}\frac{u}{r}\sum_{\ell=1}^\infty i^{\ell-1}\frac{2\ell+1}{\ell(\ell+1)}e^{i\big(\sigma_\ell +\delta_\ell -\delta_\ell(r)\big)}
\Big(e^{2i\delta_\ell(r)}-1\Big)
H^{(+)}_\ell(kr_g,kr) P^{(1)}_\ell(\cos\theta) +{\cal O}\Big(r^2_g,r_g\frac{\omega_{\tt p}^2}{\omega^2}\Big).
  \label{eq:Pi_g+p}
\end{eqnarray}

Note that this solution is valid, in principle, even inside the opaque Sun. Indeed, because of the plasma model (\ref{eq:eps}) and (\ref{eq:n-eps_n-ism}), the phase shift vanishes, $\delta_\ell=0$, and (\ref{eq:Pi-in+sl}) reduces to the plasma-free solution (\ref{eq:Pi_ie*+*=}).

The first term in (\ref{eq:Pi-in+sl}) is the Debye potential of an EM wave propagating  in a vacuum but modified by the plasma in the solar system. The second term represents the effect of the solar plasma on the propagation of the EM waves inside the termination shock, $0< r \leq R_\star$. Notice that, as the distance increases, this term approaches the form of the Debye potential $\Pi_{\tt s}$ for the scattered EM field given by (\ref{eq:Pi-s_ass}). Proper accounting for such a dependence makes it possible to compare high-precision observations conducted from different locations within the solar system.

Thus, we have identified all the Debye potentials involved in the Mie problem \cite{Mie:1908}, namely the potential $\Pi_0$ given by (\ref{eq:Pi_ie*+*=}) representing the incident EM field, the potential $\Pi_{\tt s}$ from (\ref{eq:Pi-s_ass}) describing the scattered EM field outside the termination shock, $r>R_\star$, and the potential $\Pi_{\tt in}$ from (\ref{eq:Pi-in+sl}) describing it inside the termination shock, $0<r\leq R_\star$.

\section{General solution for the EM field outside the termination shock}
\label{sec:EM-field-outside}

To describe the scattering of light by the extended solar corona, we use solutions for the Debye potential representing the scattered EM wave, (\ref{eq:Pi-s_ass}),  and the EM wave inside the termination shock boundary, (\ref{eq:Pi_g+p}).
The presence of the Sun itself is not yet captured. For this, we need to set additional boundary conditions that describe the interaction of the Sun with the incident radiation. Similarly to \cite{Turyshev-Toth:2017,Turyshev-Toth:2018-plasma}, we apply the fully absorbing boundary conditions that represent the physical size and the surface properties of the Sun \cite{Turyshev-Toth:2018-grav-shadow}.

We begin with the area that lies outside the termination shock where three regions are present, namely (i) the shadow region, (ii) the geometric optics region, and (iii) the interference region. Clearly, as far as imaging with the SGL is concerned,  the interference region is of most importance. This is where the SGL focuses light coming from a distant object, forming an image.

\subsection{Fully absorbing boundary conditions}
\label{sec:bound-c}

Boundary conditions representing the opaque Sun were introduced in \cite{Herlt-Stephani:1976} and were used in \cite{Turyshev-Toth:2017,Turyshev-Toth:2018-plasma}. Here we use these conditions again. Specifically, to set the boundary conditions, we rely on the semiclassical analogy between the partial momentum, $\ell$, and the impact parameter, $b$, that is given as $\ell=kb$ \cite{Messiah:1968,Landau-Lifshitz:1989}.

To set the boundary conditions, we require that rays with impact parameters $b\le R_\odot^\star=R_\odot +r_g$ are completely absorbed by the Sun \cite{Turyshev-Toth:2017}. Thus, the fully absorbing boundary condition signifies that all the radiation intercepted by the body of the Sun is fully absorbed by it and no reflection or coherent reemission occurs. All intercepted radiation is transformed into some other forms of energy, notably heat. Thus, we require that no scattered waves exist with impact parameter $b\ll R_\odot^\star$ or, equivalently, for $\ell \leq kR_\odot^\star$. Such formulation relies on the concept of the semiclassical impact parameter $b$ and its relationship with the partial momentum, $\ell$, as $\ell=k b$. (A relevant discussion on this relation between $\ell$ and $b$ is on p.~29 of \cite{Grandy-book-2005} with reference to \cite{vandeHulst-book-1981}.)  In terms of the boundary conditions, this means that we need to subtract the scattered waves from the incident wave for $\ell \leq kR^\star_\odot$, as was discussed in \cite{Turyshev-Toth:2017}. Furthermore, as it was shown in \cite{Turyshev-Toth:2018-grav-shadow}, the fully absorbing boundary conditions introduce a fictitious EM field that precisely compensates the incident field in the area behind the Sun. This area has the shape of a rotational hyperboloid that starts directly at the solar surface behind the Sun and extends to the vertex of the hyperboloid at $z_0=R^2_\odot/2r_g\simeq$ 547.8~AU.

\subsection{The Debye potential for the region outside the termination shock}

To implement the boundary conditions for the EM wave outside the termination shock, we realize that the total EM field in this region is given as the sum of the incident and scattered waves, $\Pi=\Pi_0+\Pi_{\tt s}$, with these two potentials given by (\ref{eq:Pi_ie*+*=}) and (\ref{eq:Pi-s_ass}), correspondingly. Also, using the asymptotic behavior of $H^{+}_L$ from (\ref{eq:free_u+}) and with the help of expression (\ref{eq:a_b_del}) for the phase shift $\delta_\ell^\star$, we notice that at large distances from the Sun we can write $H^{+}_L(kr_g,kr)e^{i\xi_b^\star}  \approx H^{+}_\ell (kr_g,kr)e^{i\delta^*_\ell}+{\cal O}(\mu^2/2kr)$ (similar to  that used in the derivations of (\ref{eq:R-L3+*})).

Accordingly, we use (\ref{eq:Pi-s_a1*0}), which represents the Debye potential in the region $r>R_\star$ and is given as
{}
\begin{eqnarray}
\Pi(r, \theta)=\Pi_0(r,\theta)+\Pi_{\tt s}(r,\theta)&=& \frac{E_0}{k^2}\frac{u}{r}\sum_{\ell=1}^\infty i^{\ell-1}\frac{2\ell+1}{\ell(\ell+1)}e^{i\sigma_\ell}
\Big\{F_\ell(kr_g,kr) + \frac{1}{2i}\big(e^{2i\delta_\ell^\star}-1\big) H^+_\ell(kr_g, kr)\Big\}P^{(1)}_\ell(\cos\theta).~~~~~~~
\label{eq:Pi-s_a1*}
\end{eqnarray}

Next, relying on the representation of the regular Coulomb function $F_\ell$ via incoming, $H^{+}_\ell$, and outgoing, $H^{-}_\ell$, waves as $F_\ell=(H^{+}_\ell-H^{-}_\ell)/2i$ (discussed in \cite{Turyshev-Toth:2017} and also by the expression given after (\ref{eq:R-L})), we may express the Debye potential (\ref{eq:Pi-s_a1*}) as
{}
\begin{eqnarray}
\Pi(r, \theta)&=& \frac{E_0}{2ik^2}\frac{u}{r}\sum_{\ell=1}^\infty i^{\ell-1}\frac{2\ell+1}{\ell(\ell+1)}e^{i\sigma_\ell}
\Big\{e^{2i\delta_\ell^\star}H^+_\ell(kr_g, kr)-H^-_\ell(kr_g, kr)\Big\}P^{(1)}_\ell(\cos\theta).
\label{eq:Pi-s_a*0}
\end{eqnarray}

This form of the combined Debye potential is convenient for implementing the fully absorbing  boundary conditions discussed in Sec.~\ref{sec:bound-c}. Specifically, subtracting from (\ref{eq:Pi-s_a*0})  the outgoing wave (i.e., $\propto H^{(+)}_\ell$) for the impact parameters $b\leq R_\odot^\star$ or equivalently for $\ell\in[1,kR_\odot^\star]$, we have
 {}
\begin{eqnarray}
\Pi(r, \theta)&=& \frac{E_0}{2ik^2}\frac{u}{r}\sum_{\ell=1}^\infty i^{\ell-1}\frac{2\ell+1}{\ell(\ell+1)}e^{i\sigma_\ell}
\Big\{e^{2i\delta_\ell}H^+_\ell(kr_g, kr)-H^-_\ell(kr_g, kr)\Big\}P^{(1)}_\ell(\cos\theta)-\nonumber\\
&&\hskip 10 pt -\, \frac{E_0}{2ik^2}\frac{u}{r}\sum_{\ell=1}^{kR_\odot^\star} i^{\ell-1}\frac{2\ell+1}{\ell(\ell+1)}e^{i\sigma_\ell}e^{2i\delta_\ell^\star}H^+_\ell(kr_g, kr)P^{(1)}_\ell(\cos\theta),
\label{eq:Pi-s_a+0}
\end{eqnarray}
or, equivalently, coming back to the form (\ref{eq:Pi-s_a1*}),
 {}
\begin{eqnarray}
\Pi(r, \theta)&=&
\Pi_0(r,\theta)+ \frac{E_0}{2ik^2}\frac{u}{r}\sum_{\ell=1}^\infty i^{\ell-1}\frac{2\ell+1}{\ell(\ell+1)}e^{i\sigma_\ell}
\big(e^{2i\delta_\ell^\star}-1\big) H^+_\ell(kr_g, kr)P^{(1)}_\ell(\cos\theta) -\nonumber\\
&&\hskip 36 pt -\,
\frac{E_0}{2ik^2}\frac{u}{r}\sum_{\ell=1}^{kR_\odot^\star} i^{\ell-1}\frac{2\ell+1}{\ell(\ell+1)}e^{i\sigma_\ell}e^{2i\delta_\ell^\star}H^+_\ell(kr_g, kr)P^{(1)}_\ell(\cos\theta).
\label{eq:Pi-s_a+}
\end{eqnarray}

This is our main result, valid for all distances outside the termination shock $r>R_\star$ and all angles. It is a rather complex expression. It requires the tools of numerical analysis to fully explore its behavior and the resulting EM field \cite{Kerker-book:1969,vandeHulst-book-1981,Grandy-book-2005}. However, in most practically important applications, we need to know the field in the forward direction. Furthermore, our main interest is to study the largest plasma impact on light propagation, which corresponds to the smallest values of the impact parameter. In this situation, we may simplify the result (\ref{eq:Pi-s_a+}) by taking into account the asymptotic behavior of the function $H^{+}_\ell(kr_g,kr)$, considering the field at large heliocentric distances, such that $kr\gg\ell$, where $\ell$ is the order of the Coulomb function (see p.~631 of \cite{Morse-Feshbach:1953}). For  $kr\rightarrow\infty $ and also for $r\gg r_{\tt t}=\sqrt{\ell(\ell+1)}/k$ (see \cite{Turyshev-Toth:2017,Turyshev-Toth:2018-plasma}), such an expression is given in the form (\ref{eq:R_solWKB+=_bar-imp}):
{}
\begin{eqnarray}
\lim_{kr\rightarrow\infty} H^{+}_\ell(kr_g,kr)&\sim&
\exp\Big[i\Big(k(r+r_g\ln 2kr)+\frac{\ell(\ell+1)}{2kr}
+\frac{[\ell(\ell+1)]^2}{24k^3r^3}+\sigma_\ell-\frac{\pi\ell}{2}\Big)\Big] +{\cal O}\big((kr)^{-5}, r_g^2\big),
\label{eq:Fass*}
\end{eqnarray}
which includes the contribution from the centrifugal potential in the radial equation (\ref{eq:R-bar-k*}) (see e.g., Appendix \ref{sec:rad_eq_wkb}, Appendix A in \cite{Turyshev-Toth:2018} or \cite{Kerker-book:1969}). In fact, expression (\ref{eq:Fass*}) extends the argument of (\ref{eq:free_u+}) to shorter distances, closer to the turning point of the potential (see the relevant discussion in Appendix~F of \cite{Turyshev-Toth:2017}). By including the extended centrifugal term in (\ref{eq:Fass*}) (i.e., shown by the terms with various powers of $\ell(\ell+1)/2kr$), we can now better describe  the bending of the trajectory of a light ray under the combined influence of gravity and plasma. (We note that, in (\ref{eq:R_solWKB+=_bar-imp}) we omitted the amplitude factor $a(\ell)$ given by (\ref{eq:sf1}). Outside the Sun, the argument of this factor is very small resulting in $a(\ell)\approx 1$. Also, one may verify that any derivative of this term produces a contribution to the amplitude of the EM wave that is $1/kr$ times smaller compared the leading terms, which is negligible.)

As a result, we may take the approximate behavior of $H^{+}_\ell $ given by  (\ref{eq:Fass*}) and use it in (\ref{eq:Pi-s_a+}) to present the solution for the Debye potential outside the termination shock, $r>R_\star$ in the following form
{}
\begin{eqnarray}
\Pi (r, \theta)&=&
\Pi_0 (r, \theta)+\frac{ue^{ik(r+r_g\ln 2kr)}}{r}\Big\{\frac{E_0}{2k^2}\sum_{\ell=1}^{kR_\odot^\star} \frac{2\ell+1}{\ell(\ell+1)}e^{i\big(2\sigma_\ell+\frac{\ell(\ell+1)}{2kr}+\frac{[\ell(\ell+1)]^2}{24k^3r^3}\big)}P^{(1)}_\ell(\cos\theta)-\nonumber\\
&&\hskip60pt -\,\frac{E_0}{2k^2}\sum_{\ell=kR_\odot^\star}^{\infty} \frac{2\ell+1}{\ell(\ell+1)}e^{i\big(2\sigma_\ell+\frac{\ell(\ell+1)}{2kr}+\frac{[\ell(\ell+1)]^2}{24k^3r^3}\big)}\big(e^{i2\delta_\ell^\star}-1\big)P^{(1)}_\ell(\cos\theta) \Big\}+{\cal O}(r^2_g,r_g\frac{\omega_{\tt p}^2}{\omega^2})=\nonumber\\
&=&\Pi_0 (r, \theta)+\Pi_{\tt bc} (r, \theta)+\Pi _{\tt p} (r, \theta).
  \label{eq:Pi_g+p0}
\end{eqnarray}

The first term in (\ref{eq:Pi_g+p0}), $\Pi_0 (r, \theta)$, is the Debye potential represents the incident EM wave propagating in the vacuum on the background of a post-Newtonian gravity field produced by a gravitational mass monopole. The solution for  $\Pi_0 (r, \theta)$ is known and is given by  (\ref{eq:Pi_ie*+*=}) in the form of infinite series with respect to partial momenta, $\ell$. For practical purposes, however, it is convenient to use an exact expression for $\Pi_0$, which was derived in \cite{Turyshev-Toth:2017} in the form
{}
\begin{eqnarray}
\Pi_0(r, \theta)&=&-\psi_0\frac{iu}{k}\frac{1-\cos\theta}{\sin\theta}\Big(e^{ikz}{}_1F_1[1+ikr_g,2,ikr(1-\cos\theta)]-e^{-ikr}{}_1F_1[1+ikr_g,2,2ikr]\Big),
\label{eq:sol-Pi0*}
\end{eqnarray}
where the constant for $\psi_0$ is given by
\begin{eqnarray}
\psi^2_0= E_0^2\,{2\pi kr_g}/({1-e^{-2\pi kr_g}}).
\label{eq:psi_02*}
\end{eqnarray}

Eq.~(\ref{eq:sol-Pi0*})  gives the Debye potential of the plasma-free wave in terms of the confluent hypergeometric function. This solution is always finite and is valid for any angle $\theta$. It allows one to describe the EM filed in the interference region of the SGL and thus to develop the wave-optical treatment of the lens.

The  EM field of the incident wave outside the interference region is derived from
(\ref{eq:sol-Pi0*}) with the help of the asymptotic expansion of the hypergeometric functions ${}_1F_1[1+ikr_g,2,ikr(1-\cos\theta)]$ and  ${}_1F_1[1+ikr_g,2,2ikr]$ at large values of argument $k(r-z)\gg1$ (see \cite{Turyshev-Toth:2017} for details). This approach allows one to  compute the asymptotic behavior of the Debye potential $\Pi_0$ from (\ref{eq:sol-Pi0*}) as
{}
\begin{eqnarray}
\Pi_0(r,\theta)&=&E_0\frac{u}{k^2r\sin\theta}\Big\{e^{ik\big(r\cos\theta-r_g\ln kr(1-\cos\theta)\big)}-
e^{ik\big(r+r_g\ln kr(1-\cos\theta)\big)+2i\sigma_0}-\nonumber\\
&-&{\textstyle\frac{1}{2}}(1-\cos\theta)\Big(
e^{-ik\big(r+r_g\ln 2kr\big)}-
e^{ik\big(r+r_g\ln 2kr\big)+2i\sigma_0}\Big)
+{\cal O}\Big(\frac{ikr_g^2}{r-z}\Big)\Big\},
\label{eq:Pi-ass}
\end{eqnarray}
where we introduced the constant $\sigma_0=\arg \Gamma(1-ikr_g)$, which for large values of $kr_g\rightarrow\infty$ is given  as  \cite{Turyshev-Toth:2017}
\begin{eqnarray}
e^{2i\sigma_0}=
\frac{\Gamma(1-ikr_g)}{\Gamma(1+ikr_g)}=e^{-2ikr_g\ln (kr_g/e)-i\frac{\pi}{2}}\Big(1+{\cal O}((kr_g)^{-1})\Big).
\label{eq:Gam_rat}
\end{eqnarray}

The second term in (\ref{eq:Pi_g+p0}), $\Pi_{\tt bc} (r, \theta)$, is due to the physical obscuration introduced by the Sun and was derived by applying the fully absorbing boundary conditions. This term is responsible for the geometric shadow behind the Sun.

The third term in (\ref{eq:Pi_g+p0}), $\Pi _{\tt p} (r, \theta)$,  quantifies the contribution of the solar plasma to the scattering of the EM as it moves  through the solar system, and evaluated at the distance $r> R_\star$. Because of the plasma model (\ref{eq:eps}), (\ref{eq:n-eps_n-ism}), the last sum in (\ref{eq:Pi_g+p0}) formally extends only to $\ell=k R_\star$ corresponding to the impact parameter equal to the distance to the termination shock. As expected, for $r>R_\star$, the phase shift $\delta_\ell=0$ and the entire plasma-scattered term vanishes.

With the solution for the Debye potential given by (\ref{eq:Pi_g+p0}), and with the help of (\ref{eq:Dr-em0})--(\ref{eq:Bp-em0}) (also see \cite{Turyshev-Toth:2017}), we may now compute the EM field in the various regions involved. Given the smallness of the ratio $(\omega_{\tt p}/\omega)^2$ ($\sim 10^{-2}$ for radio and $\sim \times 10^{-11}$ for optical wavelengths), we may neglect the distance-dependent effect of the solar plasma on the amplitude of the EM wave. This is especially true at large heliocentric distances where the effect of the plasma, behaving as $\propto 1/r^2$, on the amplitude of the EM wave is negligibly small. (If one decides to account for the plasma effect on the amplitude of the EM wave, using (\ref{eq:Dr-em0})--(\ref{eq:Bp-em0}) one would get the terms that are $1/(kr)$ times smaller than the leading terms in those expressions. Thus, any derivatives of the plasma-dependent terms present in the amplitude of these terms would provide negligible contributions.)
Thus, the plasma affects the delay of the EM wave and is fully accounted for by the solution for the Debye potentials. Therefore, we can put $\epsilon =\mu =1$ in (\ref{eq:Dr-em0})--(\ref{eq:Bp-em0}) and use the following expressions to construct the EM field in the static, spherically symmetric geometry (see details in \cite{Turyshev-Toth:2017}):
{}
\begin{align}
  \left( \begin{aligned}
{ \hat D}_r& \\
{ \hat B}_r& \\
  \end{aligned} \right) =&  \left( \begin{aligned}
\cos\phi \\
\sin\phi  \\
  \end{aligned} \right) \,e^{-i\omega t}\alpha(r, \theta), &
    \left( \begin{aligned}
{ \hat D}_\theta& \\
{ \hat B}_\theta& \\
  \end{aligned} \right) =&  \left( \begin{aligned}
\cos\phi \\
\sin\phi  \\
  \end{aligned} \right) \,e^{-i\omega t}\beta(r, \theta), &
    \left( \begin{aligned}
{ \hat D}_\phi& \\
{ \hat B}_\phi& \\
  \end{aligned} \right) =&  \left( \begin{aligned}
-\sin\phi \\
\cos\phi  \\
  \end{aligned} \right) \,e^{-i\omega t}\gamma(r, \theta),
  \label{eq:DB-sol00p*}
\end{align}
with the quantities $\alpha, \beta$ and $\gamma$ computed from the known Debye potential, $\Pi$, as
{}
\begin{eqnarray}
\alpha(r, \theta)&=&
\frac{1}{u}\Big\{\frac{\partial^2 }{\partial r^2}
\Big[\frac{r\,{\hskip -1pt}\Pi}{u}\Big]+k^2 u^4\Big[\frac{r\,{\hskip -1pt}\Pi}{u}\Big]\Big\}+{\cal O}\Big(\big(\frac{1}{u}\big)''\Big),
\label{eq:alpha*}\\
\beta(r, \theta)&=&\frac{1}{u^2r}
\frac{\partial^2 \big(r\,{\hskip -1pt}\Pi\big)}{\partial r\partial \theta}+\frac{ik\big(r\,{\hskip -1pt}\Pi\big)}{r\sin\theta},
\label{eq:beta*}\\[0pt]
\gamma(r, \theta)&=&\frac{1}{u^2r\sin\theta}
\frac{\partial \big(r\,{\hskip -1pt}\Pi\big)}{\partial r}+\frac{ik}{r}
\frac{\partial\big(r\,{\hskip -1pt}\Pi\big)}{\partial \theta}.
\label{eq:gamma*}
\end{eqnarray}

This completes the solution for the Debye potentials on the background of a spherically symmetric, static gravitational field of the Sun and steady-state, spherically symmetric solar plasma distribution. We will use (\ref{eq:DB-sol00p*})--(\ref{eq:gamma*}) to compute the relevant EM fields.

\subsection{EM field in the shadow region}

In the shadow behind the Sun (i.e, for impact parameters $b\leq R^\star_\odot$) the EM field is represented by the Debye potential of the shadow, $\Pi_{\tt sh}$, which is given as
{}
\begin{eqnarray}
\Pi_{\tt sh} (r, \theta)&=&
\Pi_0 (r, \theta)+\frac{ue^{ik(r+r_g\ln 2kr)}}{r}\frac{E_0}{2k^2}\sum_{\ell=1}^{kR_\odot^\star} \frac{2\ell+1}{\ell(\ell+1)}e^{i\big(2\sigma_\ell+\frac{\ell(\ell+1)}{2kr}+\frac{[\ell(\ell+1)]^2}{24k^3r^3}\big)}P^{(1)}_\ell(\cos\theta)+{\cal O}(r^2_g,r_g\frac{\omega_{\tt p}^2}{\omega^2}),
  \label{eq:Pi_sh}
\end{eqnarray}
where $\Pi_0 (r, \theta)$ is well represented by (\ref{eq:Pi-ass}).
As discussed in \cite{Turyshev-Toth:2017,Turyshev-Toth:2018-grav-shadow}, the potential (\ref{eq:Pi_sh}) produces no EM field. In other words, there is no light in the shadow. Furthermore, as the solar boundary is rather diffuse, there is expectation for the Poisson-Arago bright spot to be formed in this region.

\subsection{EM field outside the shadow}
\label{sec:EM-field}

In the region behind the Sun but outside the solar shadow (i.e., for light rays with impact parameters $b>R_\odot$) which includes both the geometric optics  and interference regions (in the immediate vicinity of the focal line), the EM field is derived from the Debye potential given by the remaining terms in (\ref{eq:Pi_g+p0}) to the odrer of ${\cal O}(r^2_g,r_g{\omega_{\tt p}^2}/{\omega^2})$ as
{}
\begin{eqnarray}
\Pi (r, \theta)&=& \Pi_0 (r, \theta)-\frac{ue^{ik(r+r_g\ln 2kr)}}{r}\frac{E_0}{2k^2}\sum_{\ell=kR_\odot^\star}^{\infty} \frac{2\ell+1}{\ell(\ell+1)}e^{i\big(2\sigma_\ell+\frac{\ell(\ell+1)}{2kr}+\frac{[\ell(\ell+1)]^2}{24k^3r^3}\big)}\Big(e^{i2\delta_\ell^\star}-1\Big)P^{(1)}_\ell(\cos\theta),
  \label{eq:Pi_g+!}
\end{eqnarray}
where for the geometric optics region the potential $\Pi_0 (r, \theta)$ is well represented by (\ref{eq:Pi-ass}), while for the interference region one must use the exact form of $\Pi_0 (r, \theta)$ given by (\ref{eq:sol-Pi0*}).

Expression (\ref{eq:Pi_g+!}) is our main result for the regions outside the termination shock, $r>R_\star$ and also outside the shadow region, i.e., $b\geq R_\odot^\star$. It contains all the information needed to describe the total EM field originating from an incident Coulomb-modified plane wave that passed through the region of the steady-state spherically symmetric plasma of the extended solar corona, characterized by an electron number density (\ref{eq:n-eps_n-ism}) that diminishes as $r^{-2}$ or faster.

To evaluate the total solution for the Debye potential (\ref{eq:Pi_g+!}), we present it in the following compact form:
{}
\begin{eqnarray}
{\Pi (r, \theta)}&=&
{\Pi_0 (r, \theta)}+E_0f_{\tt p}(r,\theta)\frac{ue^{ik(r+r_g\ln 2kr)}}{r},  \label{eq:Pi-g+p}
\end{eqnarray}
where the plasma scattering amplitude $f_{\tt p}(r,\theta)$ is given by
{}
\begin{eqnarray}
  f_{\tt p}(r,\theta)&=&-\frac{1}{2k^2}\sum_{\ell=kR_\odot^\star}^\infty \frac{2\ell+1}{\ell(\ell+1)}e^{i\big(2\sigma_\ell+\frac{\ell(\ell+1)}{2kr}+\frac{[\ell(\ell+1)]^2}{24k^3r^3}\big)}\Big(e^{i2\delta_\ell^\star}-1\Big)
P^{(1)}_\ell(\cos\theta) +{\cal O}(r^2_g,r_g\frac{\omega_{\tt p}^2}{\omega^2}).
  \label{eq:f-v*+}
\end{eqnarray}

We note that because of the contribution from the centrifugal potential in (\ref{eq:Fass*}), the scattering amplitude $  f_{\tt p}(r,\theta)$ is now also a function of the heliocentric distance \cite{Turyshev-Toth:2018-plasma}. This is not the case in typical problems describing nuclear and atomic scattering \cite{Messiah:1968,Landau-Lifshitz:1989,Burke:2011,Newton-book-2013}. However, as we observed in \cite{Turyshev-Toth:2017,Turyshev-Toth:2018,Turyshev-Toth:2018-plasma}, when we are interested in the trajectories of light rays, the presence of such dependence and especially the $\propto 1/r$ term in the phase of the scattering amplitude (\ref{eq:f-v*+}) allows us to properly describe the bending of the light rays in the presence of gravity together with the contrition from the dispersive medium introduced by the solar plasma.

As a result, the Debye potential for the plasma-scattered wave outside the termination shock takes the form
{}
\begin{align}
  \Pi_{\tt p} (r, \theta)=E_0f_{\tt p}(r,\theta)\frac{ue^{ik(r+r_g\ln 2kr)}}{r},
  \label{eq:Pi_ie*+8p*}
\end{align}
with the plasma scattering  amplitude  $f_{\tt p}(r,\theta)$ given by (\ref{eq:f-v*+}). We use these expressions to derive the components of the EM field produced by this wave. For this, we substitute (\ref{eq:Pi_ie*+8p*})--(\ref{eq:f-v*+})  in the expressions (\ref{eq:alpha*})--(\ref{eq:gamma*}) to derive the factors $\alpha(r,\theta), \beta(r,\theta)$ and $\gamma(\theta)$, which to the order of ${\cal O}\big(r^2_g,r_g({\omega_{\tt p}^2}/{\omega^2}),(kr)^{-5}\big)$ are computed to be:
{}
\begin{eqnarray}
\alpha(r,\theta) &=& -E_0\frac{e^{ik(r+r_g\ln 2kr)}}{uk^2r^2}\sum_{\ell=kR^\star_\odot}^\infty(\ell+{\textstyle\frac{1}{2}})e^{i\big(2\sigma_\ell+\frac{\ell(\ell+1)}{2kr}+\frac{[\ell(\ell+1)]^2}{24k^3r^3}\big)}\Big(e^{i2\delta^*_\ell}-1\Big)P^{(1)}_\ell(\cos\theta)\times\nonumber\\
&&\hskip 30pt \times\,
\Big\{u^{2}+(u^2-1)\frac{\ell(\ell+1)}{4k^2r^2}+\frac{i}{kr}\Big(1+\frac{\ell(\ell+1)}{2k^2r^2}\Big)-\frac{ikr_g}{\ell(\ell+1)} \Big\},
  \label{eq:alpha*1*}\\
\beta(r,\theta) &=& E_0\frac{ue^{ik(r+r_g\ln 2kr)}}{ikr}\sum_{\ell=kR^\star_\odot}^\infty\frac{\ell+{\textstyle\frac{1}{2}}}{\ell(\ell+1)}e^{i\big(2\sigma_\ell+\frac{\ell(\ell+1)}{2kr}+\frac{[\ell(\ell+1)]^2}{24k^3r^3}\big)}\Big(e^{i2\delta^*_\ell}-1\Big)\times\nonumber\\
&&\hskip 30pt \times\,
\Big\{\frac{\partial P^{(1)}_\ell(\cos\theta)}
{\partial \theta}\Big(1-u^{-2}
\Big(\frac{\ell(\ell+1)}{2k^2r^2}+\frac{[\ell(\ell+1)]^2}{8k^4r^4}\Big)+\frac{ir_g}{2kr^2}\Big)+\frac{P^{(1)}_\ell(\cos\theta)}{\sin\theta}
 \Big\},
  \label{eq:beta*1*}\\
\gamma(r,\theta) &=& E_0\frac{ue^{ik(r+r_g\ln 2kr)}}{ikr}\sum_{\ell=kR^\star_\odot}^\infty\frac{\ell+{\textstyle\frac{1}{2}}}{\ell(\ell+1)}e^{i\big(2\sigma_\ell+\frac{\ell(\ell+1)}{2kr}+\frac{[\ell(\ell+1)]^2}{24k^3r^3}\big)}\Big(e^{i2\delta^*_\ell}-1\Big)\times\nonumber\\
&&\hskip 30pt \times\,
\Big\{\frac{\partial P^{(1)}_\ell(\cos\theta)}
{\partial \theta}+\frac{P^{(1)}_\ell(\cos\theta)}{\sin\theta}\Big(1-u^{-2}
\Big(\frac{\ell(\ell+1)}{2k^2r^2}+\frac{[\ell(\ell+1)]^2}{8k^4r^4}\Big)+\frac{ir_g}{2kr^2}\Big)
 \Big\}.
  \label{eq:gamma*1*}
\end{eqnarray}

This is an important result as it allows us to describe the EM field in all the regions of interest for the SGL, namely the geometric optics region and the interference region.

\section{EM field in the  geometric optics region}
\label{sec:go-em-outside}

We continue our discussion by deriving the EM field in the geometric optics region outside the termination shock, which we call  the exterior geometric optics region (as opposed to the interior geometric optics region  which us situated inside the termination shock).  Specifically, we are interested in the area behind the Sun located at heliocentric distances $r>R_\star$ that are reachable by the light rays whose impact parameters are $b>R^\star_\odot$. In addition, the exterior geometric optics region is situated outside the focal region of the SGL with angles $\theta$ satisfying the condition $\theta\gg \sqrt{2r_g/r}$ \cite{Turyshev-Toth:2017}.

We note that outside the Sun the ratio $r_g/r\leq 4.25\times 10^{-6} R_\odot/r \ll1$ is very small. As a result,  for $r> R_\star$ we may treat $u(r)= 1$ and neglect the contribution from derivatives of $u(r)$ to the amplitude of the scattered EM wave in (\ref{eq:alpha*1*})--(\ref{eq:gamma*1*}). Nevertheless, we keep them for the purposes of verification and internal consistency checks.

\subsection{Solution for the function $\alpha(r,\theta)$ and the radial components of the EM field}
\label{sec:radial-comp}

We begin with the investigation of $\alpha(r,\theta)$ given by (\ref{eq:alpha*1*}). We first note that in the case of large partial momenta $\ell$ and large angles $\theta$, namely $\ell\geq kR^\star_\odot$ and $\theta\gg \sqrt{2r_g/r}$, the last two terms in the curly brackets in this expression, behaving as $\propto i/kr$ and $ikr_g/\ell(\ell+1)$, are very small compared to the two leading terms and, thus, they may be neglected (a similar conclusion was reached in \cite{Turyshev-Toth:2018-plasma}.)  As a result, we obtain the following  expression for $\alpha(r,\theta)$:
{}
\begin{eqnarray}
\alpha(r,\theta) &=& -\frac{E_0u}{k^2r^2}e^{ik(r+r_g\ln 2kr)}\hskip-4pt \sum_{\ell=kR^\star_\odot}^\infty(\ell+{\textstyle\frac{1}{2}})\Big(1+\frac{r_g}{r}\frac{\ell(\ell+1)}{4k^2r^2} \Big) e^{i\big(2\sigma_\ell+\frac{\ell(\ell+1)}{2kr}+\frac{[\ell(\ell+1)]^2}{24k^3r^3}\big)}\Big(e^{i2\delta^*_\ell}-1\Big)P^{(1)}_\ell(\cos\theta)+\nonumber\\
&& \hskip 60pt
+ \, {\cal O}(r^2_g,r_g\frac{\omega_{\tt p}^2}{\omega^2},(kr)^{-5}).
  \label{eq:alpha*1}
\end{eqnarray}

To evaluate expression (\ref{eq:alpha*1}) in the region of geometric optics and, thus, for $\theta \gg \sqrt{2r_g/r}$, we use the asymptotic representation for $P^{(1)}_l(\cos\theta)$ from \cite{Bateman-Erdelyi:1953,Korn-Korn:1968,Kerker-book:1969}, valid when $\ell\to\infty$:
{}
\begin{align}
P^{(1)}_\ell(\cos\theta)  &=
\dfrac{-\ell}{\sqrt{2\pi \ell \sin\theta}}\Big(e^{i(\ell+\frac{1}{2})\theta+i\frac{\pi}{4}}+e^{-i(\ell+\frac{1}{2})\theta-i\frac{\pi}{4}}\Big)+{\cal O}(\ell^{-\textstyle\frac{3}{2}}) ~~~~~\textrm{for}~~~~~ 0<\theta<\pi.
\label{eq:P1l<}
\end{align}

This approximation can be used to transform (\ref{eq:alpha*1}) as
{}
\begin{eqnarray}
\alpha(r, \theta)&=&\frac{E_0u}{k^2r^2}
e^{ik(r+r_g\ln 2kr)}\hskip-4pt
\sum_{\ell=kR_\odot^\star}^\infty \frac{(\ell+{\textstyle\frac{1}{2}})\sqrt{\ell}}{\sqrt{2\pi \sin\theta}}
\Big(1+\frac{r_g}{r}\frac{\ell(\ell+1)}{4k^2r^2} \Big) e^{i\big(2\sigma_\ell+\frac{\ell(\ell+1)}{2kr}+\frac{[\ell(\ell+1)]^2}{24k^3r^3}\big)}
\times\nonumber\\[-6pt]
&&\hskip 110pt \times\,
\Big(e^{i2\delta^*_\ell}-1\Big)\Big(e^{i(\ell\theta+\frac{\pi}{4})}+e^{-i(\ell\theta+\frac{\pi}{4})}\Big)+
{\cal O}\Big(r^2_g,r_g\frac{\omega_{\tt p}^2}{\omega^2},(kr)^{-5}\Big).
\label{eq:P_sum}
\end{eqnarray}

We recognize that for large $\ell\geq kR^\star_\odot$, we may replace $\ell+1\rightarrow \ell$ and $\ell+\textstyle\frac{1}{2}\rightarrow \ell$.
At this point, we may replace the sum in (\ref{eq:P_sum}) with an integral:
{}
\begin{eqnarray}
\alpha(r, \theta)&=&\frac{E_0u}{k^2r^2}e^{ik(r+r_g\ln2kr)}\hskip-4pt
\int_{\ell=kR_\odot^\star}^\infty \hskip 0pt
\frac{\ell\sqrt{\ell}d\ell}{\sqrt{2\pi \sin\theta}}
\Big(1+\frac{r_g}{r}\frac{\ell^2}{4k^2r^2} \Big) \times\nonumber\\
&& \hskip 100pt \times\,
e^{i\big(2\sigma_\ell+\frac{\ell^2}{2kr}+\frac{\ell^4}{24k^3r^3}\big)}
\Big(e^{i2\delta^*_\ell}-1\Big)\Big(e^{i(\ell\theta+\frac{\pi}{4})}+e^{-i(\ell\theta+\frac{\pi}{4})}\Big)~~~
\label{eq:Pi_s_exp1}
\end{eqnarray}
and evaluate this integral by the method of stationary phase (see \cite{Herlt-Stephani:1976,Turyshev-Toth:2017}). This method allows us to evaluate integrals of the type
{}
\begin{equation}
I=\int A(\ell)e^{i\varphi(\ell)}d\ell, \qquad
\ell\in\mathbb{R},
\label{eq:stp-1}
\end{equation}
where the amplitude $A(\ell)$ is a slowly varying function of $\ell$, while $\varphi(\ell)$ is a rapidly varying function of $\ell$.
The integral (\ref{eq:stp-1}) may be replaced, to good approximation, with a sum over the points of stationary phase, $\ell_0\in\{\ell_{1,2,..}\}$, for which $d\varphi/d\ell=0$. Defining $\varphi''=d^2\varphi/d\ell^2$, we obtain the integral
{}
\begin{equation}
I\simeq\sum_{\ell_0\in\{\ell_{1,2,..}\}} A(\ell_0)\sqrt{\frac{2\pi}{\varphi''(\ell_0)}}e^{i\big(\varphi(\ell_0)+{\textstyle\frac{\pi}{4}}\big)}.
\label{eq:stp-2}
\end{equation}

Because the scattering term $\big(e^{i2\delta^*_\ell}-1\big)$ in (\ref{eq:Pi_s_exp1}) provides two contributions to the overal expression, each with a different phase, we treat the integral (\ref{eq:Pi_s_exp1}) as the sum of two integrals: one with the contribution from the plasma phase shift $2\delta^\star_\ell$ and one without it. To demonstrate our approach, we begin with the plasma-free term in (\ref{eq:Pi_s_exp1}).

\subsubsection{Evaluating the plasma-free term}
\label{sec:no-p-r}

For the term in (\ref{eq:Pi_s_exp1}) that does not contain the plasma phase shift, $2\delta^\star_\ell$, the relevant $\ell$-dependent part of the phase is of the form \cite{Turyshev-Toth:2018-plasma}
{}
\begin{equation}
\varphi^{[0]}_{\pm}(\ell)=\pm\big(\ell\theta+\textstyle{\frac{\pi}{4}}\big)+2\sigma_\ell +\dfrac{\ell^2}{2kr}+\dfrac{\ell^4}{24k^3r^3} +{\cal O}\big((kr)^{-5}\big).
\label{eq:S-l}
\end{equation}

We recall that the Coulomb phase shift $\sigma_\ell$  has the form \cite{Abramovitz-Stegun:1965,Turyshev-Toth:2017,Turyshev-Toth:2018-grav-shadow}:
{}
\begin{equation}
\sigma_\ell=\sigma_0-\sum_{j=1}^\ell\arctan\frac{kr_g}{j}, \qquad \sigma_0=\arg\Gamma(1-ikr_g),
\label{eq:S-l-g-s}
\end{equation}
where $\sigma_0$ was evaluated in \cite{Turyshev-Toth:2017} to be
{}
\begin{eqnarray}
\sigma_0&=& -kr_g\ln \frac{kr_g}{e}-\frac{\pi}{4}.
\label{eq:sig-0*}
\end{eqnarray}
We may replace the sum in (\ref{eq:S-l-g-s}) with an integral and, for $\ell\gg kr_g$, evaluate $\sigma_\ell$ as \cite{Turyshev-Toth:2018-grav-shadow}:
{}
\begin{eqnarray}
\sigma_\ell&=& -kr_g\ln \ell.
\label{eq:sig-l*}
\end{eqnarray}
This form agrees with the other known forms of $\sigma_\ell$ \cite{Cody-Hillstrom:1970,Barata:2009ma} that are approximated for large $\ell$.

\begin{figure}[t]
\includegraphics[width=130mm]{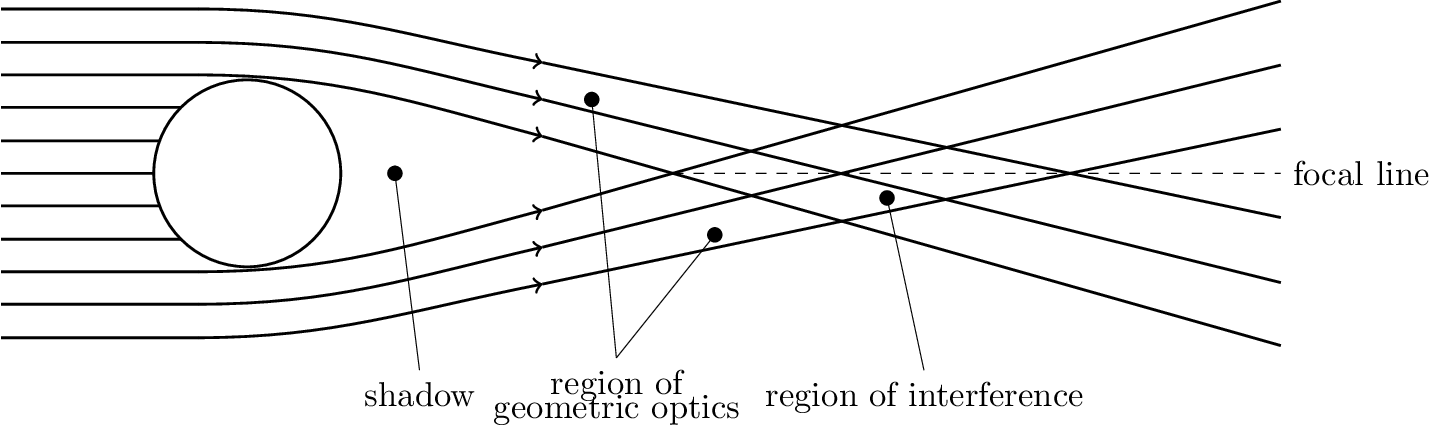}
\caption{
Three different regions of space associated with a monopole gravitational lens: the shadow, the region of geometric optics, and the region of interference (from \cite{Turyshev-Toth:2017}).\label{fig:regions}}
\end{figure}

The phase is stationary when $d\varphi^{[0]}_{\pm}/d\ell=0$, which, together with (\ref{eq:sig-l*}), implies
{}
\begin{equation}
\pm\theta-2\arctan \frac{kr_g}{\ell}+\frac{\ell}{kr}\Big(1+\frac{\ell^2}{6k^2r^2}\Big)={\cal O}\big((kr)^{-5}\big).
\label{eq:S-l-pri=}
\end{equation}
For small angles, $\sqrt{2r_g/r}\ll\theta\simeq b/r$, and large partial momenta, $\ell\simeq k R_\odot\gg kr_g$, this equation can be rewritten in the following form:
{}
\begin{equation}
\frac{\ell}{kr} = \mp {\theta}\Big(1-{\textstyle\frac{1}{6}}\theta^2\Big)+\frac{2kr_g}{\ell}+{\cal O}(\theta^5, r_g^2) \qquad {\rm or, ~equivalently,} \qquad
\frac{\ell}{kr} = \mp \sin\theta+\frac{2kr_g}{\ell}+{\cal O}(\theta^5, r_g^2).
\label{eq:S-l-pri*+}
\end{equation}
Relying on the semiclassical approximation that connects the partial momentum, $\ell$, to the impact parameter, $b$,
{}
\begin{equation}
\ell\simeq kb,
\label{eq:S-l-pri-p-g}
\end{equation}
for small angles $\theta$ (or, large distances from the sphere, $R_\odot/r<b/r\ll 1$),  we see that the points of stationary phase that must satisfy the equation are (see \cite{Turyshev-Toth:2017} for details):
{}
\begin{equation}
\frac{1}{r} = \mp\frac{ \sin{\theta}}{b}+\frac{2r_g}{b^2}+{\cal O}(\theta^5, r_g^2),
\label{eq:S-l-pri*}
\end{equation}
which describes hyperbolae that represent the geodesic trajectories of light rays in the post-Newtonian gravitational field of a mass monopole \cite{Turyshev-Toth:2017}. For an impact parameter that satisfies relation $b\geq R_\odot^\star$, these trajectories are outside the Sun, crossing from the geometric optics region behind the Sun into the interference region (see Fig.~\ref{fig:regions}).

Equation (\ref{eq:S-l-pri*+}) yields two families of solutions for  the points of stationary phase:
{}
\begin{equation}
\ell^{(1)}_{0}= \mp kr\Big(\sin\theta+\frac{2r_g}{r}\frac{1}{\sin\theta}\Big)+{\cal O}(\theta^5,r_g^2), \qquad {\rm and} \qquad
\ell^{(2)}_{0}= \pm \frac{2kr_g}{\sin\theta}+{\cal O}(\theta^5,r_g^2).
\label{eq:S-l-pri}
\end{equation}

The `$\pm$' or `$\mp$' signs in (\ref{eq:S-l-pri}) represent the families of rays propagating on opposite sides from the Sun. Also, two families of solutions represent two different waves. Thus, the family $\ell^{(1)}_{0}$ represents the incident wave with the rays whose trajectories are bent towards the Sun, obeying the eikonal approximation of geometric optics. The family $\ell^{(2)}_{0}$ describes the scattered wave, with rays that meet those of the incident wave beyond the point of their intersection with the focal line. Note that the interference region is not covered by the approximation (\ref{eq:P1l<}). The description of the interference region without plasma was given in \cite{Turyshev-Toth:2017}. In Sec.~\ref{sec:IF-region} we discuss the properties of the solution in the interference region in the presence of solar plasma. As discussed in \cite{Turyshev-Toth:2017,Turyshev-Toth:2018-grav-shadow}, the presence of both of these families of rays determine the structure of the three regions relevant for the SGL, namely the shadow, the geometric optics region and the interference region. As a result, the availability of these solutions help us develop the solution for (\ref{eq:Pi_s_exp1}).

We note that by extending the asymptotic expansion of $H^{+}_\ell(kr_g,kr)$ from (\ref{eq:Fass*}) to the order of ${\cal O}((kr)^{-(2n+1)})$ (i.e., using the WKB approximation as was done in Appendix \ref{sec:rad_eq_wkb}), the validity of the result (\ref{eq:S-l-pri}) extends to ${\cal O}(\theta^{2n+1})$. This fact was first observed in \cite{Turyshev-Toth:2018-plasma}) and used to improve the solution by including the terms of higher orders in $\theta$.

The first family of solutions of (\ref{eq:S-l-pri}), given by $\ell^{(1)}_0$,  allows us to compute the phase for the points of stationary phase (\ref{eq:S-l}) for the EM waves moving towards the interference region (a similar calculation was done in \cite{Turyshev-Toth:2018-plasma}):
{}
\begin{eqnarray}
\varphi^{[0]}_{\pm}(\ell^{(1)}_0)&=& \pm\textstyle{\frac{\pi}{4}}+kr\Big(-\textstyle{\frac{1}{2}}\theta^2+\textstyle{\frac{1}{24}}\theta^4\Big)-kr_g\ln kr(1-\cos\theta)-kr_g\ln 2kr +{\cal O}(kr\theta^6,kr_g\theta^4).
\label{eq:S-l2p}
\end{eqnarray}
To calculate $\varphi''(\ell)$ to ${\cal O}(\theta^6)$ as in (\ref{eq:S-l2p}), we need to include in the phase $\varphi^{[0]}_{\pm}(\ell)$ (\ref{eq:S-l}) another term $\propto \ell^6$, which may be taken  from  (\ref{eq:R_solWKB+=_bar-imp}). This allows us to compute $\varphi''(\ell^{(1)}_0)$:
\begin{eqnarray}
\dfrac{d^2\varphi^{[0]}_{\pm}}{d\ell^2} &=& \dfrac{1}{kr}\Big(1+\frac{\ell^2}{2k^2r^2}+\frac{3\ell^4}{8k^4r^4}+{\cal O}\big((kr)^{-6}\big)\Big)+\frac{2kr_g}{\ell^2},
\label{eq:S-l2+}
\end{eqnarray}
or, after substituting $\ell^{(1)}_0$, we have
\begin{eqnarray}
\varphi''(\ell^{(1)}_0)\equiv \dfrac{d^2\varphi^{[0]}_{\pm}}{d\ell^2} \Big|_{\ell=\ell^{(1)}_0} &=&
\dfrac{1}{kr}\Big(1+{\textstyle\frac{1}{2}}\theta^2+{\textstyle\frac{5}{24}}\theta^4+\frac{2r_g}{r\sin^2\theta}\big(1+\theta^2+{\textstyle\frac{7}{6}}\theta^4\big)+{\cal O}(\theta^6,\frac{r_g}{r}\theta^4)\Big).
\label{eq:S-l2}
\end{eqnarray}

The remaining integral is easy to evaluate using the method of stationary phase. Before we do that, we need to bring in the amplitude factor for the asymptotic expansion  $H^{+}_\ell(kr_g,kr)$ given by (\ref{eq:Fass*}). This factor, which we denote by $a(\ell)$, is readily available from (\ref{eq:R_solWKB+=_bar-imp}) in the following form:
{}
\begin{equation}
a(\ell)=\exp\Big[{\frac{\ell(\ell+1)}{4k^2r^2}}+{\frac{[\ell(\ell+1)]^2}{8k^4r^4}}\Big]+{\cal O}((kr)^{-6}).
\label{eq:sf1}
\end{equation}
Note that, if included in derivation of (\ref{eq:alpha*1*})--(\ref{eq:gamma*1*}), this term would produce corrections of the order of $1/(kr)$ smaller compared to the leading terms and, thus, negligible. In the case $\ell\gg1$ and specifically for $\ell^{(1)}_0$ is computed to be
{}
\begin{equation}
a(\ell^{(1)}_0) =1+{\textstyle\frac{1}{4}}\theta^2+{\textstyle\frac{7}{96}}\theta^4+\frac{r_g}{r}(1+{\textstyle\frac{5}{4}}\theta^2)+{\cal O}(\theta^6,\frac{r_g}{r}\theta^4).
\label{eq:sf}
\end{equation}
The fact that we did not use it in (\ref{eq:Fass*}) does not affect results of the calculations above. However, as we demonstrate below, its presence is needed to offset some of the terms that are present in the phase of (\ref{eq:S-l}). The significance of this term is realized in the fact that, for the method of stationary phase, it cancels out the contribution of the $\theta$-dependence in (\ref{eq:S-l2}), namely using result (\ref{eq:sf}) we derive
{}
\begin{equation}
a(\ell_0)\sqrt{\frac{2\pi}{\varphi''(\ell_0)}}=\sqrt{2\pi kr}\Big\{1-\frac{r_g}{r\sin^2\theta}+
\frac{r_g}{r}\big({\textstyle\frac{1}{2}}+\theta^2\big)+{\cal O}(\theta^6,\frac{r_g}{r}\theta^4)\Big\}.
\label{eq:sf*}
\end{equation}

Now, using (\ref{eq:sf*}), we have the amplitude of the integrand in (\ref{eq:Pi_s_exp1}), for $\ell \gg1$, for $\ell_0=\ell^{(1)}_0$, taking the form
{}
\begin{eqnarray}
A^{[0]}(\ell_0)a(\ell_0)\sqrt{\frac{2\pi}{\varphi''(\ell_0)}}&=&\frac{\ell_0\sqrt{\ell_0}}{\sqrt{2\pi \sin\theta}}
\Big(1+\frac{r_g}{r}\frac{\ell^2_0}{4k^2r^2} \Big) a(\ell_0)\sqrt{\frac{2\pi}{\varphi''(\ell_0)}}=\nonumber\\
&=&
(\mp1)^{\textstyle\frac{3}{2}} k^2r^2u^{-2}\sin\theta\Big(1+\frac{r_g}{r(1-\cos\theta)}+{\cal O}(\theta^4,\frac{r_g}{r}\theta^4)\Big),~~~~~~~
\label{eq:S-l3p+*}
\end{eqnarray}
where the superscript ${[0]}$ denotes the term with no plasma contribution. We can drop the $1/(ikr)$ term in the parentheses of this expression, as it is $1/(kr)$ times smaller in magnitude compared to the leading term.

As a result, the plasma-free part of the expression for $\delta\alpha^{[0]}(r,\theta)$ from (\ref{eq:Pi_s_exp1}) takes the form
{}
\begin{eqnarray}
\delta\alpha^{[0]}_\pm(r,\theta)&=&
-E_0u^{-1}\sin\theta \Big(1+\frac{r_g}{r(1-\cos\theta)}+{\cal O}(\theta^4,\frac{r_g}{r}\theta^4)\Big)e^{i\big(kr\cos\theta-kr_g\ln kr(1-\cos\theta)\big)}.
\label{eq:Pi_s_exp4+1*}
\end{eqnarray}
It is interesting that the phase of this expression is identical to the phase obtained from the equation for geodesics. The relevant results was obtained in Appendix~\ref{sec:geodesics} and \ref{sec:geom-optics} and are given by expressions (\ref{eq:X-eq4**}) and (\ref{eq:phase_t*}) correspondingly, where one has to disregard the plasma contribution.  This result agrees with that obtained in \cite{Turyshev-Toth:2017}.

We note again that by improving the asymptotic expansion of $H^{+}_\ell(kr_g,kr)$ (\ref{eq:Fass*}) (that, in a more complete form, is given by  (\ref{eq:R_solWKB+=_bar-imp})) to a higher order and extending the phase, from (\ref{eq:Fass*}), to ${\cal O}((kr)^{-(2n+1)})$ and the amplitude, $a(\ell)$, from (\ref{eq:sf}), to ${\cal O}((kr)^{-2n})$, the validity of (\ref{eq:sf*}) extends to ${\cal O}(\theta^{2n})$. If needed, this can be achieved by following the derivations presented  in Appendix \ref{sec:rad_eq_wkb}.

Now we consider the second family of solutions in (\ref{eq:S-l-pri}), given by $\ell^{(2)}_{0}$ (similar derivations were made in \cite{Turyshev-Toth:2018-grav-shadow}), which allows us to compute the stationary phase as
{}
\begin{eqnarray}
\varphi^{(2)}_{\pm}(\ell_0)&=&
\pm\textstyle{\frac{\pi}{4}}-kr_g\ln 2kr+kr_g\ln kr(1-\cos\theta)-2kr_g\ln \frac{kr_g}{e}+{\cal O}(kr_g \theta^2).
\label{eq:S-l27*+}
\end{eqnarray}
Using this result, from (\ref{eq:Pi_s_exp1}) we compute the phase of the corresponding solution (by combining the relativistic phase and the $\ell$-dependent contribution):
{}
\begin{eqnarray}
\varphi^{(2)}_{\pm}(r,\theta)&=&kr+kr_g\ln 2kr+\varphi^{(2)}_{\pm}(\ell_0)+\textstyle{\frac{\pi}{4}}=
k(r+r_g\ln kr(1-\cos\theta))+2\sigma_0+{\cal O}(kr_g \theta^2).
\label{eq:S-l27}
\end{eqnarray}

Now, using (\ref{eq:S-l2+}) and $\ell^{(2)}_{0}$ from (\ref{eq:S-l-pri}), we compute the second derivative of the phase with respect to $\ell$:
{}
\begin{equation}
\varphi''_{\pm}(\ell_0)=\frac{1}{kr}+\frac{\sin^2\theta}{2kr_g}\simeq\frac{\sin^2\theta}{2kr_g}\Big(1+\frac{2r_g}{r\sin^2\theta}\Big)+{\cal O}(\theta^5), \qquad {\rm thus, } \qquad
\sqrt{\frac{2\pi}{\varphi''(\ell_0)}}=\frac{\sqrt{4\pi kr_g}}{\sin\theta}\Big(1-\frac{r_g}{r\sin^2\theta}\Big)+{\cal O}(\theta^5).
\label{eq:S-l2202}
\end{equation}
Also, from (\ref{eq:sf1}), $a(\ell)$ is computed for $\ell^{(2)}_0$ to be $a(\ell)=1+{\cal O}(r_g^2)$.  At this point, we may evaliuate the amplitude of the integrand in (\ref{eq:Pi_s_exp1}), for $\ell \gg1$, for $\ell_0=\ell^{(2)}_0$, which is given as
{}
\begin{eqnarray}
A^{[0]}(\ell_0)a(\ell_0)\sqrt{\frac{2\pi}{\varphi''(\ell_0)}}&=&\frac{\ell_0\sqrt{\ell_0}}{\sqrt{2\pi \sin\theta}} \Big(1+\frac{r_g}{r}\frac{\ell^2_0}{4k^2r^2} \Big)a(\ell_0)\sqrt{\frac{2\pi}{\varphi''(\ell_0)}}=
(\mp1)^{\textstyle\frac{3}{2}} \frac{4k^2r^2_g}{\sin^3\theta}\Big(1-\frac{r_g}{r\sin^2\theta}\Big).~~~~
\label{eq:S-l3p+*=}
\end{eqnarray}
As a result, the plasma-free part of the expression for $\delta\alpha^{[0]}(r,\theta)$ from (\ref{eq:Pi_s_exp1}) for $\ell^{(1)}_0$ takes the form
{}
\begin{eqnarray}
\delta\alpha^{[0]}_\pm(r,\theta)&=&E_0\Big(\frac{2r_g}{r}\Big)^2\frac{1}{\sin^3\theta} e^{ik(r+r_g\ln kr(1-\cos\theta)+2i\sigma_0}
\sim {\cal O}(r_g^2).
\label{eq:Pi_rr}
\end{eqnarray}
We observe again that the phase of this expression is identical to the phase of radial wave obtained from the equation for geodesics. The relevant results were obtained in Appendix~\ref{sec:geodesics} and \ref{sec:geom-optics} and are given by (\ref{eq:X-eq4**_rad}) and (\ref{eq:phase_t-rad}) correspondingly, where one has to disregard the plasma contribution.  This result agrees with that obtained in \cite{Turyshev-Toth:2017}. Therefore, based on (\ref{eq:Pi_rr}) we conclude that to the order of ${\cal O}(r_g^2)$, there is no scattered wave in the radial direction which is consistent with the results reported in \cite{Turyshev-Toth:2017}.

The results (\ref{eq:Pi_s_exp4+1*}) and (\ref{eq:Pi_rr}) are the radial components of the EM wave corresponding to the two families of the impact parameters given  by (\ref{eq:S-l-pri}). We use these solutions to determine the EM field in the geometric optics region.

\subsubsection{Evaluating the term with plasma contribution}
\label{sec:yes-p-r}

We now turn our attention to the term in (\ref{eq:Pi_s_exp1}) that contains the contribution from the plasma-induced phase shift. The relevant $\ell$-dependent part of the phase is given as
{}
\begin{equation}
\varphi^{[\tt p]}_{\pm}(\ell)=\pm\big(\ell\theta+\textstyle{\frac{\pi}{4}}\big)+2\sigma_\ell+\dfrac{\ell^2}{2kr}+\dfrac{\ell^4}{24k^3r^3}+2\delta^\star_\ell+{\cal O}\big((kr)^{-5}\big),
\label{eq:S-l*p}
\end{equation}
with the plasma contribution clearly shown. From the definition (\ref{eq:delta-D*-av0WKB}) and (\ref{eq:a_b_del-r_mu0}), this plasma phase shift is given as
{}
\begin{eqnarray}
2\delta_\ell^*
=-\frac{4\pi e^2R_\odot}{m_ec^2k}\Big\{
\alpha_2 \frac{\pi}{2}\frac{R_\odot}{b}+
\sum_{i>2} {\alpha_i}Q^\star_{\beta_i}\Big(\frac{R_\odot}{b}\Big)^{\beta_i-1}\Big\}\equiv -\frac{2\pi e^2R_\odot}{m_ec^2k}\sum_i\frac{\alpha_i\beta_i}{\beta_i-1}B[{\textstyle\frac{1}{2}}\beta_i+{\textstyle\frac{1}{2}},{\textstyle\frac{1}{2}}]\Big(\frac{R_\odot}{b}\Big)^{\beta_i-1}.
\label{eq:a_b_del-r+-tot*}
\end{eqnarray}

The phase shift $2\delta_\ell^*$ relates to the semiclassical angle of deflection of a light ray, $\delta\theta_{\tt p}$, as $2\delta\theta_{\tt p}= {d 2\delta^*_\ell}/{d \ell}$  \cite{Newton-book-2013}. This angle may be computed from (\ref{eq:delta-D*-av0WKB}) and (\ref{eq:a_b_del-r_mu0}) by taking into account the semiclassical relation between the partial momenta, $\ell$, and the impact parameter, $b$, given as $\ell=kb$. As a result, the angle of light deflection by the solar plasma is computed to be
{}
\begin{eqnarray}
2\delta\theta_{\tt p}=\frac{d 2\delta^*_\ell}{k d b}
=\frac{4\pi e^2}{m_e\omega^2}\Big\{
\alpha_2 \frac{\pi}{2}\Big(\frac{R_\odot}{b}\Big)^2+
\sum_{i>2} {\alpha_i}\big(\beta_i-1\big)Q^\star_{\beta_i}\Big(\frac{R_\odot}{b}\Big)^{\beta_i}\Big\}\equiv
\frac{2\pi e^2}{m_e\omega^2}\sum_i \alpha_i\beta_i
B[{\textstyle\frac{1}{2}}\beta_i+{\textstyle\frac{1}{2}},{\textstyle\frac{1}{2}}]\Big(\frac{R_\odot}{b}\Big)^{\beta_i}.
\label{eq:ang*}
\end{eqnarray}
Note that expression (\ref{eq:ang*})  agrees with that derived in \cite{Giampieri:1994kj,Bertotti-Giampieri:1998} and used in a recent test of general relativity  using radio links with the Cassini spacecraft  \cite{Bertotti-etal-Cassini:2003}. Here, we provide a rigorous wave-optical treatment of the problem to establish the form of the refraction angle and the entire EM field as it propagates  through the solar system.
In fact, following \cite{Turyshev-Toth:2018-plasma}, using the phenomenological model (\ref{eq:model}) in (\ref{eq:ang*}), we estimate the plasma deflection angle, $\delta\theta_{\tt p}$, as a function of the impact parameter and the wavelength:
{}
\begin{eqnarray}
\delta\theta_{\tt p}=\Big\{
6.62\times 10^{-13}\Big(\frac{R_\odot}{b}\Big)^{16}+
2.05\times 10^{-13}\Big(\frac{R_\odot}{b}\Big)^{6}+
2.42\times 10^{-16}\Big(\frac{R_\odot}{b}\Big)^2\Big\}\Big(\frac{\lambda}{1~\mu{\rm m}}\Big)^2,
\label{eq:ang*ip}
\end{eqnarray}
which suggests that for sungrazing rays (i.e., for the rays with impact parameter $b\simeq R_\odot$), the bending angle (\ref{eq:ang*ip}) reaches the value of $\delta\theta_{\tt p}(R_\odot)=8.67\times 10^{-13} \,\big({\lambda}/{1~\mu{\rm m}}\big)^2$ rad, which is large for radio wavelengths, but negligible in optical or IR bands. For typical observing situations with reasonable Sun-Earth-probe separation angles \cite{Bertotti-Giampieri:1998,DSN-handbook-2017,Verma-etal:2013}, expression (\ref{eq:ang*}) provides a good description. This, once again, justifies the application of the eikonal approximation.

Examining (\ref{eq:ang*ip}) as a function of the impact parameter, we see that the first two terms in this expression diminish rather rapidly with the quadratic term in (\ref{eq:ang*ip}) becoming dominant after $b\simeq 8R_\odot$. However, this value of $b$ corresponds to a focal region at the heliocentric distance of $z=b^2/2r_g\sim3.5\times 10^4$~AU, which is beyond any practical interest as far as imaging with the SGL is concerned. For a focal region at $600~{\rm AU}\lesssim z\lesssim 1000~{\rm AU}$, knowledge of the properties of the solar corona at small impact parameters $1.05 R_\odot\lesssim b\lesssim 1.35 R_\odot$ is the most relevant.

Coming back to the phase (\ref{eq:S-l*p}), we see that this phase is stationary when $d\varphi^{[\tt p]}_{\pm}/d\ell=0$, which, similarly to (\ref{eq:S-l-pri*})--(\ref{eq:S-l-pri}),  implies
{}
\begin{equation}
\pm\theta -2\arctan \frac{kr_g}{\ell} +\frac{\ell}{kr}\Big(1+\frac{\ell^2}{6k^2r^2}\Big)+2\delta\theta_{\tt p}={\cal O}\big((kr)^{-5},r_g^2\big).
\label{eq:S-l-pri*p0}
\end{equation}

Similarly to (\ref{eq:S-l-pri*}), for small angles, $\theta\simeq b/r$, and large partial momenta, $\ell\simeq k R_\odot\gg kr_g$, equation (\ref{eq:S-l-pri*p0}) could be rewritten in the following from
{}
\begin{equation}
\frac{\ell}{kr} = \mp {\theta}\Big(1-{\textstyle\frac{1}{6}}\theta^2\Big)-2\delta\theta_{\tt p}+\frac{2kr_g}{\ell}+{\cal O}(\theta^5, r_g^2),
\label{eq:S-l-pri*2}
\end{equation}
or, equivalently,
{}
\begin{equation}
\frac{\ell}{kr} = \mp \sin{\theta}-2\delta\theta_{\tt p}+\frac{2kr_g}{\ell}+{\cal O}(\theta^5, r_g^2,\delta\theta_{\tt p}^3).
\label{eq:S-l-pri*2*}
\end{equation}

Equation (\ref{eq:S-l-pri*2*}) yields two families of solutions for  the points of stationary phase:
{}
\begin{equation}
\ell^{(1)}_{0}= \mp kr\Big(\sin\theta\pm 2\delta\theta_{\tt p} +\frac{2r_g}{r}\frac{1}{\sin\theta}\Big)+{\cal O}(\theta^5,r_g^2,r_g\delta\theta_{\tt p}), \qquad {\rm and} \qquad
\ell^{(2)}_{0}= \pm \frac{2kr_g}{\sin\theta}+{\cal O}(\theta^5,r_g^2,r_g\delta\theta_{\tt p}),
\label{eq:S-l-pp}
\end{equation}
where we neglected the terms  of the order of $r_g\omega^2_{\rm p}/\omega^2$ or, equivalently, the terms $\propto r_g\delta\theta_{\tt p}$.

With the results given in (\ref{eq:S-l-pp}), for the first family of solutions, we may compute the needed expressions for the value of the phase along the path of stationary phase:
{}
\begin{eqnarray}
\varphi^{[\tt p]}_{\pm}(\ell^{(1)}_0)&=&
\pm{\textstyle{\frac{\pi}{4}}}+kr\big(-{\textstyle{\frac{1}{2}}}\theta^2+{\textstyle{\frac{1}{24}}}\theta^4\big)-kr_g\ln kr(1-\cos\theta)-kr_g\ln 2kr +2\delta^\star_\ell+{\cal O}\Big(\theta^5\delta\theta_{\tt p},\delta\theta^2_{\tt p},kr_g \theta^2\Big),
\label{eq:S-l2*pd1}
\end{eqnarray}
and for the second derivative of the phase along the same path, similarly to (\ref{eq:S-l2}), from (\ref{eq:S-l2+}) we have
{}
\begin{eqnarray}
\frac{d^2\varphi^{[\tt p]}_{\pm}}{d\ell^2} &=&\dfrac{1}{kr}\Big(1+\frac{\ell^2}{2k^2r^2}+\frac{3\ell^4}{8k^4r^4}+{\cal O}((kr)^{-6})\Big)+\frac{2kr_g}{\ell^2}+\dfrac{d^2 2\delta^\star_\ell}{d\ell^2},
\label{eq:S-l2*pdd=}
\end{eqnarray}
which, for the first family of solutions, $\ell^{(1)}$, from (\ref{eq:S-l-pp}) yields:
{}
\begin{eqnarray}
\varphi''(\ell^{(1)}_0)\equiv \dfrac{d^2\varphi^{[0]}_{\pm}}{d\ell^2} \Big|_{\ell=\ell^{(1)}_0} &=&
\dfrac{1}{kr}\Big(1+{\textstyle\frac{1}{2}}\theta^2+{\textstyle\frac{5}{24}}\theta^4+\frac{2r_g}{r\sin^2\theta}\big(1+\theta^2+{\textstyle\frac{7}{6}}\theta^4\big)+{\cal O}(\theta^6,\frac{r_g}{r}\theta^4)\Big)+\dfrac{d^2 2\delta^\star_\ell}{d\ell^2}.
\label{eq:S-l2*=}
\end{eqnarray}

Using (\ref{eq:ang*}), we estimate the magnitude of the second term in this expression:
{}
\begin{eqnarray}
\dfrac{d^2 2\delta^\star_\ell}{d\ell^2}= \frac{2d \delta\theta_{\tt p}}{k d b}=
-\frac{1}{kR_\odot} \frac{2\pi e^2}{m_e\omega^2}\sum_i \alpha_i\beta^2_i
B[{\textstyle\frac{1}{2}}\beta_i+{\textstyle\frac{1}{2}},{\textstyle\frac{1}{2}}]\Big(\frac{R_\odot}{b}\Big)^{\beta_i+1}.
\label{eq:ang*-dd}
\end{eqnarray}

Similarly to \cite{Turyshev-Toth:2018-plasma}, we evaluated this quantity with
the empirical model (\ref{eq:model}). We see that for the smallest impact parameter $b=R_\odot$ this quantity takes the largest value of ${d^2 (2\delta^\star_\ell)}/{d\ell^2}=1.57\times 10^{-26}\,(\lambda/(1~\mu{\rm m})^3$. For optical wavelengths, even at the heliocentric distance of $r\simeq 6.5\times 10^3$~AU, this term is over $10^4$ times smaller than the $1/(kr)$ term in (\ref{eq:S-l2*=}), representing a small correction to $\varphi''(\ell_0)$ that may be neglected for our purposes. This is equivalent to treating the deflection angle $\delta\theta_{\tt p}$ constant, which is consistent with the eikonal (or high-energy) approximation \cite{Akhiezer-Pomeranchuk:1950,Glauber-Matthiae:1970,Semon-Taylor:1977,Sharma-etal:1988,Sharma-Somerford:1990,Sharma-Sommerford-book:2006}.

As a result, the expression for the second derivative of the phase from (\ref{eq:S-l2*=}) takes the form equivalent to (\ref{eq:S-l2}):
{}
\begin{eqnarray}
\varphi''(\ell^{(1)}_0)\equiv \dfrac{d^2\varphi^{[0]}_{\pm}}{d\ell^2} \Big|_{\ell=\ell^{(1)}_0} &=&
\dfrac{1}{kr}\Big(1+{\textstyle\frac{1}{2}}\theta^2+{\textstyle\frac{5}{24}}\theta^4+\frac{2r_g}{r\sin^2\theta}\big(1+\theta^2+{\textstyle\frac{7}{6}}\theta^4\big)+{\cal O}(\theta^6,\frac{r_g}{r}\theta^4,\delta\theta_{\tt p}^2,\frac{\delta\theta_{\tt p}}{kb})\Big).
\label{eq:S-l2*p2d}
\end{eqnarray}

The relevant, plasma-dependent part in the integral in (\ref{eq:Pi_s_exp1}) is now easy to evaluate using the method of stationary phase. Similarly to (\ref{eq:S-l3p+*}), we have the amplitude of the plasma-dependent  term in (\ref{eq:Pi_s_exp1}), evaluated to be
{}
\begin{eqnarray}
A^{[\tt p]}(\ell_0)a(\ell_0)\sqrt{\frac{2\pi}{\varphi''(\ell_0)}}&=&\frac{\ell_0\sqrt{\ell_0}}{\sqrt{2\pi \sin\theta}}
\Big(1+\frac{r_g}{r}\frac{\ell^2_0}{4k^2r^2} \Big)
a(\ell_0)\sqrt{\frac{2\pi}{\varphi''(\ell_0)}}=\nonumber\\
&=&(\mp1)^{\textstyle\frac{3}{2}}k^2r^2u^{-2}\sqrt{1\pm \frac{2\delta\theta_{\tt p}}{\sin\theta}}\sin(\theta\pm2\delta\theta_{\tt p})
\Big(1+\frac{r_g}{r(1-\cos\theta)}+{\cal O}(\theta^4,\delta\theta^2_{\tt p})\Big),
\label{eq:S-l3p+**p}
\end{eqnarray}
where the superscript ${[\tt p]}$ denotes the term due to the plasma phase shift. Using the expression relating the angle $\theta$ with the unperturbed direction  of light propagation, $\sin\theta \simeq b/r$, we may evaluate the square rooted expression:
{}
\begin{eqnarray}
\sqrt{1\pm \frac{2\delta\theta_{\tt p}}{\sin\theta}}=1\pm\frac{\delta\theta_{\tt p}}{\sin\theta}+{\cal O}(\delta\theta^2_{\tt p})\simeq 1\pm \frac{r\delta\theta_{\tt p}}{b}+{\cal O}(\delta\theta^2_{\tt p}).
\label{eq:S-sr}
\end{eqnarray}
Considering  (\ref{eq:ang*}), we see that the largest value of the bending angle, $\delta\theta_{\tt p}$,  is reached at the smallest impact parameters, $b=R_\odot$, limiting the size of this angle as  $\delta\theta_{\tt p}(R_\odot)\leq 8.65\times 10^{-13} \,\big({\lambda}/{1~\mu{\rm m}}\big)^2$ rad, resulting in the size of the ratio in (\ref{eq:S-sr}) of $r\delta \theta_{\tt p}/b\lesssim 1.02\times 10^{-7}\,\big({\lambda}/{1~\mu{\rm m}}\big)^2$, which is small for radio-wavelengths ($\lambda\sim 1$~mm), but is negligible for optical band. Treating the impact parameter as $b=\sqrt{2r_g r}$ \cite{Turyshev-Toth:2017,Turyshev-Toth:2018-grav-shadow}  and taking $\delta \theta_{\tt p}(b)$ from (\ref{eq:ang*}) together with empirical model (\ref{eq:model}), we see that for optical wavelengths the second term in (\ref{eq:S-sr}) is always below $10^{-7}$ and never becomes significant. Therefore, we omit this term from further consideration.

As a result, similarly to (\ref{eq:Pi_s_exp4+1*}), we obtain the contribution of the plasma-dependent term in  (\ref{eq:Pi_s_exp1}) in the  form
{}
\begin{eqnarray}
\delta\alpha^{[{\tt p}]}_\pm(r,\theta)&=&
E_0u^{-1}\sin\big(\theta\pm2\delta\theta_{\tt p}\big)\Big(1+\frac{r_g}{r(1-\cos\theta)}\Big)
e^{i\big(kr\cos\theta-kr_g\ln kr(1-\cos\theta)+2\delta^*_\ell\big)}+{\cal O}(\theta^4,\delta\theta^2_{\tt p},\frac{r_g}{r}\theta^4,\frac{r\delta\theta_{\tt p}}{b}\big).~~~~
\label{eq:Pi_s_exp4+1**p}
\end{eqnarray}
The phase of this expression is identical to the phase obtained from the equation for geodesics. The relevant results were obtained in Appendix~\ref{sec:geodesics} and \ref{sec:geom-optics} and are given by expressions (\ref{eq:X-eq4**}) and (\ref{eq:phase_t*}).

With the expressions
(\ref{eq:Pi_s_exp4+1*}) and (\ref{eq:Pi_s_exp4+1**p}) at hand, we may now present the  quantity $\alpha(r,\theta)$ from  (\ref{eq:Pi_s_exp1}) as
{}
\begin{eqnarray}
\alpha(r,\theta)&=&\delta\alpha^{[0]}_\pm(r,\theta)+\delta\alpha^{[\tt p]}_\pm(r,\theta)=\nonumber\\
&&\hskip -50pt \,=
E_0u^{-1}\Big(1+\frac{r_g}{r(1-\cos\theta)}\Big)\Big(
\sin\Big(\theta\pm\frac{d2\delta^\star_\ell}{d\ell}\Big)e^{i2\delta^*_\ell}-\sin\theta\Big)e^{i\big(kr\cos\theta-kr_g\ln kr(1-\cos\theta)\big)}+
{\cal O}(\theta^4,\delta\theta^2_{\tt p},\frac{r_g}{r}\theta^4,\frac{r\delta\theta_{\tt p}}{b}).~~~~~~
\label{eq:Pi_s_exp4+1*=}
\end{eqnarray}
This expression indicates that the scattered wave---which is governed by the expressions (\ref{eq:alpha*1*})--(\ref{eq:gamma*1*}) that result from the scattering amplitude (\ref{eq:f-v*+}), to first order in gravity and plasma contributions or up to the terms of the order of ${\cal O}(r_g(\omega_{\tt p}/\omega)^2)$---may be given by the difference of two waves: the wave that moves on the effective background given by both gravity and plasma and the one that moves only on the gravitational background.

Similarly to (\ref{eq:Pi_rr}), for the second family of solutions from (\ref{eq:S-l-pp}), we obtain $\delta\alpha^{[0]}_\pm(r,\theta)\sim {\cal O}(r_g^2)$. Therefore, there is no scattered wave in the radial direction:
{}
\begin{eqnarray}
\alpha_{\tt s}(r,\theta)&=&\delta\alpha^{[0]}_\pm(r,\theta)+\delta\alpha^{[\tt p]}_\pm(r,\theta)\sim{\cal O}(r^2_g).~~~
\label{eq:Pi_s_rrd}
\end{eqnarray}

Using the approach presented above, we may now evaluate the scattering factors  $\beta(r,\theta)$ and $\gamma(r,\theta)$ needed to determine the other components of the EM field.

\subsection{Evaluating the  function $\beta(r,\theta)$}
\label{sec:amp_func-beta}

To investigate the behavior $\beta(r,\theta)$ from (\ref{eq:beta*1*}) we neglect terms of the order of $\propto r_g/kr^2$ and obtain the following  expression for $\beta(r,\theta)$:
{}
\begin{eqnarray}
\beta(r, \theta)&=& E_0\frac{e^{ik(r+r_g\ln 2kr)}}{ikr}\sum_{\ell=kR_\odot^\star}^\infty \frac{(\ell+{\textstyle\frac{1}{2}})}{\ell(\ell+1)}
e^{i\big(2\sigma_\ell+\frac{\ell(\ell+1)}{2kr}+\frac{[\ell(\ell+1)]^2}{24k^3r^3}\big)}
\Big(e^{i2\delta^*_\ell}-1\Big)\times\nonumber\\
&&\hskip 46pt \times\,\Big\{
\frac{\partial P^{(1)}_\ell(\cos\theta)}{\partial \theta}
\Big(1-u^{-2}\Big(\frac{\ell(\ell+1)}{2k^2r^2}-\frac{[\ell(\ell+1)]^2}{8k^4r^4}\Big)\Big)+\frac{P^{(1)}_\ell(\cos\theta)}{\sin\theta}\Big\}.
  \label{eq:beta**1}
\end{eqnarray}

To evaluate the magnitude of the function $\beta(r, \theta)$, we  need to establish the asymptotic behavior of $P^{(1)}_{l}(\cos\theta)/\sin\theta$ and $\partial P^{(1)}_{l}(\cos\theta)/\partial \theta$. For fixed $\theta$ and $\ell\rightarrow\infty$ this behavior is given\footnote{We note that, for any large $\ell$, formulae (\ref{eq:pi-l*})--(\ref{eq:tau-l*}) are insufficient in a region close to the forward direction $(\theta=0$) and back direction ($\theta=\pi$). More precisely, (\ref{eq:pi-l*})--(\ref{eq:tau-l*}) hold for $\sin\theta\gg1/\ell$ (see discussion in \cite{Turyshev-Toth:2018}.) Nevertheless, these expressions are sufficient for our purposes as in the region of interest the latter condition is satisfied.} \cite{vandeHulst-book-1981} as (this can be obtained directly from (\ref{eq:P1l<})):
{}
\begin{eqnarray}
\frac{P^{(1)}_\ell(\cos\theta)}{\sin\theta}
&=& \Big(\frac{2\ell}{\pi\sin^3\theta}\Big)^{\frac{1}{2}} \sin\Big((\ell+{\textstyle\frac{1}{2}})\theta-{\textstyle\frac{\pi}{4}}\Big)+{\cal O}(\ell^{-\textstyle\frac{3}{2}}),
\label{eq:pi-l*}\\
\frac{dP^{(1)}_\ell(\cos\theta)}{d\theta}
&=&  \Big(\frac{2\ell^3}{\pi\sin\theta}\Big)^{\frac{1}{2}} \cos\Big((\ell+{\textstyle\frac{1}{2}})\theta-{\textstyle\frac{\pi}{4}}\Big)+{\cal O}(\ell^{-\textstyle\frac{1}{2}}).
\label{eq:tau-l*}
\end{eqnarray}

With these approximations, the function $\beta(r,\theta)$ in the region outside the geometric shadow (i.e., not on the optical axis), takes the following form:
{}
\begin{eqnarray}
\beta(r,\theta)&=&E_0\frac{e^{ik(r+r_g\ln 2kr)}}{ikr}
\sum_{\ell=kR_\odot^\star}^\infty \frac{(\ell+{\textstyle\frac{1}{2}})}{\ell(\ell+1)}
e^{i\big(2\sigma_\ell+\frac{\ell(\ell+1)}{2kr}+\frac{[\ell(\ell+1)]^2}{24k^3r^3}\big)}\Big(e^{2i\delta^*_\ell}-1\Big)
\times\nonumber\\
&&\times\,
\Big\{
 \Big(\frac{2\ell^3}{\pi\sin\theta}\Big)^{\frac{1}{2}} \Big(1-\frac{\ell(\ell+1)}{2k^2r^2}-\frac{[\ell(\ell+1)]^2}{8k^4r^4}\Big)\cos\Big((\ell+{\textstyle\frac{1}{2}})\theta-{\textstyle\frac{\pi}{4}}\Big)+\Big(\frac{2\ell}{\pi\sin^3\theta}\Big)^{\frac{1}{2}} \sin\Big((\ell+{\textstyle\frac{1}{2}})\theta-{\textstyle\frac{\pi}{4}}\Big)
\Big\}.~~~~~~~
\label{eq:S1-v0s}
\end{eqnarray}
For large $\ell\gg1$, the first term in the curly brackets in (\ref{eq:S1-v0s}) dominates, so that this expression may be given as
{}
\begin{eqnarray}
\beta(r,\theta)&=&E_0\frac{e^{ik(r+r_g\ln 2kr)}}{ikr}
\sum_{\ell=kR_\odot^\star}^\infty \frac{(\ell+{\textstyle\frac{1}{2}})}{\ell(\ell+1)}
\Big(\frac{2\ell^3}{\pi\sin\theta}\Big)^{\frac{1}{2}} \Big(1-\frac{\ell(\ell+1)}{2k^2r^2}-\frac{[\ell(\ell+1)]^2}{8k^4r^4}\Big)
\times\nonumber\\
&&\hskip 80pt \times\,
\Big(e^{2i\delta^*_\ell}-1\Big)e^{i\big(2\sigma_\ell+\frac{\ell(\ell+1)}{2kr}+\frac{[\ell(\ell+1)]^2}{24k^3r^3}\big)}
\cos\Big((\ell+{\textstyle\frac{1}{2}})\theta-{\textstyle\frac{\pi}{4}}\Big).~~~~~~~
\label{eq:S1-v0s+}
\end{eqnarray}

To evaluate $\beta(r,\theta)$ from the expression (\ref{eq:S1-v0s+}), we again use the method of stationary phase. For this, representing (\ref{eq:S1-v0s+}) in the form of an integral over $\ell$, we have:
{}
\begin{eqnarray}
\beta(r,\theta)&=&-E_0\frac{e^{ik(r+r_g\ln 2kr)}}{kr}
\int_{\ell=kR_\odot^\star}^\infty \hskip-5pt   \frac{\sqrt{\ell}d\ell}{\sqrt{2\pi\sin\theta}} \Big(1-\frac{\ell^2}{2k^2r^2}-\frac{\ell^4}{8k^4r^4}\Big)
\times\nonumber\\
&&\hskip 100pt \times\,
\Big(e^{2i\delta^*_\ell}-1\Big)
e^{i\big(2\sigma_\ell+\frac{\ell^2}{2kr}+\frac{\ell^4}{24k^3r^3}\big)}
\Big(e^{i(\ell\theta+{\textstyle\frac{\pi}{4}})}-e^{-i(\ell\theta+{\textstyle\frac{\pi}{4}})}\Big).~~~~~~~
\label{eq:S1-v0s+int*}
\end{eqnarray}

As we have done with (\ref{eq:Pi_s_exp1}), we treat this integral as a sum of  two integrals: a plasma-free and a plasma-dependent term. Expression (\ref{eq:S1-v0s+int*}) shows that the $\ell$-dependent parts of the phase have a structure identical to (\ref{eq:S-l}) and (\ref{eq:S-l*p}). Therefore, the same solutions for the points of stationary phase apply. As a result, using (\ref{eq:S-l-pri}) and (\ref{eq:S-l2}), from (\ref{eq:S1-v0s+int*}) for the part of the integral that does not depend on the plasma phase shift, $\delta^\star_\ell$ and for the first family of solutions (\ref{eq:S-l-pri}), to the order of ${\cal O}(\theta^6, ({r_g}{/r})\theta^2)$, we have
{}
\begin{eqnarray}
A^{[0]}(\ell_0)a(\ell_0)\sqrt{\frac{2\pi}{\varphi''(\ell_0)}}&=& \frac{\sqrt{\ell_0}}{\sqrt{2\pi\sin\theta}}\Big(1-\frac{\ell^2_0}{2k^2r^2}-\frac{\ell^4_0}{8k^4r^4}\Big)a(\ell_0)\sqrt{\frac{2\pi}{\varphi''(\ell_0)}}=
\sqrt{\mp1}kru^{-1}\Big(\cos\theta-\frac{r_g}{r}\Big).~~~
\label{eq:S-l3p}
\end{eqnarray}

As a result, similarly to (\ref{eq:Pi_s_exp4+1*}), the expression for the $\delta \beta^{[0]}_\pm(r,\theta)$  takes the form
{}
\begin{eqnarray}
\delta \beta^{[0]}_\pm(r,\theta)&=&
-E_0u^{-1}\Big(\cos\theta-\frac{r_g}{r}\Big) e^{ik\big(r\cos\theta-r_g\ln kr(1-\cos\theta)\big)}
+{\cal O}(\theta^6, \frac{r_g}{r}\theta^2).~~~~~
\label{eq:Pi_s_exp4+1pp}
\end{eqnarray}

Next, using the $\ell$-dependent phase (\ref{eq:S-l*p}) with the plasma phase shift included and the relevant expressions (\ref{eq:S-l2*pd1}) and (\ref{eq:S-l2*p2d}), to the order of ${\cal O}(\theta^6, \delta\theta_{\tt p}^2, ({r_g}/{r})\theta^2,{r\delta\theta_{\tt p}}/{b})$, we have:
{}
\begin{eqnarray}
A^{[\tt p]}(\ell_0)a(\ell_0)\sqrt{\frac{2\pi}{\varphi''(\ell_0)}}&=&\frac{\sqrt{\ell_0}}{\sqrt{2\pi\sin\theta}} \Big(1-\frac{\ell^2_0}{2k^2r^2}-\frac{\ell^4_0}{8k^4r^4}\Big)a(\ell_0)\sqrt{\frac{2\pi}{\varphi''(\ell_0)}}=
\sqrt{\mp1}kru^{-1}\Big(\cos\Big(\theta\pm\frac{d2\delta^\star_\ell}{d\ell}\Big)-\frac{r_g}{r}\Big).~~~~~~
\label{eq:S-l3pp}
\end{eqnarray}
Thus, the plasma-dependent term in (\ref{eq:S1-v0s+int*}), namely $\delta \beta^{[\tt p]}_\pm(r,\theta)$, takes the form
{}
\begin{eqnarray}
\delta \beta^{[\tt p]}_\pm(r,\theta)&=&
E_0u^{-1}\Big(\cos\Big(\theta\pm\frac{d2\delta^\star_\ell}{d\ell}\Big)-\frac{r_g}{r}\Big)e^{i\big(k(r\cos\theta-r_g\ln kr(1-\cos\theta))+2\delta^\star_\ell\big)}
+{\cal O}(\theta^6, \delta\theta^2_{\tt p},\frac{r_g}{r}\theta^2,\frac{r\delta\theta_{\tt p}}{b}).
\label{eq:Pi_s_exp4+1*p}
\end{eqnarray}

Using the expressions (\ref{eq:Pi_s_exp4+1pp}) and (\ref{eq:Pi_s_exp4+1*p}), we present the integral (\ref{eq:S1-v0s+int*}) as
{}
\begin{eqnarray}
\beta(r,\theta)&=&\delta \beta^{[0]}_\pm(r,\theta)+\delta \beta^{[{\tt p}]}_\pm(r,\theta)=\nonumber\\
&&\hskip -50pt \,=
\,E_0u^{-1}\Big\{\Big(\cos\Big(\theta\pm\frac{d2\delta^\star_\ell}{d\ell}\Big)-\frac{r_g}{r}\Big)e^{i2\delta^\star_\ell}
-\Big(\cos\theta-\frac{r_g}{r}\Big)
+{\cal O}(\theta^6,\delta\theta^2_{\tt p},\frac{r_g}{r}\theta^2,\frac{r\delta\theta_{\tt p}}{b})
\Big\}e^{ik\big(r\cos\theta-r_g\ln kr(1-\cos\theta)\big)}.~~~
\label{eq:Pi_s_exp4+1*=*}
\end{eqnarray}

Now we turn our attention to the second family of solutions in (\ref{eq:S-l-pri}).
Similarly to (\ref{eq:S-l3p+*=}), we have
{}
\begin{eqnarray}
A^{[0]}(\ell_0)a(\ell_0)\sqrt{\frac{2\pi}{\varphi''(\ell_0)}}&=&\frac{\sqrt{\ell}}{\sqrt{2\pi\sin\theta}} \Big(1-\frac{\ell^2}{2k^2r^2}-\frac{\ell^4}{8k^4r^4}\Big)a(\ell_0)\sqrt{\frac{2\pi}{\varphi''(\ell_0)}}=
\sqrt{\pm1} \frac{kr_g}{2\sin^2\textstyle{\frac{1}{2}}\theta}+{\cal O}(\theta^4,r^2_g),~~~~
\label{eq:S-l3p+*=2}
\end{eqnarray}
which yields the following result for $\delta \beta^{[0]}_\pm(r,\theta)$:
{}
\begin{eqnarray}
\delta \beta^{[0]}_\pm(r,\theta)&=&
-E_0\frac{r_g}{2r\sin^2\textstyle{\frac{1}{2}}\theta}e^{i\big(k(r+r_g\ln kr(1-\cos\theta))+2\sigma_0\big)}
+{\cal O}(\theta^6, \frac{r_g}{r}\theta^2).~~~~~
\label{eq:beta-2}
\end{eqnarray}

In an analogous manner, the second family of solutions from (\ref{eq:S-l-pri}) results in the plasma-dependent factor $\delta \beta^{\tt [p]}_\pm(r,\theta)$:
{}
\begin{eqnarray}
\delta \beta^{\tt [p]}_\pm(r,\theta)&=& E_0\frac{r_g}{2r\sin^2\textstyle{\frac{1}{2}}\theta}e^{i\big(k(r+r_g\ln kr(1-\cos\theta))+2\sigma_0+2\delta^\star_\ell \big)}
+{\cal O}(\theta^6, \frac{r_g}{r}\theta^2).~~~~~
\label{eq:beta-2p}
\end{eqnarray}
The phase of this expression is identical to the phase of a radial wave obtained from the equation for geodesics. The relevant results were obtained in Appendix~\ref{sec:geodesics} and \ref{sec:geom-optics} and are given by (\ref{eq:X-eq4**_rad}) and (\ref{eq:phase_t-rad}).

Finally, using the expressions (\ref{eq:beta-2})--(\ref{eq:beta-2p}), we present the integral (\ref{eq:S1-v0s+int*}) for the second family of solutions (\ref{eq:S-l-pri}) as
{}
\begin{eqnarray}
\beta(r,\theta)&=&\delta \beta^{[0]}_\pm(r,\theta)+\delta \beta^{[{\tt p}]}_\pm(r,\theta)=
E_0\frac{r_g}{2r\sin^2\textstyle{\frac{1}{2}}\theta}
\Big(e^{i2\delta^\star_\ell}-1
+{\cal O}(\theta^6,\delta\theta^2_{\tt p},\frac{r_g}{r}\theta^2,\frac{r\delta\theta_{\tt p}}{b})
\Big)e^{i\big(k(r+r_g\ln kr(1-\cos\theta))+2\sigma_0\big)}.~~~~~~~
\label{eq:Pi_s_exp4+1*=*2}
\end{eqnarray}

Thus, for the scattered wave, to accepted approximation, the plasma contribution affects only the phase of the  wave and not its amplitude or direction of its propagation.

\subsection{Evaluating the  function $\gamma(r,\theta)$}
\label{sec:amp_func-der}

To determine the remaining components of the EM field (\ref{eq:DB-sol00p*}), we need to evaluate  the behavior of the function $\gamma(r,\theta)$ from (\ref{eq:gamma*1*}) that is given in the following from:
{}
\begin{eqnarray}
\gamma(r, \theta)&=& E_0\frac{e^{ik(r+r_g\ln 2kr)}}{ikr} \sum_{\ell=kR_\odot^\star}^\infty \frac{(\ell+{\textstyle\frac{1}{2}})}{\ell(\ell+1)}
e^{i\big(2\sigma_\ell+\frac{\ell(\ell+1)}{2kr}+\frac{[\ell(\ell+1)]^2}{24k^3r^3}\big)}
\Big(e^{i2\delta^*_\ell}-1\Big)\times\nonumber\\
&&\hskip 50pt \times\,\Big\{
\frac{\partial P^{(1)}_\ell(\cos\theta)}{\partial \theta}
+\frac{P^{(1)}_\ell(\cos\theta)}{\sin\theta}\Big(1-u^{-2}\Big(\frac{\ell(\ell+1)}{2k^2r^2}-\frac{[\ell(\ell+1)]^2}{8k^4r^4}\Big)\Big)\Big\}.
  \label{eq:gamma**1}
\end{eqnarray}

To evaluate this expression, we use the asymptotic behavior of $P^{(1)}_{l}(\cos\theta)/\sin\theta$ and $\partial P^{(1)}_{l}(\cos\theta)/\partial \theta$ given by (\ref{eq:pi-l*}) and (\ref{eq:tau-l*}), correspondingly, and rely on the method of stationary phase. Similarly to (\ref{eq:S1-v0s}), we drop the second term in the curly brackets in (\ref{eq:gamma**1}). The remaining expression for $\gamma(r, \theta)$, for large partial momenta, $\ell\gg1$,  is now determined by evaluating the following integral:
{}
\begin{eqnarray}
\gamma(r, \theta)&=& -E_0\frac{e^{ik(r+r_g\ln 2kr)}}{kr} \int_{\ell=kR_\odot^\star}^\infty \hskip -5pt
\frac{\sqrt{\ell}}{\sqrt{2\pi\sin\theta}}e^{i\big(2\sigma_\ell+\frac{\ell^2}{2kr}+\frac{\ell^4}{24k^3r^3}\big)}
\Big(e^{i2\delta^*_\ell}-1\Big)
\Big(e^{i(\ell\theta+{\textstyle\frac{\pi}{4}})}-e^{-i(\ell\theta+{\textstyle\frac{\pi}{4}})}\Big).
  \label{eq:gamma**1*}
\end{eqnarray}

Clearly, this expression yields the same equation to determine the points of the stationary phase (\ref{eq:S-l}) and (\ref{eq:S-l*p}) and, thus, all the relevant results obtained in Sec.~\ref{sec:radial-comp}. Therefore,  the $\ell$-dependent amplitude of (\ref{eq:gamma**1*}), which is independent on the plasma phase shift, $A^{[0]}$, is evaluated as
{}
\begin{equation}
A^{[0]}(\ell_0)a(\ell_0)\sqrt{\frac{2\pi}{\varphi''(\ell_0)}}=
\frac{\sqrt{\ell_0}}{\sqrt{2\pi\sin\theta}}a(\ell_0)\sqrt{\frac{2\pi}{\varphi''(\ell_0)}}=\pm\sqrt{\mp1}
kru+{\cal O}(\theta^5, \frac{r_g}{r}\theta^2).
\label{eq:S-l3p+0*}
\end{equation}

Therefore, the plasma-independent part of the function $\delta \gamma^{[0]}_\pm(r,\theta)$ is given as
{}
\begin{eqnarray}
\delta \gamma^{[0]}_\pm(r,\theta)&=&
-E_0u e^{i\big(kr\cos\theta-kr_g\ln kr(1-\cos\theta)\big)}+{\cal O}(\theta^5,\frac{r_g}{r}\theta^2).~~~~~
\label{eq:Pi_s_exp4+1gg}
\end{eqnarray}

Similarly, we have the following expression for the amplitude $A^{[\tt p]}$, which, with (\ref{eq:S-l2*pd1}) and (\ref{eq:S-l2*p2d}), is evaluated to be
{}
\begin{equation}
A^{[\tt p]}(\ell_0)a(\ell_0)\sqrt{\frac{2\pi}{\varphi''(\ell_0)}}=
\frac{\sqrt{\ell_0}}{\sqrt{2\pi\sin\theta}}a(\ell_0)\sqrt{\frac{2\pi}{\varphi''(\ell_0)}}=\pm\sqrt{\mp1}
kru
+{\cal O}(\theta^5, \frac{r_g}{r}\theta^2,\frac{r\delta\theta_{\tt p}}{b}).
\label{eq:S-l3p0**+pp}
\end{equation}

Therefore, the plasma-dependent term in (\ref{eq:gamma**1*}), $\delta \gamma^{[\tt p]}_\pm(r,\theta)$, takes the form
{}
\begin{eqnarray}
\delta \gamma^{[\tt p]}_\pm(r,\theta)&=&
E_0ue^{i\big(kr\cos\theta-kr_g\ln kr(1-\cos\theta)+2\delta^\star_\ell\big)}
+{\cal O}(\theta^6, \delta\theta^2_{\tt p},\frac{r\delta\theta_{\tt p}}{b},\frac{r_g}{r}\theta^2).
\label{eq:Pi_s_exp4+1*pg}
\end{eqnarray}

We may now use the expressions (\ref{eq:Pi_s_exp4+1gg}) and (\ref{eq:Pi_s_exp4+1*pg}) to present the integral (\ref{eq:gamma**1*}) as
{}
\begin{eqnarray}
\gamma(r,\theta)&=&\delta\gamma^{[0]}_\pm(r,\theta)+\delta\gamma^{[\tt p]}_\pm(r,\theta)=
E_0u\Big(
e^{i2\delta^*_\ell}-1\Big)e^{ikr\cos\theta-kr_g\ln kr(1-\cos\theta)}+{\cal O}(\theta^6,\delta\theta^2_{\tt p},\frac{r\delta\theta_{\tt p}}{b}).
\label{eq:Pi_s_exp4+}
\end{eqnarray}

Now, for the second family of solutions (\ref{eq:S-l-pri}), we get the following results for the plasma-independent term:
{}
\begin{equation}
A^{[0]}(\ell_0)a(\ell_0)\sqrt{\frac{2\pi}{\varphi''(\ell_0)}}=
\frac{\sqrt{\ell_0}}{\sqrt{2\pi\sin\theta}}a(\ell_0)\sqrt{\frac{2\pi}{\varphi''(\ell_0)}}=\sqrt{\pm1} \frac{kr_g}{2\sin^2\textstyle{\frac{1}{2}}\theta}+{\cal O}(\theta^4,r^2_g).
\label{eq:S-l3p+0*2}
\end{equation}
which yields a result for $\delta \gamma^{[0]}_\pm(r,\theta)$ that is identical to (\ref{eq:beta-2}):
{}
\begin{eqnarray}
\delta \gamma^{[0]}_\pm(r,\theta)&=&
-E_0\frac{r_g}{2r\sin^2\textstyle{\frac{1}{2}}\theta}e^{i\big(k(r+r_g\ln kr(1-\cos\theta))+2\sigma_0\big)}
+{\cal O}(\theta^6, \frac{r_g}{r}\theta^2).~~~~~
\label{eq:gamma-2}
\end{eqnarray}

The result for the plasma-dependent factor $\delta \gamma^{\tt [p]}_\pm(r,\theta)$ is identical to (\ref{eq:beta-2p}), namely:
{}
\begin{eqnarray}
\delta \gamma^{\tt [p]}_\pm(r,\theta)&=& E_0\frac{r_g}{2r\sin^2\textstyle{\frac{1}{2}}\theta}e^{i\big(k(r+r_g\ln kr(1-\cos\theta))+2\sigma_0+2\delta^\star_\ell \big)}
+{\cal O}(\theta^6, \frac{r_g}{r}\theta^2).~~~~~
\label{eq:gamma-2p}
\end{eqnarray}

As a result, using the expressions (\ref{eq:gamma-2}) and (\ref{eq:gamma-2p}), we present the integral (\ref{eq:S1-v0s+int*}) for the second family of solutions (\ref{eq:S-l-pri})  as follows
{}
\begin{eqnarray}
\gamma(r,\theta)&=&\delta \gamma^{[0]}_\pm(r,\theta)+\delta \gamma^{[{\tt p}]}_\pm(r,\theta)=
E_0\frac{r_g}{2r\sin^2\textstyle{\frac{1}{2}}\theta}
\Big(
e^{i2\delta^\star_\ell} -1
+{\cal O}(\theta^6,\delta\theta^2_{\tt p},\frac{r_g}{r}\theta^2,\frac{r\delta\theta_{\tt p}}{b})
\Big)e^{i\big(k(r+r_g\ln kr(1-\cos\theta))+2\sigma_0\big)}.~~~~~~
\label{eq:gamma_scat}
\end{eqnarray}

At this point, we have all the necessary ingredients to present the ultimate solution for the scattered EM field in the eikonal approximation.

\subsection{Solution for the EM field outside the termination shock}
\label{sec:EM-fieldsol}

To determine the components of the EM field, we use the expressions that we obtained for the functions $\alpha(r,\theta)$, $\beta(r,\theta)$ and $\gamma(r,\theta)$, which are given by (\ref{eq:Pi_s_exp4+1*=}), (\ref{eq:Pi_s_exp4+1*=*}) and (\ref{eq:Pi_s_exp4+}), correspondingly, and substitute them in (\ref{eq:DB-sol00p*}). As a result, we establish the solution for the scattered EM field in the region outside the termination shock boundary, which to the order of ${\cal O}(\theta^5,\delta\theta^2_{\tt p},({r_g}/{r})\theta^4,r_g^2,{r\delta\theta_{\tt p}}/{b})$ has the from:
{}
\begin{eqnarray}
\left( \begin{aligned}
{  \hat D}_r^{\tt p}& \\
{  \hat B}_r^{\tt p}& \\
  \end{aligned} \right) &=&  E_0u^{-1} \Big(1+\frac{r_g}{r(1-\cos\theta)}\Big)
\Big(
\sin\Big(\theta\pm\frac{d2\delta^\star_\ell}{d\ell}\Big)
e^{i2\delta^*_\ell}-\sin\theta \Big)
\left( \begin{aligned}
\cos\phi \\
\sin\phi  \\
  \end{aligned} \right)
  e^{i\big(k(r\cos\theta-r_g\ln kr(1-\cos\theta))-\omega t\big)},
  \label{eq:DB-r*+*p2}\\
  \left( \begin{aligned}
{ \hat D}^{\tt p}_\theta& \\
{ \hat B}^{\tt p}_\theta& \\
  \end{aligned} \right) &=&
  E_0u^{-1}
\Big(\Big(\cos\big(\theta\pm\frac{d2\delta^\star_\ell}{d\ell}\big)-\frac{r_g}{r}\Big)
 e^{i2\delta^*_\ell}-  \Big(\cos\theta-\frac{r_g}{r}\Big)
 \Big)
    \left( \begin{aligned}
\cos\phi \\
\sin\phi  \\
  \end{aligned} \right)
  e^{i\big(k(r\cos\theta-r_g\ln kr(1-\cos\theta))-\omega t\big)},
  \label{eq:DB-t-pl=p1}\\
\left( \begin{aligned}
{ \hat D}^{\tt p}_\phi& \\
{ \hat B}^{\tt p}_\phi& \\
  \end{aligned} \right) &=&
  E_0u
 \Big(  e^{i2\delta^*_\ell}- 1 \Big)
  \left( \begin{aligned}
-\sin\phi \\
\cos\phi  \\
  \end{aligned} \right)e^{i\big(k(r\cos\theta-r_g\ln kr(1-\cos\theta))-\omega t\big)}.
  \label{eq:DB-t-pl=p2}
\end{eqnarray}
Clearly, when plasma is absent the entire EM field given by (\ref{eq:DB-r*+*p2})--(\ref{eq:DB-t-pl=p2}) vanishes. Note that the phases and the amplitude factors of the terms above are consistent with those found with equation of the geodesics with and without presence of the plasma, as identified in Appendices~\ref{sec:geodesics} and \ref{sec:geom-optics}. In fact, the total scattered EM field given by (\ref{eq:DB-r*+*p2})--(\ref{eq:DB-t-pl=p2}) is shown as the difference of two waves propagating in different backgrounds: with and without the plasma. The resulting EM field given above describes the total effect of the solar plasma on the incident EM wave. At the same time, the EM field of the incident wave is produced by the Debye potential $\Pi_0$ from (\ref{eq:Pi-ass}) and
is given as \cite{Turyshev-Toth:2017}:
{}
\begin{eqnarray}
  \left( \begin{aligned}
{  \hat D}^{\tt (0)}_r& \\
{  \hat B}^{\tt (0)}_r& \\
  \end{aligned} \right) &=& E_0u^{-1}
\sin\theta \Big(1+\frac{r_g}{r(1-\cos\theta)}\Big)
\left( \begin{aligned}
\cos\phi \\
\sin\phi  \\
  \end{aligned} \right)
  e^{i\big(k(r\cos\theta-r_g\ln kr(1-\cos\theta))-\omega t\big)},
  \label{eq:DB-t-pl=p10} \\
    \left( \begin{aligned}
{  \hat D}^{\tt (0)}_\theta& \\
{  \hat B}^{\tt (0)}_\theta& \\
  \end{aligned} \right) &=& E_0u^{-1}
\Big(\cos\theta-\frac{r_g}{r}\Big)
\left( \begin{aligned}
\cos\phi \\
\sin\phi  \\
  \end{aligned} \right) e^{i\big(k(r\cos\theta-r_g\ln kr(1-\cos\theta))-\omega t\big)},
   \label{eq:DB-th=p10} \\
   \left( \begin{aligned}
{  \hat D}^{\tt (0)}_\phi& \\
{  \hat B}^{\tt (0)}_\phi& \\
  \end{aligned} \right) &=&E_0u
  \left( \begin{aligned}
-\sin\phi \\
\cos\phi  \\
  \end{aligned} \right) \,e^{i\big(k(r\cos\theta-r_g\ln kr(1-\cos\theta))-\omega t\big)}.
  \label{eq:DB-t-pl=p20}
\end{eqnarray}

Finally, in accord with (\ref{eq:Pi-g+p}), the total EM field is given by the sum of the incident and scattered EM waves given by (\ref{eq:DB-t-pl=p10})--(\ref{eq:DB-t-pl=p20}) and (\ref{eq:DB-r*+*p2})--(\ref{eq:DB-t-pl=p2}), correspondingly. Thus, computing the total field as ${\vec D}={\vec D}^{\tt (0)}+{\vec D}^{\tt p}$ and ${\vec B}={\vec B}^{\tt (0)}+{\vec B}^{\tt p}$, then, up to terms of ${\cal O}(\theta^5,\delta\theta^2_{\tt p},r_g\theta^4/r,r_g^2,{r\delta\theta_{\tt p}}/{b})$ the components of this field have the following form:
{}
\begin{eqnarray}
  \left( \begin{aligned}
{  \hat D}_r& \\
{  \hat B}_r& \\
  \end{aligned} \right) &=&
  E_0u^{-1}
\sin\Big(\theta\pm\frac{d2\delta^\star_\ell}{d\ell}\Big)\Big(1+\frac{r_g}{r(1-\cos\theta)}\Big) \left( \begin{aligned}
\cos\phi \\
\sin\phi  \\
  \end{aligned} \right)
  e^{i\big(kr\cos\theta-kr_g\ln kr(1-\cos\theta)+2\delta^*_\ell-\omega t\big)},~~~~~~
  \label{eq:DB-t-pl=pV0}\\
  \left( \begin{aligned}
{  \hat D}_\theta& \\
{  \hat B}_\theta& \\
  \end{aligned} \right) &=&
  E_0u^{-1}
  \Big(\cos\big(\theta\pm\frac{d2\delta^\star_\ell}{d\ell}\big)-\frac{r_g}{r}\Big)\left( \begin{aligned}
\cos\phi \\
\sin\phi  \\
  \end{aligned} \right)e^{i\big(kr\cos\theta-kr_g\ln kr(1-\cos\theta)+2\delta^*_\ell-\omega t\big)},
  \label{eq:DB-t-pl=pV1}\\
\left( \begin{aligned}
{  \hat D}^{\tt }_\phi& \\
{  \hat B}^{\tt }_\phi& \\
  \end{aligned} \right) &=&
  E_0u
\left( \begin{aligned}
-\sin\phi \\
\cos\phi  \\
  \end{aligned} \right) \,e^{i\big(kr\cos\theta-kr_g\ln kr(1-\cos\theta)+2\delta^*_\ell-\omega t\big)}.
  \label{eq:DB-t-pl=pV2}
\end{eqnarray}

We recall that in the case when gravity is involved, there are two waves that characterize the scattering process in the region of geometric optics: the incident wave given by (\ref{eq:DB-t-pl=p10})--(\ref{eq:DB-t-pl=p20}) and the scattered wave, which was computed in \cite{Turyshev-Toth:2017} (see equations (49)--(50) therein) and is given as
{}
\begin{eqnarray}
    \left( \begin{aligned}
{  \hat D}^{\tt (0)}_\theta& \\
{  \hat B}^{\tt (0)}_\theta& \\
  \end{aligned} \right)_{\tt \hskip -4pt s} =   \left( \begin{aligned}
{  \hat B}^{\tt (0)}_\phi& \\
-{  \hat D}^{\tt (0)}_\phi& \\
  \end{aligned} \right)_{\tt \hskip -4pt s}= E_0
 \frac{r_g}{2r\sin^2\frac{\theta}{2}}
 \left( \begin{aligned}
\cos\phi \\
\sin\phi  \\
  \end{aligned} \right)e^{i\big(k(r+r_g\ln kr(1-\cos\theta))+2\sigma_0-\omega t\big)},
  \qquad
    \left( \begin{aligned}
{  \hat D}^{\tt (0)}_r& \\
{  \hat B}^{\tt (0)}_r& \\
  \end{aligned} \right)_{\tt \hskip -4pt s} &=& {\cal O}(r^2_g).~~~
   \label{eq:scat-th}
\end{eqnarray}

We may compute the total scattered EM field in the geometric optics region behind the Sun. Similarly to (\ref{eq:DB-t-pl=pV0})--(\ref{eq:DB-t-pl=pV2})  we add the corresponding components of the plasma-free field (\ref{eq:scat-th}) and those that account for the plasma-induced phase shift and given by (\ref{eq:Pi_s_rrd}), (\ref{eq:Pi_s_exp4+1*=*2}) and (\ref{eq:gamma_scat}). Computing ${\vec D}_{\tt s}={\vec D}^{(0)}_{\tt s}+{\vec D}^{\tt (p)}_{\tt s}$ and ${\vec B}_{\tt s}={\vec B}^{(0)}_{\tt s}+{\vec B}^{\tt (p)}_{\tt s}$, we have
{}
\begin{eqnarray}
    \left( \begin{aligned}
{  \hat D}_\theta& \\
{  \hat B}_\theta& \\
  \end{aligned} \right)_{\tt \hskip -4pt s} =   \left( \begin{aligned}
{  \hat B}_\phi& \\
-{  \hat D}_\phi& \\
  \end{aligned} \right)_{\tt \hskip -4pt s}= E_0
 \frac{r_g}{2r\sin^2\frac{\theta}{2}}
 \left( \begin{aligned}
\cos\phi \\
\sin\phi  \\
  \end{aligned} \right)
  e^{i\big(k(r+r_g\ln kr(1-\cos\theta))+2\delta^\star_\ell+2\sigma_0-\omega t\big)},
  \qquad
    \left( \begin{aligned}
{  \hat D}_r& \\
{  \hat B}_r& \\
  \end{aligned} \right)_{\tt \hskip -4pt s} &=& {\cal O}(r^2_g).
   \label{eq:scat-th-tot}
\end{eqnarray}

Therefore, the total EM field behind a very large sphere, $\lambda\ll R_\odot$, embedded in the spherically symmetric plasma distribution, has the structure similar to the incident EM wave. However its phase and propagation direction are affected by the delay introduced by the plasma in the solar system. The EM field outside the termination shock takes a very simple form that depends on the plasma phase shift, $\delta^*_\ell$. This phase shift given by (\ref{eq:s-d1}) is clearly showing its dependence on the solar impact parameter. Eqs.~(\ref{eq:DB-t-pl=pV0})--(\ref{eq:DB-t-pl=pV2}) account for this contribution.  The resulting expression for the phase of the wave is well known and corresponds to that described by the equation of geodesics, as derived in Appendices \ref{sec:geodesics} and \ref{sec:geom-optics}, namely by (\ref{eq:X-eq4**}) and (\ref{eq:phase_t*}). Similarly, the phase of the expressions (\ref{eq:scat-th-tot}) is consistent with that of a radial geodesic as given by (\ref{eq:X-eq4**_rad}) and (\ref{eq:phase_t-rad}), also see \cite{Turyshev-Toth:2017} for discussion. As such, they are consistent with the expressions for the phase of the EM wave moving through the solar plasma derived by other authors  \cite{Giampieri:1994kj,Bertotti-Giampieri:1998,Bisnovatyi-Kogan:2015dxa}.

This completes the derivation for the EM field in the region of geometric optics outside the termination shock.

\subsection{Diffraction of light within the heliosphere}
\label{sec:EM-field-inner}
\label{sec:Dp-inside}

To establish the solution for the Debye potential inside the termination shock, we need to implement the fully absorbing boundary conditions, as we did in Sec.~\ref{sec:EM-field}. To do this, we identically rewrite (\ref{eq:Pi-in+sl}) using a representation of the Coulomb function $F_\ell(kr_g,kr)$ via incoming and outgoing waves, $H^{+}_\ell(kr_g,kr)$ and $H^{-}_\ell(kr_g,kr)$:
{}
\begin{equation}
\Pi_{\tt in} (r, \theta)=
\frac{E_0}{2ik^2}\frac{u}{r}\sum_{\ell=1}^\infty i^{\ell-1}\frac{2\ell+1}{\ell(\ell+1)}e^{i\big(\sigma_\ell +\delta_\ell -\delta_\ell(r)\big)}  \Big\{
e^{2i\delta_\ell(r)}H^{+}_\ell(kr_g,kr)-H^{-}_\ell(kr_g,kr)\Big\} P^{(1)}_\ell(\cos\theta) +{\cal O}(r^2_g,r_g\frac{\omega_{\tt p}^2}{\omega^2}).~~~~
  \label{eq:Pi_g+p1}
\end{equation}
By removing from this expression the outgoing waves corresponding to the impact parameters $b\leq R^\star_\odot$ or, equivalently, for  $\ell\in[1,kR^\star_\odot]$, we implement the fully absorbing boundary conditions that account for the physical properties of the solar surface. The resulting expression has the form
{}
\begin{eqnarray}
\Pi_{\tt in}(r, \theta)&=&
\frac{E_0}{2ik^2}\frac{u}{r}\sum_{\ell=1}^\infty i^{\ell-1}\frac{2\ell+1}{\ell(\ell+1)}e^{i\big(\sigma_\ell +\delta_\ell -\delta_\ell(r)\big)}  \Big\{
e^{2i\delta_\ell(r)}H^{+}_\ell(kr_g,kr)-H^{-}_\ell(kr_g,kr)\Big\} P^{(1)}_\ell(\cos\theta) -\nonumber\\
&-&
\frac{E_0}{2ik^2}\frac{u}{r}\sum_{\ell=1}^{kR^\star_\odot} i^{\ell-1}\frac{2\ell+1}{\ell(\ell+1)}e^{i\big(\sigma_\ell+\delta_\ell-\delta_\ell(r)\big)}e^{2i\delta_\ell(r)}H^{+}_\ell(kr_g,kr) P^{(1)}_\ell(\cos\theta)
+{\cal O}(r^2_g,r_g\frac{\omega_{\tt p}^2}{\omega^2}).
  \label{eq:Pi_g+p1_bc}
\end{eqnarray}

This is the final solution for the EM field that travels in the vicinity of the Sun in the presence of solar gravity and solar plasma.  This solution may be given in the following equivalent form:
{}
\begin{eqnarray}
\Pi_{\tt in}(r, \theta)&=&
\frac{E_0}{k^2}\frac{u}{r}\sum_{\ell=1}^\infty i^{\ell-1}\frac{2\ell+1}{\ell(\ell+1)}e^{i\big(\sigma_\ell +\delta_\ell -\delta_\ell(r)\big)} F_\ell(kr_g,kr) P^{(1)}_\ell(\cos\theta)-\nonumber\\
&-&\frac{E_0}{2ik^2}\frac{u}{r}\sum_{\ell=1}^{kR^\star_\odot} i^{\ell-1}\frac{2\ell+1}{\ell(\ell+1)}e^{i\big(\sigma_\ell +\delta_\ell -\delta_\ell(r)\big)}
H^{+}_\ell(kr_g,kr) P^{(1)}_\ell(\cos\theta) +\nonumber\\
&+&\frac{E_0}{2ik^2}\frac{u}{r}\sum_{\ell=kR^\star_\odot}^\infty i^{\ell-1}\frac{2\ell+1}{\ell(\ell+1)}e^{i\big(\sigma_\ell +\delta_\ell -\delta_\ell(r)\big)}
\Big(e^{2i\delta_\ell(r)}-1\Big)
H^{+}_\ell(kr_g,kr) P^{(1)}_\ell(\cos\theta) +{\cal O}(r^2_g,r_g\frac{\omega_{\tt p}^2}{\omega^2}).~~~
  \label{eq:Pi_g+p22+}
\end{eqnarray}

The first term in (\ref{eq:Pi_g+p22+}) is the Debye potential of the pure gravity case derived in \cite{Turyshev-Toth:2017}, modified by the presence of plasma in the solar system. The second term is responsible for the geometric shadow cast by the Sun.  The third term represents the impact of solar plasma on the propagation of the EM field outside the Sun, but within the distance to the termination shock $R^\star_\odot\leq r \leq R_\star$.

Expression (\ref{eq:Pi_g+p1_bc})  is our solution for the Debye potential representing the EM field that travels through the solar system in the presence of the solar plasma. Reinstating the Coulomb function $F_\ell(kr_g,kr)$ and, similarly to (\ref{eq:Pi_g+p0}), taking into account the asymptotic behavior of the function $H^{+}_\ell(kr_g,kr)$ given by (\ref{eq:Fass*}), expression (\ref{eq:Pi_g+p22+}) representing the solution  for the Debye potential within the solar system takes the form
{}
\begin{eqnarray}
\Pi_{\tt in}(r, \theta)&=&
\frac{E_0}{k^2}\frac{u}{r}\sum_{\ell=1}^\infty i^{\ell-1}\frac{2\ell+1}{\ell(\ell+1)}e^{i\big(\sigma_\ell +\delta_\ell -\delta_\ell(r)\big)} F_\ell(kr_g,kr) P^{(1)}_\ell(\cos\theta)+\nonumber\\
&+&\frac{e^{ik(r+r_g\ln 2kr)}}{r}\frac{uE_0}{2k^2}\Big\{\sum_{\ell=1}^{kR^\star_\odot} \frac{2\ell+1}{\ell(\ell+1)}e^{i\big(2\sigma_\ell +\delta_\ell -\delta_\ell(r)+\frac{\ell(\ell+1)}{2kr}+\frac{[\ell(\ell+1)]^2}{24k^3r^3}\big)}P^{(1)}_\ell(\cos\theta) -\nonumber\\
&-&\sum_{\ell=kR^\star_\odot}^\infty \frac{2\ell+1}{\ell(\ell+1)}e^{i\big(2\sigma_\ell +\delta_\ell -\delta_\ell(r)+\frac{\ell(\ell+1)}{2kr}+\frac{[\ell(\ell+1)]^2}{24k^3r^3}\big)}
\Big(e^{2i\delta_\ell(r)}-1\Big)P^{(1)}_\ell(\cos\theta) \Big\}+{\cal O}(r^2_g,r_g\frac{\omega_{\tt p}^2}{\omega^2}).
  \label{eq:Pi_g+p22}
\end{eqnarray}

This is our main result for the Debye potential representing the EM field in the geometric optics region situated inside the termination shock, $0<r\leq R_\star$ (i.e., interior geometric optics region). It describes the propagation of monochromatic EM waves on the background of monopole gravity and that of a generic steady-state spherically symmetric plasma, for which the number density diminishes as $r^{-2}$ or faster. The first term in (\ref{eq:Pi_g+p22}) is the Debye potential of the incident wave modified by the plasma as the wave propagates through the solar system. The second term is for the geometric shadow behind the Sun (similar to that discussed in \cite{Turyshev-Toth:2018,Turyshev-Toth:2018-grav-shadow}), also modified by the plasma. The third term represents the ongoing scattering of the EM field as it propagates through the spherically symmetric distribution of the extended solar corona, given at a particular heliocentric position within solar system, $R^\star_\odot\leq r \leq R_\star$.

Note that because of the plasma model (\ref{eq:eps}) and (\ref{eq:n-eps_n-ism}), the last sum in (\ref{eq:Pi_g+p22}) formally extends only to $\ell=k R_\star$, corresponding to the impact parameter equal to the distance to the termination shock. For $r>R_\star$, not only does the vanishing phase shift, $\delta_\ell=0$, essentially eliminate this term, but this distance is also outside the boundary that characterizes the inner region, as for $r>R_\star$ we enter the domain of the scattered wave discussed in Sec.~\ref{sec:go-em-outside}.

To discuss the diffraction of light in the solar system, we refer to the solution for the Debye potential $\Pi$ given by (\ref{eq:Pi_g+p22}). Each term in (\ref{eq:Pi_g+p22}) has the contribution of the ongoing plasma phase shift given as $\delta^*_\ell -\delta_\ell(r)$, where  $\delta^*_\ell$  and $\delta_\ell(r)$  are given by (\ref{eq:a_b_del}) and (\ref{eq:a_b_del-r}), correspondingly, with eikonal phase shifts for the short-range plasma potential $\xi_b(r)$ and $\xi_b^\star$ given by (\ref{eq:delta-D*-av0WKB+}) and (\ref{eq:delta-D*-av0WKB}), respectively. To evaluate these terms, we derive the differential plasma-induced phase shift occurring as the wave travels through the heliocentric ranges  $R^\star_\odot \leq r\leq R_\star$. Defining
{}
\begin{eqnarray}
\delta^*_\ell -\delta_\ell(r)=\xi_b^\star-\xi_b(r)\equiv
\delta \xi_b(r),
\label{eq:delta-diff0}
\end{eqnarray}
from (\ref{eq:delta-D*-av0WKB+}) and (\ref{eq:delta-D*-av0WKB}) we compute
{}
\begin{eqnarray}
\delta \xi_b(r)=
-\frac{2\pi e^2R_\odot}{m_ec^2k}
\sum_{i>2} {\alpha_i}\Big(\frac{R_\odot}{b}\Big)^{\beta_i-1}\Big\{Q_{\beta_i}^\star-Q_{\beta_i}\big(\sqrt{r^2-b^2}\big)\Big\}.
\label{eq:delta-diff}
\end{eqnarray}

As we can see, the differential phase shift  $\delta \xi_b(r)$ is independent of either $L$ or $\ell$ and is a function of the heliocentric distance only. Thus, in all the terms of (\ref{eq:Pi_g+p22}) we may move the factor $\exp\big[{i(\delta^*_\ell -\delta_\ell(r))\big]}\equiv \exp\big[i\delta \xi_b(r)\big]$  outside the summation over $\ell$. With this, and using $\Pi_0 (r, \theta)$ from (\ref{eq:Pi_ie*+*=}), the Debye potential $\Pi$ from (\ref{eq:Pi_g+p22})  takes the form
{}
\begin{eqnarray}
\Pi_{\tt in}(r, \theta)&=&e^{i\delta \xi_b(r)}\Big[
\Pi_0 (r, \theta)+
\frac{ue^{ik(r+r_g\ln 2kr)}}{r}\frac{E_0}{2k^2}\Big\{\sum_{\ell=1}^{kR^\star_\odot} \frac{2\ell+1}{\ell(\ell+1)}e^{i\big(2\sigma_\ell +\frac{\ell(\ell+1)}{2kr}+\frac{[\ell(\ell+1)]^2}{24k^3r^3}\big)}P^{(1)}_\ell(\cos\theta) -\nonumber\\
&&\hskip 30pt
-\,\sum_{\ell=kR^\star_\odot}^\infty \frac{2\ell+1}{\ell(\ell+1)}e^{i\big(2\sigma_\ell +\frac{\ell(\ell+1)}{2kr}+\frac{[\ell(\ell+1)]^2}{24k^3r^3}\big)}
\Big(e^{2i\delta_\ell(r)}-1\Big)P^{(1)}_\ell(\cos\theta) \Big\}+{\cal O}(r^2_g,r_g\frac{\omega_{\tt p}^2}{\omega^2})\Big].
  \label{eq:Pi_g+p22*+}
\end{eqnarray}

Clearly, $\delta \xi_b(r)$ is significant only in the immediate vicinity of the Sun, where $r\simeq R_\odot$, but it falls off rapidly for larger distances.
Using the phenomenological model (\ref{eq:model}), we estimate the magnitude of the differential phase shift (\ref{eq:delta-diff}). For this, with the help of (\ref{eq:delta-etas}) and (\ref{eq:delta-etas-m}), expression (\ref{eq:delta-diff}) takes the from
{}
\begin{eqnarray}
\delta \xi_b(r)=
-\Big\{586.17\Big(\frac{R_\odot}{b}\Big)^{15}\Big(Q_{16}^\star-Q_{16}\big(\sqrt{r^2-b^2}\big)\Big)+303.87\Big(\frac{R_\odot}{b}\Big)^{5}\Big(Q_{6}^\star-Q_{6}\big(\sqrt{r^2-b^2}\big)\Big)\Big\}\Big(\frac{\lambda}{1~\mu{\rm m}}\Big).
\label{eq:delta-diff-mod}
\end{eqnarray}

Examining this expression, we see that it reaches its largest value for the smallest impact parameter of $b\simeq R_\odot$. However, even for radio waves passing that close to the Sun, the phase shift (\ref{eq:delta-diff-mod}) results in a practically negligible effect. Evaluating for $\lambda \simeq 1$~cm, the delay introduced by (\ref{eq:delta-diff-mod}) at $r=10R_\odot$ is $\delta d_b= \delta \xi_b(r) (\lambda/2\pi)\simeq 1 \lambda$ and rapidly diminishes as $r$ increases.  In fact, at heliocentric distances beyond  $r\simeq 20R_\odot$, even for such rather long wavelengths, the differential phase shift  introduced  by (\ref{eq:delta-diff-mod}) is totally negligible.

As a result,  we may set $\delta \xi_b(r)=0$ in (\ref{eq:Pi_g+p22*+}), making it equivalent to the solution for the Debye potential given by (\ref{eq:Pi_g+p0}). With this, all the results that we obtained earlier in Sec.~\ref{sec:go-em-outside} for the region outside the termination shock, $r> R_\star$, may be extended  to cover also the ranges within the termination shock, $R_\odot \leq r\leq R_\star$.  Thus, we have a compete solution for the the EM field in the geometrical optics and shadow regions of the Sun.  We now turn out attention to the region of  most importance for the SGL: the interference region.

\section{EM field in the interference region}
\label{sec:IF-region}

We are interested in the area behind the Sun, reachable by light rays with impact parameters $b>R_\odot^\star$. The focal region of the SGL begins where $r>R_\star$ and  $0\leq \theta\simeq \sqrt{2r_g/r}$. The EM field in this region is derived from the Debye potential (\ref{eq:Pi-g+p})--(\ref{eq:f-v*+}) and is given by the factors $\alpha(r,\theta)$, $\beta(r,\theta)$ and $\gamma(r,\theta)$ from (\ref{eq:alpha*1*})--(\ref{eq:gamma*1*}), which we now calculate.

\subsection{The function $\alpha(r,\theta)$ and the radial components of the EM field}
\label{sec:alpha-IF}

We begin with the investigation of $\alpha(r,\theta)$, given by (\ref{eq:alpha*1*}) as
{}
\begin{eqnarray}
\alpha(r,\theta) &=& -E_0\frac{e^{ik(r+r_g\ln 2kr)}}{uk^2r^2}\sum_{\ell=kR^\star_\odot}^\infty(\ell+{\textstyle\frac{1}{2}})e^{i\big(2\sigma_\ell+\frac{\ell(\ell+1)}{2kr}+\frac{[\ell(\ell+1)]^2}{24k^3r^3}\big)}\Big(e^{i2\delta^*_\ell}-1\Big)P^{(1)}_\ell(\cos\theta)\times\nonumber\\
&&\hskip 30pt \times\,
\Big\{u^{2}+(u^2-1)\frac{\ell(\ell+1)}{4k^2r^2}+\frac{i}{kr}\Big(1+\frac{\ell(\ell+1)}{2k^2r^2}\Big)-\frac{ikr_g}{\ell(\ell+1)} \Big\}
+{\cal O}\Big(r^2_g,r_g\frac{\omega_{\tt p}^2}{\omega^2},(kr)^{-5}\Big).
  \label{eq:alpha*IF}
\end{eqnarray}

To evaluate expression (\ref{eq:alpha*IF}) in the interference region and for $0\leq \theta \simeq \sqrt{2r_g/r}$, we use the asymptotic representation for $P^{(1)}_l(\cos\theta)$ from \cite{Bateman-Erdelyi:1953,Korn-Korn:1968,Kerker-book:1969}, valid when $\ell\to\infty$:
{}
\begin{eqnarray}
P^{(1)}_\ell(\cos\theta)&=& \frac{\ell+{\textstyle\frac{1}{2}}}{\cos{\textstyle\frac{1}{2}}\theta}J_1\big((\ell+{\textstyle\frac{1}{2}})2\sin{\textstyle\frac{1}{2}}\theta\big).
\label{eq:pi-l0}
\end{eqnarray}

This approximation may be used to transform (\ref{eq:alpha*IF}) as
 {}
\begin{eqnarray}
\alpha(r,\theta) &=& -E_0\frac{e^{ik(r+r_g\ln 2kr)}}{uk^2r^2\cos{\textstyle\frac{1}{2}}\theta}\sum_{\ell=kR^\star_\odot}^\infty(\ell+{\textstyle\frac{1}{2}})^2e^{i\big(2\sigma_\ell+\frac{\ell(\ell+1)}{2kr}+\frac{[\ell(\ell+1)]^2}{24k^3r^3}\big)}\Big(e^{i2\delta^*_\ell}-1\Big)J_1\big((\ell+{\textstyle\frac{1}{2}})2\sin{\textstyle\frac{1}{2}}\theta\big)\times\nonumber\\
&&\hskip 30pt \times\,
\Big\{u^{2}+(u^2-1)\frac{\ell(\ell+1)}{4k^2r^2}+\frac{i}{kr}\Big(1+\frac{\ell(\ell+1)}{2k^2r^2}\Big)-\frac{ikr_g}{\ell(\ell+1)} \Big\}
+{\cal O}\Big(r^2_g,r_g\frac{\omega_{\tt p}^2}{\omega^2},(kr)^{-5}\Big).
  \label{eq:alpha*IF*}
\end{eqnarray}

At this point, we may replace the sum in (\ref{eq:alpha*IF*}) with an integral (accounting for the fact that $\ell\gg1$ and keeping the terms up to ${\cal O}(\theta)$) to be evaluated with the method of stationary phase:
 {}
\begin{eqnarray}
\alpha(r,\theta) &=& -E_0\frac{e^{ik(r+r_g\ln 2kr)}}{uk^2r^2}\int_{\ell=kR^\star_\odot}^\infty  \ell^2 d\ell e^{i\big(2\sigma_\ell+\frac{\ell^2}{2kr}+\frac{\ell^4}{24k^3r^3}\big)}\Big(e^{i2\delta^*_\ell}-1\Big)J_1\big(\ell \theta\big)\times\nonumber\\
&&\hskip 30pt \times\,
\Big\{u^{2}+(u^2-1)\frac{\ell^2}{4k^2r^2}+\frac{i}{kr}\Big(1+\frac{\ell^2}{2k^2r^2}\Big)-\frac{ikr_g}{\ell^2} \Big\}
+{\cal O}\Big(r^2_g,r_g\frac{\omega_{\tt p}^2}{\omega^2},(kr)^{-5},\theta^2\Big).
  \label{eq:alpha*IF*int}
\end{eqnarray}
As before, we evaluate this integral treating plasma-independent and plasma-dependent terms separately.

\subsubsection{The plasma-independent part of $\alpha(r,\theta)$}
\label{sec:IF-nop}

In evaluating the plasma-independent part, we see that the $\ell$-dependent phase in this expression is given as
{}
\begin{eqnarray}
\varphi^{[0]}(\ell)&=& 2\sigma_\ell+\frac{\ell^2}{2kr}+\frac{\ell^4}{24k^3r^3}+{\cal O}((kr)^{-5})=-2kr_g\ln \ell+\frac{\ell^2}{2kr}+\frac{\ell^4}{24k^3r^3}+{\cal O}((kr)^{-5}).
\label{eq:IF-phase}
\end{eqnarray}

The phase is stationary when $d\varphi^{[0]}(\ell)/d\ell=0$, resulting in
{}
\begin{eqnarray}
-\frac{2kr_g}{\ell}+\frac{\ell}{kr}+\frac{\ell^3}{6k^3r^3}={\cal O}((kr)^{-5})\qquad \Rightarrow\qquad
\ell^4 +6k^2r^2 \ell^2 -12 k^4r^3 r_g={\cal O}((kr)^{-2}).
\label{eq:IF-phase2}
\end{eqnarray}
We may now solve this equation for $\ell^2(r_g)$, keeping only the terms of the first power of $r_g$. Requiring that in the absence of gravity no rays would  reach the focal area or $\lim_{r_g\rightarrow0}\ell^2(r_g) \rightarrow0$, we have only one solution given as
{}
\begin{eqnarray}
\ell^2=k^22r_gr +{\cal O}((kr)^{-1}) \qquad {\rm or } \qquad
\ell_0=k\sqrt{2r_gr}.
\label{eq:IF-phase3}
\end{eqnarray}
This solution represents the smallest partial momenta for the light trajectories to reach a particular heliocentric distance, $r$, on the focal line of the SGL. It is consistent with the solution to the equation for geodesics (see Appendix~\ref{sec:geodesics}) which yields the solution for the impact parameter of $b=\sqrt{2r_gr}$. In addition, we also choose such that $\ell$ is positive.

Solution  (\ref{eq:IF-phase3})  allows us to compute the stationary phase (\ref{eq:IF-phase}) as
{}
\begin{eqnarray}
\varphi^{[0]}(\ell_0)&=& -kr_g\ln 2kr+\sigma_0 +{\textstyle\frac{\pi}{2}}.
\label{eq:IF-phase00}
\end{eqnarray}

Using (\ref{eq:IF-phase}), we compute the relevant $\varphi''_{\pm}(\ell)$ as below
\begin{eqnarray}
\dfrac{d^2\varphi^{[0]}_{}}{d\ell^2} &=& \dfrac{1}{kr}\Big(1+\frac{\ell^2}{2k^2r^2}+{\cal O}\big((kr)^{-4}\big)\Big)+\frac{2kr_g}{\ell^2}
\qquad \Rightarrow \qquad
\dfrac{d^2\varphi^{[0]}_{}}{d\ell^2} = \dfrac{2}{kr}\Big(1+\frac{r_g}{2r}+{\cal O}\big(r_g^2\big)\Big).
\label{eq:IF-phase*}
\end{eqnarray}

The amplitude factor for the asymptotic expansion  $H^{+}_\ell(kr_g,kr)$ from  (\ref{eq:Fass*}), denoted by $a(\ell)$, which is given by (\ref{eq:sf1}). This factor for $\ell_0$ from (\ref{eq:IF-phase3}) is computed to be
{}
\begin{equation}
a(\ell_0)=1+{\frac{r_g}{2r}}+{\cal O}(r_g^2).
\label{eq:IF-phase*5}
\end{equation}

Using results (\ref{eq:IF-phase*}) and (\ref{eq:IF-phase*5}) we derive
{}
\begin{equation}
a(\ell_0)\sqrt{\frac{2\pi}{\varphi''(\ell_0)}}=\sqrt{\pi kr}\Big\{1+\frac{r_g}{4r}+{\cal O}(r_g^2)\Big\}.
\label{eq:IF-phase*6}
\end{equation}

Now, using (\ref{eq:IF-phase*6}), we have the amplitude of the integrand in (\ref{eq:alpha*IF*int}), for $\ell$ from (\ref{eq:IF-phase3}), taking the form
{}
\begin{eqnarray}
A^{[0]}(\ell_0)a(\ell_0)\sqrt{\frac{2\pi}{\varphi''(\ell_0)}}&=&\ell^2_0\Big\{u^{2}+(u^2-1)\frac{\ell^2_0}{4k^2r^2}+\frac{i}{kr}\Big(1+\frac{\ell^2_0}{2k^2r^2}\Big)-\frac{ikr_g}{\ell^2_0} \Big\}J_1\big(\ell \theta\big)a(\ell_0)\sqrt{\frac{2\pi}{\varphi''(\ell_0)}}=\nonumber\\
&=&
k^22r_gr \sqrt{\pi kr}\Big(1+\frac{5r_g}{4r} +{\cal O}(r_g^2, (kr)^{-1})\Big)J_1\big(k\sqrt{2r_gr} \theta\big),~~~~~
\label{eq:IF-phase*7}
\end{eqnarray}
where the superscript ${[0]}$ denotes the term with no plasma contribution. As before, we dropped the $i/(kr)$ terms inside the parentheses, as these terms are very small compared to the leading terms.

As a result, the plasma-free part of the expression for $\delta\alpha^{[0]}(r,\theta)$ given by (\ref{eq:alpha*IF*int}) takes the form
{}
\begin{eqnarray}
\delta\alpha^{[0]}_\pm(r,\theta)&=&
iE_0
\sqrt{\frac{2r_g}{r} }\sqrt{2\pi kr_g}e^{i\sigma_0}J_1\big(k\sqrt{2r_gr} \theta\big)e^{ikr}\Big(1+{\cal O}(r_g^2, (kr)^{-1})\Big).
\label{eq:IF-phase*7*}
\end{eqnarray}

\subsubsection{Evaluating the plasma-dependent part of $\alpha(r,\theta)$}
\label{sec:IF-p}

For the plasma-dependent term in (\ref{eq:alpha*IF*int}), the $\ell$-dependent phase is given as
{}
\begin{eqnarray}
\varphi^{\tt [p]}(\ell)&=& 2\sigma_\ell+\frac{\ell^2}{2kr}+\frac{\ell^4}{24k^3r^3}+2\delta^\star_\ell +{\cal O}((kr)^{-5})=-2kr_g\ln \ell+\frac{\ell^2}{2kr}+\frac{\ell^4}{24k^3r^3}+2\delta^\star_\ell +{\cal O}((kr)^{-5}).
\label{eq:IFp-phase}
\end{eqnarray}

Considering (\ref{eq:IFp-phase}), we see that the points of stationary phase, where $d\varphi^{\tt [p]}_\ell/d\ell=0$, are given by the equation
{}
\begin{equation}
-\frac{2kr_g}{\ell}+\frac{\ell}{kr}+\frac{\ell^3}{6k^3r^3}+2\delta\theta_{\tt p}={\cal O}((kr)^{-5}), \qquad \text{or} \qquad
-2kr_g+\frac{\ell^2}{kr}+\frac{\ell^4}{6k^3r^3}+2\delta\theta_{\tt p}\ell={\cal O}((kr)^{-5}).
\label{eq:IFp-phase*1}
\end{equation}
As we saw in Sec.~\ref{sec:IF-nop}, the partial momenta for the points of stationary phase is $\ell \propto k\sqrt{2r_gr}$, which makes  the $\ell^4$ term in (\ref{eq:IFp-phase*1}) of ${\cal O}(r_g^2)$. Thus, we may neglect the term $\ell^4/6k^3r^3$ and solve the remaining quadratic equation for $\ell$:
{}
\begin{equation}
\ell^2+2kr\delta\theta_{\tt p}\ell-2k^2r_gr={\cal O}((kr)^{-2}).
\label{eq:IFp-phase*2}
\end{equation}
If we require that, in the limit when plasma is absent or when $\delta\theta_{\tt p}\rightarrow0$, the partial momenta are to coincide with those obtained earlier, namely (\ref{eq:IF-phase3}), then there is only one solution of (\ref{eq:IFp-phase*2}) (similar to that discussed in \cite{Deguchi-Watson:1987}):
{}
\begin{equation}
\ell=kr\Big(\sqrt{\frac{2r_g}{r}+\delta\theta_{\tt p}^2}-\delta\theta_{\tt p}\Big).
\label{eq:IFp-phase*3}
\end{equation}
Note that this solution correctly represents another situation where in the absence of gravity, the rays do not reach the focal line, or, in other words, the focal line is reached only by the ray with $\ell=0$, which is blocked by the Sun.

As in this paper we only treat effects linear in plasma contribution, the terms of ${\cal O}(\delta\theta_{\tt p}^2)$ must be neglected, which brings (\ref{eq:IFp-phase*3}) to the following form, valid for $\sqrt{{2r_g}/{r}}> \delta\theta_{\tt p}$:
{}
\begin{equation}
\ell\simeq
kr\Big(\sqrt{\frac{2r_g}{r}}-\delta\theta_{\tt p}\Big)+{\cal O}(\delta\theta_{\tt p}^2)=
k\Big(\sqrt{2r_gr}-r\delta\theta_{\tt p}\Big)+{\cal O}(\delta\theta_{\tt p}^2).
\label{eq:IFp-phase*4}
\end{equation}
This expression represents the combined effects of gravity and plasma on the light rays traveling towards the focal area. From the left side of these two expressions, we see that, as gravity works by bending light by the angle of $\sqrt{{2r_g}/{r}}$ towards the focal line, plasma ``unbends'' these rays by the amount of  $\delta\theta_{\tt p}$. Similarly, the right side of these two expressions tells the same story using the concept of the impact parameters. To reach the focal line at the heliocentric distance $r$, rays must have the impact parameter $b=\sqrt{2r_gr}$. In the presence of plasma, to reach the same distance $r$, the impact parameter must be smaller by $r\delta\theta_{\tt p}$, consistent with our description of the effect.

Although (\ref{eq:IFp-phase*3}) ensures that $\ell$ is always positive for any $\sqrt{{2r_g}/{r}}$ and $\delta\theta_{\tt p}$, equation (\ref{eq:IFp-phase*4}), where the quadratic term $\delta\theta_{\tt p}^2$ is neglected, suggests that $\ell>0$ only if we require $\sqrt{{2r_g}/{r}}> \delta\theta_{\tt p}$ or, equivalently, $ \sqrt{2r_gr}\geq r\delta\theta_{\tt p}$. This is the result of our approximation, where we consider only terms linear with respect to $\delta\theta_{\tt p}$. We keep this observation in mind and use (\ref{eq:IFp-phase*3}) to guide us when interpreting the results.

We compute the stationary phase (\ref{eq:IFp-phase}) for the values of $\ell$ given by (\ref{eq:IFp-phase*4}):
{}
\begin{equation}
\varphi^{\tt [p]}(\ell_0)= -kr_g\ln 2kr+\sigma_0 +{\textstyle\frac{\pi}{2}}+2\delta^\star_\ell.
\label{eq:IFp-phase*5}
\end{equation}
Computing the second derivative of the phase (\ref{eq:IFp-phase}), $\varphi (\ell)$, with respect to $\ell$, we have:
\begin{eqnarray}
\dfrac{d^2\varphi^{\tt [p]}_{}}{d\ell^2} &=& \dfrac{1}{kr}\Big(1+\frac{\ell^2}{2k^2r^2}+{\cal O}\big((kr)^{-4}\big)\Big)+\frac{2kr_g}{\ell^2}+\dfrac{d^2 2\delta^\star_b}{d\ell^2},
\label{eq:S-l2+p}
\end{eqnarray}
where, similarly to (\ref{eq:S-l2*=}),  the last term could be neglected.
Now, using $\ell$ from (\ref{eq:IFp-phase*4}), we have
{}
\begin{equation}
\varphi_{}^{\tt [p]}{}''(\ell_0)=\dfrac{2}{kr}\Big(1+\frac{r_g}{2r}
+{\cal O}\big(r_g^2,\delta\theta_{\tt p}\sqrt{\frac{2r_g}{r}}\big)\Big), \qquad
\sqrt{\frac{2\pi}{\varphi''(\ell_0)}}=\sqrt{\pi kr}\Big(1-\frac{r_g}{4r}
+{\cal O}\big(r_g^2,\delta\theta_{\tt p}^2, \delta\theta_{\tt p}\sqrt{\frac{2r_g}{r}}
\big)\Big).
\label{eq:S-l220-g}
\end{equation}

The amplitude factor $a(\ell)$ from (\ref{eq:sf1}) is computed to be
{}
\begin{equation}
a(\ell)=1+{\frac{r_g}{2r}}+{\cal O}(r_g^2,r_g\delta\theta_{\tt p}).
\label{eq:IF-phase*5*2}
\end{equation}

Using results (\ref{eq:S-l220-g}) and (\ref{eq:IF-phase*5}) we derive
{}
\begin{equation}
a(\ell_0)\sqrt{\frac{2\pi}{\varphi''(\ell_0)}}=\sqrt{\pi kr}\Big\{1+\frac{r_g}{4r}
+{\cal O}\big(r_g^2,\delta\theta_{\tt p}^2, \delta\theta_{\tt p}\sqrt{\frac{2r_g}{r}}
\big)\Big\}.
\label{eq:IFp-phase*6}
\end{equation}

As a result, using (\ref{eq:IFp-phase*6}), we have the amplitude of the integrand in (\ref{eq:alpha*IF*int}), for $\ell$ from (\ref{eq:IFp-phase*4}), is taking the form
{}
\begin{eqnarray}
A^{\tt [p]}(\ell_0)a(\ell_0)\sqrt{\frac{2\pi}{\varphi''(\ell_0)}}&=&\ell^2_0 \Big\{u^{2}+(u^2-1)\frac{\ell^2_0}{4k^2r^2}+\frac{i}{kr}\Big(1+\frac{\ell^2_0}{2k^2r^2}\Big)-\frac{ikr_g}{\ell^2_0} \Big\}J_1\big(\ell \theta\big)a(\ell_0)\sqrt{\frac{2\pi}{\varphi''(\ell_0)}}=\nonumber\\
&=&
k^22r_gr \sqrt{\pi kr}\Big(1+{\textstyle\frac{5}{4}}\frac{r_g}{r}
+{\cal O}\big(r_g^2, (kr)^{-1},\delta\theta_{\tt p}\sqrt{\frac{2r_g}{r}}
\big)\Big)J_1\Big(k\big(\sqrt{2r_gr}-r\delta\theta_{\tt p}\big) \theta\Big),~~~~~
\label{eq:IFp-phase*7}
\end{eqnarray}
where the superscript ${\tt [p]}$ denotes the term that includes the plasma contribution. As before, we  dropped the $i/(kr)$ terms in the parentheses of this expression, as these terms are very small compared to the leading terms.

As a result, the plasma-dependent part of the expression for $\delta\alpha^{[0]}(r,\theta)$ given by (\ref{eq:alpha*IF*int}) takes the form
{}
\begin{eqnarray}
\delta\alpha^{\tt [p]}_\pm(r,\theta)&=&
-iE_0
\sqrt{\frac{2r_g}{r} }\sqrt{2\pi kr_g}e^{i\sigma_0}J_1\Big(k\big(\sqrt{2r_gr}-r\delta\theta_{\tt p}\big) \theta\Big)e^{i(kr+2\delta^\star_\ell)}\Big(1+{\cal O}(r_g^2, (kr)^{-1},
\delta\theta_{\tt p}\sqrt{\frac{2r_g}{r}}
)\Big).
\label{eq:IFp-phase*7*}
\end{eqnarray}

Finally, the entire $\alpha(r,\theta)$ term from (\ref{eq:alpha*IF*int}) may be given as
{}
\begin{eqnarray}
\delta\alpha(r,\theta)&=&\delta\alpha^{\tt [0]}_\pm(r,\theta)+\delta\alpha^{\tt [p]}_\pm(r,\theta)=\nonumber\\
&&\hskip -50pt
=\,-iE_0
\sqrt{\frac{2r_g}{r} }\sqrt{2\pi kr_g}e^{i\sigma_0}\Big\{J_1\Big(k\big(\sqrt{2r_gr}-r\delta\theta_{\tt p}\big) \theta\Big)e^{i2\delta^\star_\ell}-J_1\Big(k\sqrt{2r_gr} \theta\Big)+
{\cal O}\big(r_g^2, (kr)^{-1},
\delta\theta_{\tt p}\sqrt{\frac{2r_g}{r}}
\big)\Big\}e^{ikr}.~~
\label{eq:alpha*IF*int-*}
\end{eqnarray}

We can use the same approach to compute the remaining two scattering
factors, $\beta(r,\theta)$ and $\gamma(r,\theta)$.

\subsection{The function $\beta(r,\theta)$ and the $\theta$-components of the EM field}
\label{sec:beta-IF}

The $\beta(r,\theta)$ function is given by (\ref{eq:beta*1*}) in the following form:
{}
\begin{eqnarray}
\beta(r,\theta) &=& E_0\frac{e^{ik(r+r_g\ln 2kr)}}{ikr}\sum_{\ell=kR^\star_\odot}^\infty\frac{\ell+{\textstyle\frac{1}{2}}}{\ell(\ell+1)}e^{i\big(2\sigma_\ell+\frac{\ell(\ell+1)}{2kr}+\frac{[\ell(\ell+1)]^2}{24k^3r^3}\big)}\Big(e^{i2\delta^*_\ell}-1\Big)\times\nonumber\\
&&\hskip 30pt \times\,
\Big\{\frac{\partial P^{(1)}_\ell(\cos\theta)}
{\partial \theta}\Big(1-u^{-2}
\Big(\frac{\ell(\ell+1)}{2k^2r^2}+\frac{[\ell(\ell+1)]^2}{8k^4r^4}\Big)+\frac{ir_g}{2kr^2}\Big)+\frac{P^{(1)}_\ell(\cos\theta)}{\sin\theta}
 \Big\}.
  \label{eq:beta-IF}
\end{eqnarray}

To evaluate the magnitude of the function $\beta(r,\theta)$, we  need to establish the asymptotic behavior of the Legendre polynomials $P^{(1)}_{l}(\cos\theta)$ in the relevant regime. The asymptotic formulae for the Legendre polynomials  if $w=(\ell+{\textstyle\frac{1}{2}})\theta$ is fixed and $\ell$ goes to $\infty$ are \cite{vandeHulst-book-1981}:
{}
\begin{eqnarray}
\frac{P^{(1)}_\ell(\cos\theta)}{\sin\theta}&=& {\textstyle\frac{1}{2}}\ell(\ell+1)\Big(J_0(w)+J_2(w)\Big),
\label{eq:pi-l}
\qquad
\frac{dP^{(1)}_\ell(\cos\theta)}{d\theta}= {\textstyle\frac{1}{2}}\ell(\ell+1)\Big(J_0(w)-J_2(w)\Big).
\label{eq:tau-l}
\end{eqnarray}

For any large $\ell$, formulae (\ref{eq:pi-l*})--(\ref{eq:tau-l*}) are insufficient in a region close to the forward direction $(\theta=0$) and back direction ($\theta=\pi$). In the forward region they are complemented by the asymptotic formulae (\ref{eq:tau-l}). Similar formulae may be used for the  back region. More precisely, the formulae (\ref{eq:pi-l*})--(\ref{eq:tau-l*}) hold for $\sin\theta\gg1/\ell$, and those  given by (\ref{eq:tau-l}) hold for $\theta\ll1$. The overlapping domain is $1/\ell\ll\sin\theta\ll 1$. For our discussion of the SGL, expressions (\ref{eq:tau-l}) are more appropriate as they describe the EM field at or near the optical axis where  $\theta\approx 0$; however, when needed, we use (\ref{eq:pi-l*})--(\ref{eq:tau-l*}) to describe the EM field at small, but finite angles away from the SGL's optical axis.

Using (\ref{eq:tau-l}), we transform (\ref{eq:beta-IF}) as follows:
{}
\begin{eqnarray}
\beta(r,\theta) &=& E_0\frac{e^{ik(r+r_g\ln 2kr)}}{ikr}\sum_{\ell=kR^\star_\odot}^\infty(\ell+{\textstyle\frac{1}{2}})e^{i\big(2\sigma_\ell+\frac{\ell(\ell+1)}{2kr}+\frac{[\ell(\ell+1)]^2}{24k^3r^3}\big)}\Big(e^{i2\delta^*_\ell}-1\Big)\times\nonumber\\
&&\hskip -30pt \times\,
\Big\{J_0\big((\ell+{\textstyle\frac{1}{2}})\theta\big)- {\textstyle\frac{1}{2}}\Big(J_0\big((\ell+{\textstyle\frac{1}{2}})\theta\big)-J_2\big((\ell+{\textstyle\frac{1}{2}})\theta\big)\Big)\Big(u^{-2}
\Big(\frac{\ell(\ell+1)}{2k^2r^2}+\frac{[\ell(\ell+1)]^2}{8k^4r^4}\Big)-\frac{ir_g}{2kr^2}\Big) \Big\}.
  \label{eq:beta*IF*}
\end{eqnarray}

Now we may replace the sum in (\ref{eq:beta*IF*}) with an integral (accounting for the fact that $\ell\gg1$ and keeping terms up to the order of $\propto \theta$):
 {}
\begin{eqnarray}
\beta(r,\theta) &=& E_0\frac{e^{ik(r+r_g\ln 2kr)}}{ikr}\int_{\ell=kR^\star_\odot}^\infty \ell d\ell e^{i\big(2\sigma_\ell+\frac{\ell^2}{2kr}+\frac{\ell^4}{24k^3r^3}\big)}\Big(e^{i2\delta^*_\ell}-1\Big)\times\nonumber\\
&&\hskip 60pt \times\,
\Big\{J_0\big(\ell\theta\big)- {\textstyle\frac{1}{2}}\Big(J_0\big(\ell\theta\big)-J_2\big(\ell\theta\big)\Big)\Big(u^{-2}
\Big(\frac{\ell^2}{2k^2r^2}+\frac{\ell^4}{8k^4r^4}\Big)-\frac{ir_g}{2kr^2}\Big) \Big\}.
  \label{eq:beta*IF*int}
\end{eqnarray}
We evaluate this integral with the method of stationary phase, treating plasma-independent and plasma-dependent terms separately.

As the $\ell$-dependent phase in (\ref{eq:beta*IF*int}) is the same as (\ref{eq:alpha*IF*int}), corresponding results  obtained in Secs.  \ref{sec:IF-nop} and \ref{sec:IF-p} are also applicable here.
In fact, the same solutions for the points of stationary phase apply. As a result, using (\ref{eq:IF-phase3}) and (\ref{eq:IF-phase*6}), from (\ref{eq:beta*IF*int}) for the part of the integral that does not depend on the plasma phase shift, $\delta^\star_\ell$, we have
{}
\begin{eqnarray}
A^{[0]}(\ell_0)a(\ell_0)\sqrt{\frac{2\pi}{\varphi''(\ell_0)}}&=& \ell_0\Big\{J_0\big(\ell_0\theta\big)- {\textstyle\frac{1}{2}}\Big(J_0\big(\ell_0\theta\big)-J_2\big(\ell_0\theta\big)\Big)\Big(u^{-2}
\Big(\frac{\ell^2_0}{2k^2r^2}+\frac{\ell^4_0}{8k^4r^4}\Big)-\frac{ir_g}{2kr^2}\Big) \Big\}a(\ell_0)\sqrt{\frac{2\pi}{\varphi''(\ell_0)}}=\nonumber\\
&=&kr\sqrt{2\pi kr_g}
\Big\{J_0\big(k\sqrt{2r_gr}\theta\big) +{\cal O}\big(\frac{r_g}{r}, r_g^2\big)\Big\}.~~~
\label{eq:beta*IF*=1}
\end{eqnarray}

As a result, the expression for $\delta \beta^{[0]}_\pm(r,\theta)$  in the interference region takes the form
{}
\begin{eqnarray}
\delta \beta^{[0]}(r,\theta)&=& -E_0\sqrt{2\pi kr_g}e^{i\sigma_0}
J_0\big(k\sqrt{2r_gr}\theta\big)e^{ikr}\Big(1+{\cal O}\big(\frac{r_g}{r}, r_g^2\big)\Big).
\label{eq:beta*IF*=2}
\end{eqnarray}

Next, we evaluate the plasma-dependent term in (\ref{eq:beta*IF*int}). Using the  expressions (\ref{eq:IFp-phase*4}) and (\ref{eq:IFp-phase*6}), from (\ref{eq:beta*IF*int}) we have:
{}
\begin{eqnarray}
A^{[\tt p]}(\ell_0)a(\ell_0)\sqrt{\frac{2\pi}{\varphi''(\ell_0)}}&=&
kr\sqrt{2\pi kr_g}
\Big(1-
\frac{\delta\theta_{\tt p}}{\sqrt{2r_g/r}}\Big)
\Big\{J_0\Big(k\big(\sqrt{2r_gr}-r\delta\theta_{\tt p}\big) \theta\Big)+{\cal O}\big(\frac{r_g}{r}, r_g^2\big)\Big\}.~~~~
\label{eq:beta*IF*=3}
\end{eqnarray}
Thus, the term in (\ref{eq:beta*IF*int}) that depends on the  contribution from the plasma-induced phase shift takes the form
{}
\begin{eqnarray}
\delta \beta^{[\tt p]}_\pm(r,\theta)&=&E_0\sqrt{2\pi kr_g}e^{i\sigma_0}\Big(1-
\frac{\delta\theta_{\tt p}}{\sqrt{2r_g/r}}\Big)
J_0\Big(k\big(\sqrt{2r_gr}-r\delta\theta_{\tt p}\big) \theta\Big)e^{i(kr+2\delta^\star_\ell)}\Big(1+{\cal O}\big(\frac{r_g}{r}, r_g^2\big)\Big).
\label{eq:beta*IF*=4}
\end{eqnarray}

Using the expressions (\ref{eq:beta*IF*=2}) and (\ref{eq:beta*IF*=4}), we present the integral (\ref{eq:beta*IF*int}) as
{}
\begin{eqnarray}
\beta(r,\theta)&=&\delta \beta^{[0]}(r,\theta)+\delta \beta^{[{\tt p}]}(r,\theta)=
 \nonumber\\
&=&
E_0\sqrt{2\pi kr_g}e^{i\sigma_0}\Big\{\Big(1-
\frac{\delta\theta_{\tt p}}{\sqrt{2r_g/r}}\Big)J_0\Big(k\big(\sqrt{2r_gr}-r\delta\theta_{\tt p}\big) \theta\Big)e^{i2\delta^\star_\ell}-
J_0\Big(k\sqrt{2r_gr} \theta\Big)+{\cal O}\big(\frac{r_g}{r}, r_g^2\big)\Big\}e^{ikr}.~~~~~~
\label{eq:beta*IF*int-*}
\end{eqnarray}

\subsection{The function $\gamma(r,\theta)$ and the $\phi$-components of the EM field}
\label{sec:gamma-IF}

The $\phi$-components for the EM field is given by the factor   $\gamma(r,\theta)$, which, from (\ref{eq:gamma*1*}) is given as
{}
\begin{eqnarray}
\gamma(r,\theta) &=& E_0\frac{e^{ik(r+r_g\ln 2kr)}}{ikr}\sum_{\ell=kR^\star_\odot}^\infty\frac{\ell+{\textstyle\frac{1}{2}}}{\ell(\ell+1)}e^{i\big(2\sigma_\ell+\frac{\ell(\ell+1)}{2kr}+\frac{[\ell(\ell+1)]^2}{24k^3r^3}\big)}\Big(e^{i2\delta^*_\ell}-1\Big)\times\nonumber\\
&&\hskip 30pt \times\,
\Big\{\frac{\partial P^{(1)}_\ell(\cos\theta)}
{\partial \theta}+\frac{P^{(1)}_\ell(\cos\theta)}{\sin\theta}\Big(1-u^{-2}
\Big(\frac{\ell(\ell+1)}{2k^2r^2}+\frac{[\ell(\ell+1)]^2}{8k^4r^4}\Big)+\frac{ir_g}{2kr^2}\Big)\Big\}.
  \label{eq:gamma*sum}
\end{eqnarray}

Similarly to the discussion in the preceding Section \ref{sec:beta-IF}, we use (\ref{eq:pi-l*})--(\ref{eq:tau-l*}) and transform (\ref{eq:gamma*sum}) to the integral, while also taking $\ell\gg1$:
{}
\begin{eqnarray}
\gamma(r,\theta) &=& E_0\frac{e^{ik(r+r_g\ln 2kr)}}{ikr}\int_{\ell=kR^\star_\odot}^\infty\ell d\ell e^{i\big(2\sigma_\ell+\frac{\ell^2}{2kr}+\frac{\ell^4}{24k^3r^3}\big)}\Big(e^{i2\delta^*_\ell}-1\Big)\times\nonumber\\
&&\hskip 65pt \times\,
\Big\{J_0(\ell\theta)-{\textstyle\frac{1}{2}}\Big(J_0(\ell\theta)+J_2(\ell\theta)\Big)\Big(u^{-2} \Big(\frac{\ell^2}{2k^2r^2}+\frac{\ell^4}{8k^4r^4}\Big)-\frac{ir_g}{2kr^2}\Big)\Big\}.
  \label{eq:gamma*IF*int}
\end{eqnarray}
We evaluate this integral with the method of stationary phase, again treating plasma-independent and plasma-dependent terms separately.
As a result, we have
{}
\begin{eqnarray}
A^{[0]}(\ell_0)a(\ell_0)\sqrt{\frac{2\pi}{\varphi''(\ell_0)}}&=& \ell_0\Big\{J_0\big(\ell_0\theta\big)- {\textstyle\frac{1}{2}}\Big(J_0\big(\ell_0\theta\big)+J_2\big(\ell_0\theta\big)\Big)\Big(u^{-2}
\Big(\frac{\ell^2_0}{2k^2r^2}+\frac{\ell^4_0}{8k^4r^4}\Big)-\frac{ir_g}{2kr^2}\Big) \Big\}a(\ell_0)\sqrt{\frac{2\pi}{\varphi''(\ell_0)}}=\nonumber\\
&=&kr\sqrt{2\pi kr_g}
\Big\{J_0\big(k\sqrt{2r_gr}\theta\big) +{\cal O}\big(\frac{r_g}{r}, r_g^2\big)\Big\}.~~~
\label{eq:gamma*IF*=1}
\end{eqnarray}

Thus, the expression for $\delta \gamma^{[0]}_\pm(r,\theta)$  in the interference region takes the form
{}
\begin{eqnarray}
\delta \gamma^{[0]}(r,\theta)&=& -E_0\sqrt{2\pi kr_g}e^{i\sigma_0}
J_0\big(k\sqrt{2r_gr}\theta\big)e^{ikr}\Big(1+{\cal O}\big(\frac{r_g}{r}, r_g^2\big)\Big).
\label{eq:gamma*IF*=2}
\end{eqnarray}

We evaluate the plasma-dependent term in (\ref{eq:gamma*IF*int}). Using the relevant expressions (\ref{eq:IFp-phase*4}), (\ref{eq:IFp-phase*6}) from (\ref{eq:gamma*IF*int}), we have:
{}
\begin{eqnarray}
A^{[\tt p]}(\ell_0)a(\ell_0)\sqrt{\frac{2\pi}{\varphi''(\ell_0)}}&=&
kr\sqrt{2\pi kr_g}
\Big(1-
\frac{\delta\theta_{\tt p}}{\sqrt{2r_g/r}}\Big)
\Big\{J_0\Big(k\big(\sqrt{2r_gr}-r\delta\theta_{\tt p}\big) \theta\Big)+{\cal O}\big(\frac{r_g}{r}, r_g^2\big)\Big\}.~~~~
\label{eq:gamma*IF*=3}
\end{eqnarray}
Thus, the term in (\ref{eq:gamma*IF*int}) that depends on the  contribution of the plasma-induced phase shift takes the form
{}
\begin{eqnarray}
\delta \gamma^{[\tt p]}_\pm(r,\theta)&=&\sqrt{2\pi kr_g}e^{i\sigma_0}\Big(1-
\frac{\delta\theta_{\tt p}}{\sqrt{2r_g/r}}\Big)
J_0\Big(k\big(\sqrt{2r_gr}-r\delta\theta_{\tt p}\big) \theta\Big)e^{i(kr+2\delta^\star_\ell)}\Big(1+{\cal O}\big(\frac{r_g}{r}, r_g^2\big)\Big).
\label{eq:gamma*IF*=4}
\end{eqnarray}

Using the expressions (\ref{eq:gamma*IF*=2}) and (\ref{eq:gamma*IF*=4}), we present the integral (\ref{eq:gamma*IF*int}) as
{}
\begin{eqnarray}
\gamma(r,\theta)&=&\delta \gamma^{[0]}(r,\theta)+\delta \gamma^{[{\tt p}]}(r,\theta)=
 \nonumber\\
&=&
\sqrt{2\pi kr_g}e^{i\sigma_0}\Big\{\Big(1-
\frac{\delta\theta_{\tt p}}{\sqrt{2r_g/r}}\Big)J_0\Big(k\big(\sqrt{2r_gr}-r\delta\theta_{\tt p}\big) \theta\Big)e^{i2\delta^\star_\ell}-
J_0\Big(k\sqrt{2r_gr} \theta\Big)+{\cal O}\big(\frac{r_g}{r}, r_g^2\big)\Big\}e^{ikr}.~~~~~
\label{eq:gamma*IF*int-*}
\end{eqnarray}

\subsection{The EM field in the interference region}
\label{sec:amp_func-IF}

Now we are ready to present the components of the EM field in the interference region in the presence of plasma. We do that by using the expressions that we obtained for the functions $\alpha(r,\theta)$, $\beta(r,\theta)$ and $\gamma(r,\theta)$, which are given by (\ref{eq:alpha*IF*int-*}), (\ref{eq:beta*IF*int-*}) and (\ref{eq:gamma*IF*int-*}), correspondingly, and substitute them in (\ref{eq:DB-sol00p*}). As a result, we establish the solution for the scattered EM field in the region outside the termination shock boundary, up to terms of ${\cal O}(r_g^2,\delta\theta_{\tt p}\sqrt{{2r_g}/{r}},(kr)^{-1},\delta\theta^2_{\tt p})$:
{}
\begin{eqnarray}
\left( \begin{aligned}
{  \hat D}_r^{\tt p}& \\
{  \hat B}_r^{\tt p}& \\
  \end{aligned} \right) &=&  -iE_0
\sqrt{\frac{2r_g}{r} }\sqrt{2\pi kr_g}e^{i\sigma_0}\Big\{J_1\Big(k\big(\sqrt{2r_gr}-r\delta\theta_{\tt p}\big) \theta\Big)e^{i2\delta^\star_\ell}-J_1\Big(k\sqrt{2r_gr} \theta\Big)
\Big\}e^{i(kr-\omega t)}
\left( \begin{aligned}
\cos\phi \\
\sin\phi  \\
  \end{aligned} \right),
  \label{eq:IF-DBr}\\
  \left( \begin{aligned}
{ \hat D}^{\tt p}_\theta& \\
{ \hat B}^{\tt p}_\theta& \\
  \end{aligned} \right) &=&
  E_0
\sqrt{2\pi kr_g}e^{i\sigma_0}\Big\{\Big(1-
  \frac{\delta\theta_{\tt p}}{\sqrt{2r_g/r}}\Big)J_0\Big(k\big(\sqrt{2r_gr}-r\delta\theta_{\tt p}\big) \theta\Big)e^{i2\delta^\star_\ell}-
J_0\Big(k\sqrt{2r_gr} \theta\Big)
\Big\}e^{i(kr-\omega t)}\left( \begin{aligned}
\cos\phi \\
\sin\phi  \\
  \end{aligned} \right),
  \label{eq:IF-DBth}\\
\left( \begin{aligned}
{ \hat D}^{\tt p}_\phi& \\
{ \hat B}^{\tt p}_\phi& \\
  \end{aligned} \right) &=&
  E_0
 \sqrt{2\pi kr_g}e^{i\sigma_0}\Big\{\Big(1-
  \frac{\delta\theta_{\tt p}}{\sqrt{2r_g/r}}\Big)J_0\Big(k\big(\sqrt{2r_gr}-r\delta\theta_{\tt p}\big) \theta\Big)e^{i2\delta^\star_\ell}-
J_0\Big(k\sqrt{2r_gr} \theta\Big)
\Big\}e^{i(kr-\omega t)}\left( \begin{aligned}
-\sin\phi \\
\cos\phi  \\
  \end{aligned} \right).~~~~~
  \label{eq:IF-DBph}
\end{eqnarray}

The EM field produced by the Debye potential $\Pi_0$ the wave in the interference region in the absence of plasma was given in   \cite{Turyshev-Toth:2017} in the following form:
{}
\begin{eqnarray}
  \left( \begin{aligned}
{  \hat D}^{\tt (0)}_r& \\
{  \hat B}^{\tt (0)}_r& \\
  \end{aligned} \right) &=& -iE_0
\sqrt{\frac{2r_g}{r} }\sqrt{2\pi kr_g}e^{i\sigma_0}J_1\Big(k\sqrt{2r_gr}\theta\Big)e^{i(kr-\omega t)}\left( \begin{aligned}
\cos\phi \\
\sin\phi  \\
  \end{aligned} \right),
  \label{eq:IF-DBr0} \\
    \left( \begin{aligned}
{  \hat D}^{\tt (0)}_\theta& \\
{  \hat B}^{\tt (0)}_\theta& \\
  \end{aligned} \right) &=& E_0
\sqrt{2\pi kr_g}e^{i\sigma_0}J_0\Big(k\sqrt{2r_gr} \theta\Big)e^{i(kr-\omega t)}\left( \begin{aligned}
\cos\phi \\
\sin\phi  \\
  \end{aligned} \right) ,
   \label{eq:IF-DBth0} \\
   \left( \begin{aligned}
{  \hat D}^{\tt (0)}_\phi& \\
{  \hat B}^{\tt (0)}_\phi& \\
  \end{aligned} \right) &=&E_0
\sqrt{2\pi kr_g}e^{i\sigma_0}J_0\Big(k\sqrt{2r_gr} \theta\Big)e^{i(kr-\omega t)}\left( \begin{aligned}
-\sin\phi \\
\cos\phi  \\
  \end{aligned} \right).
  \label{eq:IF-DBph0}
\end{eqnarray}

The total field in accord with (\ref{eq:Pi-g+p}) is given by the sum of (\ref{eq:IF-DBr})--(\ref{eq:IF-DBph}) and (\ref{eq:IF-DBr0})--(\ref{eq:IF-DBph0}) up to the terms of the order of ${\cal O}(\theta^2,\delta\theta^2_{\tt p},r_g^2,\delta\theta_{\tt p}\sqrt{{2r_g}/{r}},(kr)^{-1})$ has the form:
{}
\begin{eqnarray}
  \left( \begin{aligned}
{  \hat D}_r& \\
{  \hat B}_r& \\
  \end{aligned} \right) &=&  \left( \begin{aligned}
{  \hat D}^{\tt (0)}_r+{  \hat D}^{\tt p}_r& \\
{  \hat B}^{\tt (0)}_r+{  \hat B}^{\tt p}_r& \\
  \end{aligned} \right) = -iE_0
\sqrt{\frac{2r_g}{r} }\sqrt{2\pi kr_g}e^{i\sigma_0}J_1\Big(k\big(\sqrt{2r_gr}-r\delta\theta_{\tt p}\big) \theta\Big)e^{i(kr+2\delta^\star_\ell -\omega t)}
\left( \begin{aligned}
\cos\phi \\
\sin\phi  \\
  \end{aligned} \right),~~~~~
    \label{eq:DB-tot-rr}\\
  \left( \begin{aligned}
{  \hat D}_\theta& \\
{  \hat B}_\theta& \\
  \end{aligned} \right) &=&  \left( \begin{aligned}
{  \hat D}^{\tt (0)}_\theta+{  \hat D}^{\tt p}_\theta& \\
{  \hat B}^{\tt (0)}_\theta+{  \hat B}^{\tt p}_\theta& \\
  \end{aligned} \right) = E_0
 \sqrt{2\pi kr_g}e^{i\sigma_0}\Big(1-
 \frac{\delta\theta_{\tt p}}{\sqrt{2r_g/r}}\Big)J_0\Big(k\big(\sqrt{2r_gr}-r\delta\theta_{\tt p}\big) \theta\Big)e^{i(kr+2\delta^\star_\ell-\omega t)}\left( \begin{aligned}
\cos\phi \\
\sin\phi  \\
  \end{aligned} \right),
  \label{eq:DB-tot-th}\\
\left( \begin{aligned}
{  \hat D}^{\tt }_\phi& \\
{  \hat B}^{\tt }_\phi& \\
  \end{aligned} \right) &=&\left( \begin{aligned}
{  \hat D}^{\tt (0)}_\phi+{  \hat D}^{\tt p}_\phi& \\
{  \hat B}^{\tt (0)}_\phi+{  \hat B}^{\tt p}_\phi& \\
  \end{aligned} \right) =E_0
  \sqrt{2\pi kr_g}e^{i\sigma_0}\Big(1-
  \frac{\delta\theta_{\tt p}}{\sqrt{2r_g/r}}\Big)J_0\Big(k\big(\sqrt{2r_gr}-r\delta\theta_{\tt p}\big) \theta\Big)e^{i(kr+2\delta^\star_\ell-\omega t)}\left( \begin{aligned}
-\sin\phi \\
\cos\phi  \\
  \end{aligned} \right).~~~~~
  \label{eq:DB-tot-ph}
\end{eqnarray}

The radial component of the EM field (\ref{eq:DB-tot-rr}) is negligibly small compared to the other two components, which is consistent with the fact that while passing through the solar plasma the EM wave preserves its transverse structure.

Expressions (\ref{eq:DB-tot-rr})--(\ref{eq:DB-tot-ph}) describe the EM field in the interference region of the SGL  in the spherical coordinate system.  To study this field on the image plane, we need to transform (\ref{eq:DB-tot-rr})--(\ref{eq:DB-tot-ph})  to a cylindrical coordinate system \cite{Herlt-Stephani:1976,Turyshev-Toth:2017}. To do that, we follow the approach demonstrated in  \cite{Turyshev-Toth:2017}, where instead of spherical coordinates $(r,\theta,\phi)$, we introduced a cylindrical coordinate system $(\rho,\phi,z)$, more convenient for these purposes.  In the region $r \gg r_g$, this was done by defining $R=ur = r+{r_g}/{2}+{\cal O}(r_g^2)$ and introducing the coordinate transformations $ \rho=R\sin\theta,$ $ z=R\cos\theta$, which, from (\ref{eq:metric-gen}), result in the following line element:
{}
\begin{eqnarray}
ds^2&=&u^{-2}c^2dt^2-u^2\big(dr^2+r^2(d\theta^2+\sin^2\theta d\phi^2)\big)=u^{-2}c^2dt^2-\big(d\rho^2+\rho^2d\phi^2+nu^2dz^2\big)+{\cal O}(r_g^2).
\label{eq:cyl_coord}
\end{eqnarray}

As a result, using (\ref{eq:DB-tot-rr})--(\ref{eq:DB-tot-ph}), for a high-frequency EM wave (i.e., neglecting terms $\propto(kr)^{-1}$) and for $r\gg r_g$, we derive the field near the optical axis, which up to terms of ${\cal O}(\rho^2/z^2)$, takes the form
{}
\begin{eqnarray}
  \left( \begin{aligned}
{E}_z& \\
{H}_z& \\
  \end{aligned} \right) &=&{\cal O}\Big(\frac{\rho}{z}\Big),
    \label{eq:DB-sol-z}\\
    \left( \begin{aligned}
{E}_\rho& \\
{H}_\rho& \\
  \end{aligned} \right) &=&
 E_0
  \sqrt{2\pi kr_g}e^{i\sigma_0}\Big(1-
  \frac{\delta\theta_{\tt p}}{\sqrt{2r_g/r}}\Big)J_0\Big(k\sqrt{2r_gr}\Big(1-\frac{\delta\theta_{\tt p}}{\sqrt{2r_g/r}}\Big) \theta\Big)e^{i(kr+2\delta^\star_\ell-\omega t)}
 \left( \begin{aligned}
 \cos\phi& \\
 \sin\phi& \\
  \end{aligned} \right),
  \label{eq:DB-sol-rho}\\
    \left( \begin{aligned}
{E}_\phi& \\
{H}_\phi& \\
  \end{aligned} \right) &=&
E_0 \sqrt{2\pi kr_g}e^{i\sigma_0}\Big(1-
\frac{\delta\theta_{\tt p}}{\sqrt{2r_g/r}}\Big)J_0\Big(k\sqrt{2r_gr}\Big(1-\frac{\delta\theta_{\tt p}}{\sqrt{2r_g/r}}\Big) \theta\Big)e^{i(kr+2\delta^\star_\ell-\omega t)}
 \left( \begin{aligned}
 -\sin\phi& \\
 \cos\phi& \\
  \end{aligned} \right),
  \label{eq:DB-ph*}
\end{eqnarray}
where $r=\sqrt{z^2+\rho^2}=z(1+{\rho^2}/{2z^2})=z+{\cal O}(\rho^2/z)$) and $\theta=\rho/z+{\cal O}(\rho^2/z^2)$.
Note that these expressions were obtained  using the approximations (\ref{eq:pi-l}) and are valid for forward scattering when $\theta\approx 0$, or when $\rho\leq r_g$.

\subsection{Plasma contribution to image formation}
\label{sec:SGL-imaging}

Using the result (\ref{eq:DB-sol-z})--(\ref{eq:DB-ph*}), we may now compute the energy flux at the image region of the SGL. The relevant components of the time-averaged Poynting vector for the EM field in the image volume, as a result, may be given in the following form (see \cite{Turyshev-Toth:2017} for details):
{}
\begin{eqnarray}
{\bar S}_z&=&\frac{c}{8\pi}E_0^2
\frac{4\pi^2}{1-e^{-4\pi^2 r_g/\lambda}}
\frac{r_g}{\lambda}\, \Big(1-\frac{\delta\theta_{\tt p}}{\sqrt{2r_g/z}}\Big)^2J^2_0\Big(2\pi\frac{\rho}{\lambda}\Big(\sqrt{\frac{2r_g}{z}}-\delta\theta_{\tt p}\Big)\Big),
\label{eq:S_z*6z}
\end{eqnarray}
with ${\bar S}_\rho= {\bar S}_\phi=0$ for any practical purposes. Also, we recognized that the following convenient expression is valid
{}
\begin{eqnarray}
k\sqrt{2r_gr}\,\theta=2\pi\frac{\rho}{\lambda}\sqrt{\frac{2r_g}{z}}+{\cal O}(\rho^2/z).
\label{eq:J0}
\end{eqnarray}
Therefore, the non-vanishing component of the amplification vector $ {\vec \mu}$, defined as ${\vec \mu}={\vec {\bar S}}/|{\vec{\bar S}}_0|$ where $|{\bar {\vec S}}_0|=(c/8\pi)E_0^2$ is the time-averaged Poynting vector of the wave propagating in empty spacetime, takes the form
{}
\begin{eqnarray}
{\bar \mu}_z&=&
\frac{4\pi^2}{1-e^{-4\pi^2 r_g/\lambda}}
\frac{r_g}{\lambda}\, \Big(1-\frac{\delta\theta_{\tt p}}{\sqrt{2r_g/z}}\Big)^2J^2_0\Big(2\pi\frac{\rho}{\lambda}\sqrt{\frac{2r_g}{z}} \Big(1-
\frac{\delta\theta_{\tt p}}{\sqrt{2r_g/z}}\Big)\Big),
\label{eq:S_z*6z-mu}
\end{eqnarray}
where the argument of the Bessel function to first order in $\delta\theta_{\tt p}$ is from (\ref{eq:IFp-phase*3}) with $\delta\theta_{\tt p}$ itself is  given by  (\ref{eq:ang*}).

 At this point, it is instructive to reinstate the full dependence of the critical partial momenta  $\ell_0$ from (\ref{eq:IFp-phase*3}) on the plasma deflection angle $\delta\theta_{\tt p}$ and, by repeating some of the plasma-related derivations given in Secs.~\ref{sec:beta-IF}--\ref{sec:gamma-IF}, to present the result (\ref{eq:S_z*6z-mu}) in the following more informative form:
{}
\begin{eqnarray}
{\bar \mu}_z&=&
\frac{4\pi^2}{1-e^{-4\pi^2 r_g/\lambda}}
\frac{r_g}{\lambda}\, {\cal F}^2_{\tt pg}J^2_0\Big(2\pi\frac{\rho}{\lambda}
\sqrt{\frac{2r_g}{z}}
{\cal F}_{\tt pg}\Big), \qquad {\rm where} \qquad {\cal F}_{\tt pg}=\Big(1+\frac{\delta\theta^2_{\tt p}}{\delta\theta^2_{\tt g}}\Big)^\frac{1}{2}-\frac{\delta\theta_{\tt p}}{\delta\theta_{\tt g}}\geq 0,
\label{eq:S_z*6z-mu+}
\end{eqnarray}
with $\delta\theta_{\tt g}=\sqrt{2r_g/z}=2r_g/b$ being the Einstein deflection angle due to the gravitational monopole. This result, to first order, is valid for any values of $\delta\theta_{\tt p}$ and $\delta\theta_{\tt g}$ and is very helpful to understand the impact of plasma on the optical properties of the SGL. While Eqs.~(\ref{eq:S_z*6z-mu}) and (\ref{eq:S_z*6z-mu+}) yield similar results when $\delta\theta_{\tt p}\ll\delta\theta_{\tt g}$, reinstating the dependence, from (\ref{eq:IFp-phase*3}), on $\delta\theta_{\tt p}^2/\delta\theta_{\tt g}^2$, helps better understand the behavior of the amplification factor, ${\bar \mu}_z$, at longer wavelaengths.

As we can see from (\ref{eq:S_z*6z-mu+}), the plasma contribution to the optical properties of the SGL is governed by the factor ${\cal F}_{\tt pg}$, which, in the absence of plasma, is ${\cal F}_{\tt pg}=1$. For estimation purposes, we rely on (\ref{eq:ang*ip}), which is the result of evaluating the generic expression for the plasma deflection angle $\delta\theta_{\tt p}$ (\ref{eq:ang*}) for the values given by the phenomenological model (\ref{eq:model}). Then, by using $\delta\theta_{\tt g}=2r_g/b=8.49\times 10^{-6}\,(R_\odot/b)$, we estimate the ratio of the two deflection angles as:
{}
\begin{eqnarray}
\frac{\delta\theta_{\tt p}}{\delta\theta_{\tt g}}=\Big\{
7.80\times 10^{-8}\Big(\frac{R_\odot}{b}\Big)^{15}+
2.41\times 10^{-8}\Big(\frac{R_\odot}{b}\Big)^{5}+
2.85\times 10^{-11}\Big(\frac{R_\odot}{b}\Big)\Big\}\Big(\frac{\lambda}{1~\mu{\rm m}}\Big)^2.
\label{eq:ang*ip+g}
\end{eqnarray}

Examining (\ref{eq:ang*ip+g}) as a function of the impact parameter, we see that for sungrazing rays passing by the Sun with impact parameter $b\simeq R_\odot$, this ratio reaches its largest value of $\delta\theta_{\tt p}/\delta\theta_{\tt g}=1.02\times 10^{-7} \,\big({\lambda}/{1~\mu{\rm m}}\big)^2$, which may be quite significant for microwave and longer wavelengths \cite{Turyshev-Andersson:2002}. For a wave with $\lambda\simeq 3$~mm passing that close to the Sun, the plasma contribution approaches that due to the gravitational bending, $\delta\theta_{\tt p}/\delta\theta_{\tt g}\sim 0.92$. As a result, the factor ${\cal F}_{\tt pg}$ from (\ref{eq:S_z*6z-mu+}) decreases to  ${\cal F}_{\tt pg}\sim 0.44$, which, as seen from (\ref{eq:S_z*6z-mu+}), leads to reducing the light amplification of the SGL to only ${\cal F}^2_{\tt pg}\sim 0.19$ compared to its value for the  plasma-free case and broadening the PSF by a  factor of ${\cal F}^{-1}_{\tt pg}\sim 2.28$, thus, reducing the angular resolution of the SGL in this case by the same amount. For the wavelength $\lambda\simeq 3$~cm, the ratio (\ref{eq:ang*ip+g}) increases to $\delta\theta_{\tt p}/\delta\theta_{\tt g}\sim 91.8$, which reduces the light amplification by a factor of ${\cal F}^2_{\tt pg}\sim 2.97\times 10^{-5}$ compared to the plasma-free case and degrading the resolution by ${\cal F}^{-1}_{\tt pg}\sim184$ times.  Further increasing the wavelength to $\lambda\simeq 30$~cm leads to an obliteration of the optical properties of the SGL, where light amplification is reduced by a factor of $2.97\times 10^{-9}$ compared to the plasma-free case, with angular resolution degraded by $1.84\times 10^{5}$ times.

At the same time, one can clearly see from (\ref{eq:ang*ip+g}) that for optical or IR bands, say for $\lambda\simeq 1~\mu$m or less, the ratio (\ref{eq:ang*ip+g}) is exceedingly small and may be neglected which results in ${\cal F}_{\tt pg}= 1$ for waves in this part of the EM spectrum. This conclusion opens the way for using the SGL for imaging and spectroscopic applications of faint, distant targets.

\section{Discussion and Conclusions}
\label{sec:disc}

Conceptually, the direct imaging of exoplanets is quite straightforward: we simply seek to detect photons from a planet that moves on the background of its parent star.  Emissions from an exoplanet can generally be separated into two sources: stellar emission reflected by the planet's surface or its atmosphere, and thermal emission, which may be either intrinsic thermal emission or emission resulting from heating by the parent star. The reflected light has a spectrum that is broadly similar to that of the star, with additional features arising from the planetary surface or atmosphere. Therefore, for sunlike stars, this reflected emission generally peaks at optical or near optical wavelengths, which are the focus of our present paper.

Although exoplanets are quite faint, it is the proximity of the much brighter stellar source that presents the most severe practical obstacle for direct observation. In the case of the SGL, light from the parent star is typically focused many tens of thousands of kilometers away from the focal line that corresponds to the instantaneous position of the exoplanet. Therefore, light contamination due to the parent star is not a problem when imaging with the SGL \cite{Turyshev-etal:2018}.

We studied the propagation of a monochromatic EM wave on the background of a spherically symmetric gravitational field produced by a gravitational mass monopole described in the first post-Newtonian approximation of the general theory of relativity taken in the harmonic gauge \cite{Turyshev-Toth:2013} and the solar corona represented by the free electron plasma distribution described by a generic, spherically symmetric power law model for the electron number density (\ref{eq:n-eps_n-ism}). We used a generalized model for the solar plasma, which covers the entire solar system from the solar photosphere to the termination shock (i.e., valid for heliocentric distances of $0\leq r \leq R_\star$, first introduced in \cite{Turyshev-Toth:2018-plasma}). We considered the linear combination of gravity and plasma effects, neglecting interaction between the two. This approximation is valid in the solar system environment. Our results, within the required accuracy, do not depend on the actual value of $R_\star$, and as such, deviations from spherical symmetry by the termination shock boundary bear no relevance.

In Sec.~\ref{sec:em-waves-gr+pl}, we solved Maxwell's equations on the background of the solar system, which includes the static gravitational field of the solar monopole and the presence of solar plasma. We used the Mie approach to decompose the Maxwell equations and to present the solution in terms of Debye potentials.  We were able to carry out the variable decomposition of the set of the relevant Maxwell equations and reduce the entire problem to solving the radial equation in the presence of an arbitrary power law potential, representing the plasma.

In Sec.~\ref{sec:sol-EM-Deb} we used the eikonal approximation, valid for all the regions of interest, to solve for the radial function. We established the solution for the EM wave in the exterior region of the solar system (i.e., the region beyond the termination shock, $r>R_\star$) given by (\ref{eq:Pi-degn-sol-in}) and also in the interior region ($0\leq r\leq R_\star$), given by (\ref{eq:Pi-degn-sol-s}). We then used the boundary (continuity) conditions (\ref{eq:bound_cond-expand1+})--(\ref{eq:bound_cond-expand4+}) to match these two solutions at the boundary represented by the termination shock. We established a compact, closed form solution to the boundary value problem in the form of the Debye potentials representing the EM field outside and inside the termination shock boundary, given by (\ref{eq:Pi-s_a1*0}) and (\ref{eq:Pi_g+p}), respectively. Next, we implemented fully absorbing conditions representing the opaque Sun, thus establishing solutions for every region of interest for imaging with the SGL, both outside (\ref{eq:Pi_g+p0}) and inside (\ref{eq:Pi_g+p22}) the termination shock. The resulting Debye potentials fully capture the physics of the EM wave propagation in the complex environment of the solar system. These solutions are new and extend previously known results into the regime where gravity and plasma are both present.

In Sec.~\ref{sec:EM-field-outside} we studied the general solution for the EM field outside the termination shock. We derived the expression for the Debye potential for the plasma-scattered wave outside the termination shock (\ref{eq:Pi_ie*+8p*}). This result is then used to investigate the EM field in all the regions behind the Sun, namely the region of the solar shadow, the geometric optics region and the interference region.

In Sec.~\ref{sec:go-em-outside} we studied the EM field in the region of geometric optics outside the termination shock. We demonstrated that the presence of the solar plasma affects all characteristics of the incident unpolarized light, including the direction of the EM wave propagation, its amplitude and its phase. We observed that the combination of the eikonal approximation and the method of stationary phase results in the expression for the phase of the EM wave that is identical to the one that is usually found by applying the equation of geodesics. This similarly confirmed the validity of our results.  Our approach also allowed us to derive the magnitude of the EM wave as it moves through the refractive medium of the solar system. We also studied the EM field in the interior region of the solar system and investigated the EM field in the geometric optics region inside the termination shock.  We demonstrated that the results obtained in the exterior region are directly applicable for this region as well.  We note that our solution for this region may have immediate practical applications, as it allows for  proper accounting for the effect of solar plasma on modern-day astronomical observations and the tracking of interplanetary spacecraft.

In Sec.~\ref{sec:IF-region}, we focused our attention on the interference region, and investigated the optical properties of the SGL.  We have shown that the presence of the solar plasma leads to a reduction of the light amplification of the SGL and to a broadening of its point-spread function.  Although its presence affects the optical properties of the SGL, its contribution is negligible for optical and IR wavelengths. On the other hand, plasma severely reduces both the light amplification of the SGL and its resolution for wavelengths longer than $\lambda\gtrsim 1$ cm. In general, the steady-state component of the solar plasma uniformly pushes the gravitational caustic \cite{Turyshev-Toth:2017} away from the Sun, but does not introduce additional optical aberrations, leaving the image quality unaffected. Thus, although prospective observations will be conducted through the most intense region of the solar corona, the SGL may be used for imaging of exoplanets at optical and near IR wavelengths \cite{vonEshleman:1979,Labeyrie:1994,Turyshev-Toth:2017}.  We have shown that the signals received from those faint targets are not affected by the refraction in the solar corona at the level of any practical importance.

The steady-state, spherically symmetric component of the solar plasma affects the optical properties of the SGL, especially for microwave or longer wavelengths. It leads to a defocusing, which should not affect the size nor the position of the caustic line, except for the distance to the beginning of the focal line. Such plasma behavior does not induce aberrations \cite{Koechlin-etal:2005} leaving the PSF of the SGL unchanged. What may cause aberrations are deviations from spherical symmetry in the solar corona electron number density  (\ref{eq:n-eps_n-ism}). In a conservative estimate, we consider the upper limit of the index variations being as large as the steady-state component \cite{Giampieri:1994kj}, and varying temporally. Temporal variability in the plasma may introduce additional aberrations. Unpredictable variations must be treated as noise, and accounted for with standard observational techniques \cite{Huber-etal:2013,Lang-book:2009}. Short term temporal variability in the plasma may be accounted for by relying, for instance, on longer integration times, which will be required to reduce the shot noise contribution in any case. One may also rely on the differential Doppler technique \cite{Bertotti-Giampieri:1998,Bertotti-etal-Cassini:2003}, which would allow the plasma contribution to be greatly reduced, by more than three orders of magnitude. In addition to temporal variability of the solar atmosphere, two further physical optics effects, namely spectral broadening and angular broadening, may come into play. However, discussion of these effects is beyond the scope of present paper.

In this paper, we relied on the spherical symmetry to capture the largest terms, representing the realistic field distributions in the solar system.  An almost identical approach may be used to account for any nonsphericity that may be present either in the gravitational field or in the plasma distribution, or else would be introduced by imprecise spacecraft navigation and trajectory determination. Thus, the $1/r$ or $1/r^2$ terms may be included by applying the model that is already developed here. One would have to redefine the the $r_g$ and $\mu^2$ parameters in (\ref{eq:R-bar-k*2}). Similar analysis could be performed to account for higher order terms from the Schwarzschild solution, notably those $\propto r^2_g$. If quadrupole terms (i.e., terms in the potential that behave as $1/r^3$) are present, one can use a spheroidal coordinate system to solve the Maxwell equations. For higher order non-sphericity, given that for the solar system those terms are very small, one may develop a perturbation approach with respect to appropriately defined small parameters.

Concluding, we emphasize that the approach presented here may be extended on a more general case of an extended Sun \cite{Roxburgh:2001,Mecheri-2004,Turyshev:2012nw,Park-etal:2017} and an arbitrary model of the solar plasma with a weak latitude dependence \cite{Muhleman-etal:1977,Muhleman-Anderson:1981}. In addition, the effect on the central caustic of the SGL due to outer solar planets should be taken into account. Similarly to microlensing searches for exoplanets (see \cite{Chung-etal:2005,Gaudi:2012} and references therein), this effect may be important when Jupiter, Saturn or Neptune are very close the optical axis of the SGL, thus providing an additional signal. Finally, one has to evaluate the effect of the solar corona on the photometric signal to noise (SNR) ratio, where the corona's contribution could impact the integration time for observations with the SGL.
This work is on-going and will be reported elsewhere.

\begin{acknowledgments}
This work in part was performed at the Jet Propulsion Laboratory, California Institute of Technology, under a contract with the National Aeronautics and Space Administration.

\end{acknowledgments}


\appendix

\section{Representing Maxwell's equations in terms of Debye potentials}
\label{app:Debye}

Following  \cite{Born-Wolf:1999}, in  \cite{Turyshev-Toth:2017} we represented Maxwell's equations in terms of Debye potentials in the plasma-free case, but in the presence of a static gravitational monopole taken in the first post-Newtonian approximation of the general theory of relativity. In this Appendix, we incorporate the contribution of the solar plasma into our description.

To investigate the propagation of light in the vicinity of the Sun, we consider the metric (\ref{eq:metric-gen}), together with (\ref{eq:eps}) and (\ref{eq:n-eps_n-ism}) and use the approach developed in Appendix E of \cite{Turyshev-Toth:2017}.
We consider the propagation of an EM wave in the vacuum, where no sources or currents exist, i.e., $j^k\equiv (\rho,{\vec j})=0$. This allows us to present the vacuum form of Maxwell's equations (\ref{eq:max-set1})--(\ref{eq:max-set2}),
presenting them for the steady-state, spherically symmetric plasma distribution as
{}
\begin{eqnarray}
{\rm curl}\,{\vec D}&=&-  \mu u^2\frac{1}{c}\frac{\partial {\vec B}}{\partial t}+{\cal O}(G^2),
\qquad ~{\rm div}\big(\epsilon u^2\,{\vec D}\big)={\cal O}(G^2),
\label{eq:rotE_fl*}\\[3pt]
{\rm curl}\,{\vec B}&=& \epsilon u^2\frac{1}{c}\frac{\partial {\vec D}}{\partial t}+{\cal O}(G^2),
\qquad \quad \,
{\rm div }\big(\mu u^2\,{\vec B}\big)={\cal O}(G^2),
\label{eq:rotH_fl*}
\end{eqnarray}
where the differential operators ${\rm curl}$  and ${\rm div}$ are now with respect to the 3-dimensional Euclidean flat metric.

Assuming, as usual, the time dependence of the field in the form $\exp(-i\omega t)$, where $k = \omega/c$, the time-independent parts of the electric and magnetic vectors satisfy Maxwell's
equations, Eq. (\ref{eq:rotE_fl*})--(\ref{eq:rotH_fl*}) for a static and spherically symmetric gravitational field and steady-state, spherically symmetric plasma in their time-independent form:
\begin{eqnarray}
{\rm curl}\,{\vec D}&=&ik\mu u^2 {\vec B}+{\cal O}(r_g^2),
\qquad ~~~~~~{\rm div}\big(\epsilon u^2\,{\vec D}\big)={\cal O}(r_g^2),
\label{eq:rotE_fl*+}\\[3pt]
{\rm curl}\,{\vec B}&=& -ik\epsilon u^2{\vec D}+{\cal O}(r_g^2),
\qquad \quad \,
{\rm div }\big(\mu u^2\,{\vec B}\big)={\cal O}(r_g^2),
\label{eq:rotH_fl*+}
\end{eqnarray}
where $u=1+r_g/2r+{\cal O}(r_g^2, r^{-3})$ as given by (\ref{eq:pot_w_1**}).
In spherical polar coordinates Maxwell's field equations (\ref{eq:rotE_fl*+})--(\ref{eq:rotH_fl*+}), to ${\cal O}(r_g^2,r_g\omega^2_p/\omega^2)$, become
{}
\begin{eqnarray}
-ik\epsilon u^2{\hat {D}}_r&=&\frac{1}{r^2\sin\theta}\Big(
\frac{\partial}{\partial \theta}(r\sin\theta \hat {  B}_\phi)-\frac{\partial}{\partial \phi}(r\hat {  B}_\theta)\Big),
\label{eq:Dr}\\[3pt]
-ik\epsilon u^2{\hat { D}}_\theta&=&\frac{1}{r\sin\theta}
\Big(
\frac{\partial \hat {  B}_r}{\partial \phi}-\frac{\partial}{\partial r}(r\sin\theta \hat {  B}_\phi\Big),
\label{eq:Dt}\\[3pt]
-ik\epsilon u^2{\hat { D}}_\phi&=&\frac{1}{r}
\Big(
\frac{\partial}{\partial r}(r \hat {  B}_\theta)-\frac{\partial \hat {  B}_r}{\partial \theta}\Big),
\label{eq:Dp}\\[3pt]
ik\mu u^2{\hat { B}}_r&=&\frac{1}{r^2\sin\theta}\Big(\frac{\partial}{\partial \theta}(r\sin\theta \hat {  D}_\phi)-\frac{\partial}{\partial \phi}(r\hat {  D}_\theta)\Big),
\label{eq:Br}\\[3pt]
ik\mu u^2{\hat { B}}_\theta&=&
\frac{1}{r\sin\theta}
\Big(\frac{\partial \hat {  D}_r}{\partial \phi}-\frac{\partial}{\partial r}(r\sin\theta \hat {  D}_\phi)\Big),
\label{eq:Bt}\\[3pt]
ik\mu u^2{\hat { B}}_\phi&=&\frac{1}{r}
\Big(\frac{\partial}{\partial r}(r \hat {  D}_\theta)-\frac{\partial \hat {  D}_r}{\partial \theta}\Big),
\label{eq:Bp}
\end{eqnarray}
where $({\hat D}_r, {\hat D}_\theta,{\hat D}_\phi)$, $({\hat B}_r, {\hat B}_\theta,{\hat B}_\phi)$ are the physical components of the EM field, $({\vec D}, {\vec B})$, in the presence of the metric (\ref{eq:metric-gen}), with $u$ from (\ref{eq:pot_w_1**}). For details, see \textsection 84 and \textsection 90 in Ref.~\cite{Landau-Lifshitz:1988} and also Sec. II.A and Appendices A, E in Ref.~\cite{Turyshev-Toth:2017}.

We represent the solution of equations (\ref{eq:Dr})--(\ref{eq:Bp})  as a superposition of two linearly independent fields $\big({}^e{\vec {  D}}, {}^e{\vec {  B}}\big)$ and $\big({}^m{\vec {  D}}, {}^m{\vec {  B}}\big)$,  such that
{}
\begin{eqnarray}
{}^e{\hskip -2pt}{\hat {  D}}_r &=& {\hat {  D}}_r, \qquad
{}^e{\hskip -2pt}{\hat {  B}}_r=0, \\
\label{eq:electr}
\hskip 18pt
{}^m{\hskip -2pt}{\hat {  D}}_r&=&0, \qquad ~\, {}^m{\hskip -2pt}{\hat {  B}}_r={\hat B}_r.
\label{eq:magnet}
\end{eqnarray}

With $\hat {  B}_r={}^e{\hskip -2pt}\hat {  B}_r=0$, Eqs.~(\ref{eq:Dt}) and (\ref{eq:Dp}) become
{}
\begin{eqnarray}
ik\epsilon u^2\,{}^e{\hskip -2pt}{\hat {  D}}_\theta&=&\frac{1}{r}
\frac{\partial}{\partial r}
\big(r \,{}^e{\hskip -2pt}\hat {  B}_\phi\big),
\label{eq:Dt*}\\
ik\epsilon u^2\,{}^e{\hskip -2pt}{\hat { D}}_\phi&=&-\frac{1}{r}
\frac{\partial}{\partial r}\big(r \,{}^e{\hskip -2pt}\hat {  B}_\theta\big).
\label{eq:Dp*}
\end{eqnarray}

Substituting these relationships into (\ref{eq:Bt}) and (\ref{eq:Bp}), we obtain
{}
\begin{eqnarray}
\frac{\partial}{\partial r}\Big[\frac{1}{\epsilon u^2}\frac{\partial}{\partial r}\big(r\,{}^e{\hskip -2pt}{\hat { B}}_\theta\big)\Big]+
k^2\mu u^2(r\,{}^e{\hskip -2pt}{\hat {  B}}_\theta)&=&-
\frac{ik}{\sin\theta}\frac{\partial \,{}^e{\hskip -2pt}\hat {  D}_r}{\partial \phi},
\label{eq:Bp+}\\
\frac{\partial}{\partial r}\Big[\frac{1}{\epsilon u^2}\frac{\partial}{\partial r}\big(r\,{}^e{\hskip -2pt}{\hat {  B}}_\phi\big)\Big]+
k^2\mu u^2(r\,{}^e{\hskip -2pt}{\hat {  B}}_\phi) &=&
ik\frac{\partial \,{}^e{\hskip -2pt}\hat {  D}_r}{\partial \theta}.
\label{eq:Bt+}
\end{eqnarray}

From ${\rm div} (\mu u^2{}^e{\vec { B}})=0 $ given by (\ref{eq:rotH_fl*+}) and relying on the spherical symmetry of $\epsilon$ and $\mu$ and, using $\,{}^e{\hskip -2pt}\hat { B}_r=0$ from (\ref{eq:electr}), we have
{}
\begin{eqnarray}
\frac{\partial}{\partial \theta}\big(\sin\theta \,{}^e{\hskip -2pt}\hat {  B}_\theta\big)+
\frac{\partial \,{}^e{\hskip -2pt}\hat {  B}_\phi}{\partial \phi}&=&0,
\label{eq:divB_fl+}
\end{eqnarray}
which ensures that the remaining equation (\ref{eq:Br}) is satisfied. Indeed, after substitution from (\ref{eq:Dt*}) and (\ref{eq:Dp*}), (\ref{eq:Br}) becomes
{}
\begin{eqnarray}
\frac{1}{r^2\sin\theta}\Big(\frac{\partial}{\partial \theta}\big(r\sin\theta \,{}^e{\hskip -2pt}\hat {  D}_\phi\big)-\frac{\partial}{\partial \phi}\big(r\,{}^e{\hskip -2pt}\hat {  D}_\theta\big)\Big)=-
\frac{1}{ikr^2\sin\theta}\frac{1}{\epsilon u^2}\frac{\partial}{\partial r}\Big[r\Big(\frac{\partial}{\partial \theta}\big(\sin\theta \,{}^e{\hskip -2pt}\hat {  B}_\theta\big)+\frac{\partial {}^e{\hskip -2pt}\hat {  B}_\phi}{\partial \phi}\Big)\Big]={\cal O}\big(r_g^2, r_g\frac{\omega^2_p}{\omega^2}\big),
\label{eq:Br+}
\end{eqnarray}
which is satisfied because of (\ref{eq:divB_fl+}). Strictly similar considerations apply to the complementary case with ${}^m{\hskip -2pt}\hat { D}_r=0$ as shown in (\ref{eq:magnet}).

The solution with vanishing radial magnetic field is called the {\it electric wave} (or transverse magnetic wave) and that with vanishing radial electric field is called the {\it magnetic wave} (or transverse electric wave). We show that they may each be derived from a scalar potential, ${}^e{\hskip -1pt}\Pi$ and ${}^m{\hskip -1pt}\Pi$, respectively. These are known as the Debye potentials.

It follows from (\ref{eq:Br}), since ${}^e{\hskip -2pt}\hat {  B}_r=0$, that ${}^e{\hskip -2pt}\hat {  D}_\phi$ and ${}^e{\hskip -2pt}\hat {  D}_\theta$ may be represented in terms of a gradient of a scalar:
{}
\begin{eqnarray}
{}^e{\hskip -2pt}\hat {  D}_\phi=\frac{1}{r\sin\theta}
\frac{\partial U}{\partial \phi},\qquad
{}^e{\hskip -2pt}\hat {  D}_\theta=\frac{1}{r}
\frac{\partial U}{\partial \theta}.
\label{eq:Dp-Dt}
\end{eqnarray}
If we now put
{}
\begin{eqnarray}
 U=\frac{1}{\epsilon u^2}\frac{\partial }{\partial r}\big(r\,{}^e{\hskip -1pt}\Pi\big),
\label{eq:Pi}
\end{eqnarray}
then we have, from (\ref{eq:Dp-Dt}),
{}
\begin{eqnarray}
{}^e{\hskip -2pt}\hat {  D}_\theta=\frac{1}{\epsilon u^2r}
\frac{\partial^2 \big(r\,{}^e{\hskip -1pt}\Pi\big)}{\partial r\partial \theta},
\qquad
{}^e{\hskip -2pt}\hat {  D}_\phi=\frac{1}{\epsilon u^2r\sin\theta}
\frac{\partial^2 \big(r\,{}^e{\hskip -1pt}\Pi\big)}{\partial r\partial \phi}.
\label{eq:Dp-Dt+}
\end{eqnarray}
It can be seen that (\ref{eq:Dt*}) and (\ref{eq:Dp*}) are satisfied by
{}
\begin{eqnarray}
{}^e{\hskip -2pt}{\hat {  B}}_\phi&=&\frac{ik}{r}\frac{\partial \big(r \,{}^e{\hskip -1pt}\Pi\big)}{\partial \theta},
\qquad
{}^e{\hskip -2pt}{\hat {  B}}_\theta=-\frac{ik}{r\sin\theta}
\frac{\partial \big(r \,{}^e{\hskip -1pt}\Pi\big)}{\partial \phi}.
\label{eq:Bt*}
\end{eqnarray}
If we substitute both equations from (\ref{eq:Bt*}) into (\ref{eq:Dr}), we obtain
\begin{eqnarray}
\,{}^e{\hskip -2pt}{\hat {  D}}_r&=&-\frac{1}{\epsilon u^2r^2\sin\theta}\Big[\frac{\partial}{\partial \theta}\Big(\sin\theta \frac{\partial (r\,{}^e{\hskip -1pt}\Pi)}{\partial \theta}\Big)+\frac{1}{\sin\theta}\frac{\partial^2 (r\,{}^e{\hskip -1pt}\Pi)}{\partial \phi^2}\Big].
\label{eq:Dr*+}
\end{eqnarray}

Substitution from (\ref{eq:Bt*}) and (\ref{eq:Dr*+}) into (\ref{eq:Bp+}) and (\ref{eq:Bt+}) gives two equations, the first of which expresses the vanishing of the $\phi$ derivative, the second the vanishing of the $\theta$ derivative of the same expression on the left-hand side. These equations may, therefore, be satisfied by equating this expression to zero, which gives
\begin{eqnarray}
\epsilon u^2\frac{\partial}{\partial r}\Big[\frac{1}{\epsilon u^2} \frac{\partial (r\,{}^e{\hskip -1pt}\Pi)}{\partial r}\Big]+\frac{1}{r^2\sin\theta}\frac{\partial}{\partial \theta}\Big(\sin\theta \frac{\partial (r\,{}^e{\hskip -1pt}\Pi)}{\partial \theta}\Big)+
\frac{1}{r^2\sin^2\theta}\frac{\partial^2 (r\,{}^e{\hskip -1pt}\Pi)}{\partial \phi^2}+\epsilon\mu\, k^2u^4(r\,{}^e{\hskip -1pt}\Pi)={\cal O}(r_g^2,r_g\frac{\omega^2_p}{\omega^2}).
\label{eq:Pi-eq}
\end{eqnarray}

Defining $'=\partial /\partial r$, this equation may be rewritten as
\begin{align}
\frac{1}{r^2}\frac{\partial }{\partial r}\Big(r^2\frac{\partial}{\partial r} \Big[\frac{\,{}^e{\hskip -1pt}\Pi}{\sqrt{\epsilon}u}\Big]\Big)+
\frac{1}{r^2\sin\theta}\frac{\partial}{\partial \theta}\Big(\sin\theta \frac{\partial}{\partial \theta} \Big[\frac{\,{}^e{\hskip -1pt}\Pi}{\sqrt{\epsilon}u}\Big]\Big)+
\frac{1}{r^2\sin^2\theta}\frac{\partial^2 }{\partial \phi^2}\Big[\frac{\,{}^e{\hskip -1pt}\Pi}{\sqrt{\epsilon}u}\Big]+\Big(\epsilon\mu\, k^2u^4-\sqrt{\epsilon }u\big(\frac{1}{\sqrt{\epsilon }u}\big)''
\Big)\Big[\frac{\,{}^e{\hskip -1pt}\Pi}{\sqrt{\epsilon }u}\Big]=0,
\label{eq:Pi-eq+weq}
\end{align}
which is the wave equation for the quantity ${\,{}^e{\hskip -1pt}\Pi}/{\sqrt{\epsilon}u}$, in the form
\begin{eqnarray}
\Big(\Delta+\epsilon\mu\, k^2u^4-\sqrt{\epsilon}u\big(\frac{1}{\sqrt{\epsilon}u}\big)''\Big)\Big[\frac{\,{}^e{\hskip -1pt}\Pi}{\sqrt{\epsilon}u}\Big]={\cal O}\big(r_g^2,r_g\frac{\omega^2_p}{\omega^2}\big).
\label{eq:Pi-eq+wew1}
\end{eqnarray}
The equation for ${\,{}^m{\hskip -1pt}\Pi}/{\sqrt{\mu}u}$ is identical to (\ref{eq:Pi-eq+wew1}), with $\epsilon$ and $\mu$ swapped.

In a similar way, we may consider the magnetic wave, and find that this wave can be derived from a potential ${}^m{\hskip -1pt}\Pi$ that satisfies the same differential equation (\ref{eq:Pi-eq}) (or (\ref{eq:Pi-eq+wew1})) as ${}^e{\hskip -1pt}\Pi$. The complete solution of Maxwell's field equations in terms of the electric and magnetic Debye potentials, ${}^e\Pi$ and ${}^m\Pi$,  is obtained by adding the two fields (see similar derivations in Appendix E of \cite{Turyshev-Toth:2017} and Appendix A of \cite{Turyshev-Toth:2018-plasma}). This gives
{}
\begin{eqnarray}
{\hat { D}}_r=\,{}^e{\hskip -2pt}{\hat {  D}}_r+\,{}^m{\hskip -2pt}{\hat {  D}}_r&=& \frac{\partial}{\partial r}\Big[\frac{1}{\epsilon u^2} \frac{\partial (r\,{}^e{\hskip -1pt}\Pi)}{\partial r}\Big]+\mu\,k^2 u^2(r\,{}^e{\hskip -1pt}\Pi)=\nonumber\\
&=&
\frac{1}{\sqrt{\epsilon }u}\Big\{\frac{\partial^2 }{\partial r^2}
\Big[\frac{r\,{}^e{\hskip -1pt}\Pi}{\sqrt{\epsilon }u}\Big]+\Big(\epsilon\mu\, k^2 u^4-\sqrt{\epsilon}u\big(\frac{1}{\sqrt{\epsilon }u}\big)''\Big)\Big[\frac{r\,{}^e{\hskip -1pt}\Pi}{\sqrt{\epsilon }u}\Big]\Big\}=\nonumber\\
&=&
-\frac{1}{\epsilon u^2r^2\sin\theta}\Big[\frac{\partial}{\partial \theta}\Big(\sin\theta \frac{\partial (r\,{}^e{\hskip -1pt}\Pi)}{\partial \theta}\Big)+\frac{1}{\sin\theta}\frac{\partial^2 (r\,{}^e{\hskip -1pt}\Pi)}{\partial \phi^2}\Big],
\label{eq:Dr-em}\\[3pt]
{\hat { D}}_\theta=\,{}^e{\hskip -2pt}{\hat { D}}_\theta+\,{}^m{\hskip -2pt}{\hat { D}}_\theta&=&\frac{1}{\epsilon  u^2r}
\frac{\partial^2 \big(r\,{}^e{\hskip -1pt}\Pi\big)}{\partial r\partial \theta}+\frac{ik}{r\sin\theta}
\frac{\partial\big(r\,{}^m{\hskip -1pt}\Pi\big)}{\partial \phi},
\label{eq:Dt-em}\\[3pt]
{\hat {  D}}_\phi=\,{}^e{\hskip -2pt}{\hat {  D}}_\phi+\,{}^m{\hskip -2pt}{\hat {  D}}_\phi&=&\frac{1}{\epsilon  u^2r\sin\theta}
\frac{\partial^2 \big(r\,{}^e{\hskip -1pt}\Pi\big)}{\partial r\partial \phi}-\frac{ik}{r}
\frac{\partial\big(r\,{}^m{\hskip -1pt}\Pi\big)}{\partial \theta},
\label{eq:Dp-em}\\[3pt]
{\hat {  B}}_r=\,{}^e{\hskip -2pt}{\hat {  B}}_r+\,{}^m{\hskip -2pt}{\hat {  B}}_r&=&\frac{\partial}{\partial r}\Big[\frac{1}{\mu u^2} \frac{\partial (r\,{}^m{\hskip -1pt}\Pi)}{\partial r}\Big]+\epsilon\,k^2 u^2(r\,{}^m{\hskip -1pt}\Pi)=
\nonumber\\
&=&\frac{1}{\sqrt{\mu}u}\Big\{\frac{\partial^2}{\partial r^2}\Big[\frac{r\,{}^m{\hskip -1pt}\Pi}{\sqrt{\mu}u}\Big]+\Big(\epsilon\mu\,k^2 u^4-\sqrt{\mu}u\big(\frac{1}{\sqrt{\mu}u}\big)''\Big)\Big[\frac{r\,{}^mu{\hskip -1pt}\Pi}{\sqrt{\mu}u}\Big]\Big\}=\nonumber\\
&=&
-\frac{1}{\mu u^2r^2\sin\theta}\Big[\frac{\partial}{\partial \theta}\Big(\sin\theta \frac{\partial (r\,{}^m{\hskip -1pt}\Pi)}{\partial \theta}\Big)+\frac{1}{\sin\theta}\frac{\partial^2 (r\,{}^m{\hskip -1pt}\Pi)}{\partial \phi^2}\Big],~~~
\label{eq:Br-em}\\[3pt]
{\hat {  B}}_\theta=\,{}^e{\hskip -2pt}{\hat {  B}}_\theta+\,{}^m{\hskip -2pt}{\hat {  B}}_\theta&=&-\frac{ik}{r\sin\theta}
\frac{\partial\big(r\,{}^e{\hskip -1pt}\Pi\big)}{\partial \phi}+\frac{1}{\mu u^2r}
\frac{\partial^2 \big(r\,{}^m{\hskip -1pt}\Pi\big)}{\partial r\partial \theta},
\label{eq:Bt-em}\\[3pt]
{\hat {  B}}_\phi=\,{}^e{\hskip -2pt}{\hat {  B}}_\phi+\,{}^m{\hskip -2pt}{\hat {  B}}_\phi&=&\frac{ik}{r}
\frac{\partial\big(r\,{}^e{\hskip -1pt}\Pi\big)}{\partial \theta}+\frac{1}{\mu u^2r\sin\theta}
\frac{\partial^2 \big(r\,{}^m{\hskip -1pt}\Pi\big)}{\partial r\partial \phi},
\label{eq:Bp-em}
\end{eqnarray}
where the potentials ${}^e\Pi$ and ${}^m\Pi$ both satisfy the following wave equations, valid to ${\cal O}(r_g^2,r_g\omega^2_p/\omega^2)$:
\begin{eqnarray}
\Big(\Delta+\epsilon\mu\,k^2u^4-\sqrt{\epsilon}u\big(\frac{1}{\sqrt{\epsilon}u}\big)''\Big)\Big[\frac{\,{}^e{\hskip -1pt}\Pi}{\sqrt{\epsilon}u}\Big]=0,
\label{eq:Pi-eq+wew1*+-app}
\qquad
\Big(\Delta+\epsilon\mu\,k^2u^4-\sqrt{\mu}u\big(\frac{1}{\sqrt{\mu}u}\big)''\Big)\Big[\frac{\,{}^m{\hskip -1pt}\Pi}{\sqrt{\mu}u}\Big]=0.
\label{eq:Pi-eq+wmw1*-app}
\end{eqnarray}
Also, for convenience, we gave three different but equivalent forms for the radial components for the EM field.

Finally, for the components ${\hat D}_\theta, {\hat D}_\phi$ and ${\hat B}_\theta, {\hat B}_\phi$ to be continuous over spherical surface at the termination shock, $r=R_\star$ , it is evidently sufficient that the four quantities
\begin{eqnarray}
\epsilon (r\,{}^e{\hskip -1pt}\Pi), \qquad \mu (r\,{}^m{\hskip -1pt}\Pi),
\qquad \frac{\partial (r\,{}^e{\hskip -1pt}\Pi)}{\partial r},
\qquad \frac{\partial (r\,{}^m{\hskip -1pt}\Pi)}{\partial r},
\label{eq:bound_cond}
\end{eqnarray}
shall also be {\it continuous} over this surface. Thus, our boundary conditions also split into independent conditions on $\,{}^e{\hskip -1pt}\Pi$ and $\,{}^m{\hskip -1pt}\Pi$. Our diffraction problem is thus reduced to the problem of finding two mutually independent solutions of the equations (\ref{eq:Pi-eq}) (or, equivalently, (\ref{eq:Pi-eq+wew1*+-app})) with prescribed boundary conditions.

\section{Light propagation in weak and static gravity and plasma}
\label{sec:geodesics-phase}

\subsection{Light paths in weak and static gravity in the presence of plasma}
\label{sec:geodesics}

To investigate the propagation of light in the vicinity of the Sun, we consider the metric (\ref{eq:metric-gen}) with $u$ given by (\ref{eq:pot_w_1**}). To account for the presence of plasma with a refractive index $n=\sqrt{\epsilon\mu}$, following \cite{Synge-book-1960},
we rescale the speed of light  as $c\rightarrow c/n=c/\sqrt{\epsilon\mu}$, which leads to the following modification of (\ref{eq:metric-gen}):
\begin{eqnarray}
ds^2&=&n^{-1}u^{-2}c^2dt^2-nu^2\big(dr^2+r^2(d\theta^2+\sin^2\theta d\phi^2)\big),
\label{eq:metric-gen*}
\end{eqnarray}
where, for non-magnetic media, the static index of refraction $n^2= \epsilon ({\vec r})$ from (\ref{eq:eps}) together with the electron number density for the steady-state part of the solar corona from (\ref{eq:n-eps_n-ism}) and (\ref{eq:n_n-ss}), is given as
\begin{equation}
n^2= 1 - \frac{\omega_{\tt p}^{ 2}}{\omega^2},
\qquad {\rm with} \qquad
\omega_{\tt p}^2=\frac{4\pi e^2}{m_e}\sum_i \alpha_i \Big(\frac{R_\odot}{r}\Big)^{\beta_i}.
 \label{eq:n-eps}
\end{equation}

In the refractive medium of the solar plasma and in the weak gravitational field of the Sun, we may represent the trajectory of a light ray as a linear superposition of two perturbations: one introduced by gravity and the other one due to plasma. Thus, to first order in $G$ and ${\omega_{\tt p}^2}/{\omega^2}$, the trajectory of a light ray may be given as
{}
\begin{eqnarray}
x^\alpha(t)&=&x^\alpha_{0}+k^\alpha c(t-t_0)+x^\alpha_{\tt G}(t)+x^\alpha_{\tt p}(t)+{\cal O}(G^2,G\frac{\omega_{\tt p}^2}{\omega^2}),
\label{eq:x-Newt}
\end{eqnarray}
where $k^\alpha$ is the unit vector in the unperturbed direction of the light ray's propagation, while $x^\alpha_{\tt G}(t)$ and $x^\alpha_{\tt p}(t)$ are the post-Newtonian and plasma terms, correspondingly. We define the four-dimensional wavevector in curved spacetime as usual:
{}
\begin{eqnarray}
K^m=\frac{dx^m}{d\lambda}=\frac{dx^0}{d\lambda}\big(1,\frac{dx^\alpha}{dx^0}\big)= K^0\big(1,\kappa^\alpha\big),
\label{eq:K-def}
\end{eqnarray}
where $\lambda$ is the parameter along the ray's path and $\kappa^\alpha={dx^\alpha}/{dx^0}$ is the unit vector in that direction, i.e., $\kappa_\epsilon \kappa^\epsilon=-1$. From (\ref{eq:x-Newt}) we see that the unit vector $\kappa^\alpha$ may be represented  as $\kappa^\alpha=k^\alpha +k^\alpha_{\tt G}(t)+k^\alpha_{\tt p}(t)+{\cal O}(G^2,G{\omega_{\tt p}^2}/{\omega^2}),$ where $k^\alpha_{\tt G}(t)=dx^\alpha_{\tt G}/{dx^0}$ is the post-Newtonian perturbation and $k^\alpha_{\tt p}(t)=dx^\alpha_{\tt p}/{dx^0}$ is that due to plasma.
The wavevector obeys the geodesic equation: ${dK^m}/{d\lambda}+\Gamma^m_{kl}K^mK^l=0,$ which, for temporal and spatial components, yields
{}
\begin{eqnarray}
\frac{dK^0}{d\lambda}-2K^0K^\epsilon\Big(c^{-2}\partial_\epsilon U-\frac{1}{4\omega^2}\partial_\epsilon\omega_{\tt p}^2\Big)&=&{\cal O}\big(G^2,G\frac{\omega_{\tt p}^2}{\omega^2}\big),
\label{eq:K0}\\
\frac{dK^\alpha}{d\lambda}+2K^\alpha K^\epsilon\Big(c^{-2}\partial_\epsilon U-\frac{1}{4\omega^2}\partial_\epsilon\omega_{\tt p}^2\Big)+\Big((K^0)^2-K_\epsilon K^\epsilon\Big)\Big(c^{-2}\partial^\alpha U-\frac{1}{4\omega^2}\partial^\alpha\omega_{\tt p}^2\Big)&=&{\cal O}\big(G^2,G\frac{\omega_{\tt p}^2}{\omega^2}\big).
\label{eq:K-eq}
\end{eqnarray}
Equation (\ref{eq:K0}) is an integral of motion due to energy conservation (as the metric (\ref{eq:metric-gen}) is independent on time). Indeed, we can present it as
{}
\begin{eqnarray}
\frac{dK^0}{d\lambda}-K^0K^\epsilon\Big(2c^{-2}\partial_\epsilon U-\frac{1}{2\omega^2}\partial_\epsilon\omega_{\tt p}^2\Big)&=&
\frac{d}{d\lambda}\Big(g_{00}\frac{dx^0}{d\lambda}\Big)+{\cal O}\big(G^2,G\frac{\omega_{\tt p}^2}{\omega^2}\big)={\cal O}\big(G^2,G\frac{\omega_{\tt p}^2}{\omega^2}\big).~~~
\label{eq:K0-eq01}
\end{eqnarray}
Therefore, in the static field energy is conserved, and we have the following integral of motion:
{}
\begin{eqnarray}
g_{00}\frac{dx^0}{d\lambda}={\rm const}+{\cal O}\big(G^2,G\frac{\omega_{\tt p}^2}{\omega^2}\big) \quad~~~ \Rightarrow \quad~~~
x^0= ct=k^0\lambda+x^0_{\tt G}(\lambda)+x^0_{\tt p}(\lambda)+{\cal O}\big(G^2,G\frac{\omega_{\tt p}^2}{\omega^2}\big),
\label{eq:k0_s}
\end{eqnarray}
where $x^0_{\tt G}(\lambda)$ is the post-Newtonian correction and $x^0_{\tt p}(\lambda)$ is that due to plasma.
We recall that the wavevector $K^m$ is a null vector, which, to first order in $G$ and ${\omega_{\tt p}^2}/{\omega^2}$, and with $K^0=k^0+{\cal O}(G,{\omega_{\tt p}^2}/{\omega^2})$ yields the relation $K_mK^m=0=(k^0)^2\big(1+\gamma_{\epsilon\beta}k^\epsilon k^\beta+{\cal O}\big(G,{\omega_{\tt p}^2}/{\omega^2})\big)$. Then, Eq.~(\ref{eq:K-eq}) becomes
{}
\begin{eqnarray}
\frac{dK^\alpha}{d\lambda}+2(k^0)^2\big(k^\alpha k^\epsilon-\gamma^{\alpha\epsilon}k_\mu k^\mu\big)\Big(c^{-2}\partial_\epsilon U-\frac{1}{4\omega^2}\partial_\epsilon\omega_{\tt p}^2\Big)&=&{\cal O}\big(G^2,G\frac{\omega_{\tt p}^2}{\omega^2}\big).
\label{eq:K-eq2}
\end{eqnarray}

We can now represent (\ref{eq:K-eq2}) in terms of derivatives with respect to time $x^0$. First we have
{}
\begin{eqnarray}
\frac{dK^\alpha}{d\lambda}=(K^0)^2\frac{d^2x^\alpha}{dx^0{}^2}+\frac{dK^0}{d\lambda}\frac{dx^\alpha}{dx^0}.
\label{eq:X-eq]}
\end{eqnarray}
Substituting (\ref{eq:X-eq]}) into (\ref{eq:K-eq2}) and using (\ref{eq:K0}), we have
{}
\begin{eqnarray}
\frac{d^2x^\alpha}{dx^0{}^2}+2\big(k^\alpha k^\epsilon-\gamma^{\alpha\epsilon}k_\mu k^\mu\big)\Big(c^{-2}\partial_\epsilon U-\frac{1}{4\omega^2}\partial_\epsilon\omega_{\tt p}^2\Big)&=&
-2k^\alpha k^\epsilon\Big(c^{-2}\partial_\epsilon U-\frac{1}{4\omega^2}\partial_\epsilon\omega_{\tt p}^2\Big)+{\cal O}\big(G^2,G\frac{\omega_{\tt p}^2}{\omega^2}\big).
\label{eq:X-eq][}
\end{eqnarray}
Remember that for light $ds^2=0$. Then, from the fact that rays of light move along light cones, the following expression is valid $g_{mn}({dx^m}/{dx^0})({dx^n}/{dx^0})=0=1+k_\epsilon k^\epsilon +{\cal O}(G,{\omega_{\tt p}^2}/{\omega^2})$, which for (\ref{eq:X-eq][}) yields
{}
\begin{eqnarray}
\frac{d^2x^\alpha}{dx^0{}^2}=-2\big(\gamma^{\alpha\epsilon}+2k^\alpha k^\epsilon\big)
\Big(c^{-2}\partial_\epsilon U-\frac{1}{4\omega^2}\partial_\epsilon\omega_{\tt p}^2\Big)+{\cal O}\big(G^2,G\frac{\omega_{\tt p}^2}{\omega^2}\big).
\label{eq:X-eq2}
\end{eqnarray}

To continue, we examine the unperturbed part of (\ref{eq:x-Newt}) and representing it as
{}
\begin{eqnarray}
x^\alpha(t)&=&x^\alpha_{0}+k^\alpha c(t-t_0)+{\cal O}(G,{\omega_{\tt p}^2}/{\omega^2})=
[{\vec k}\times[{\vec x}_{0}\times{\vec k}]]^\alpha+k^\alpha \big(({\vec k}\cdot {\vec x}_{0})+c(t-t_0)\big)+{\cal O}(G,{\omega_{\tt p}^2}/{\omega^2}).
\label{eq:x-Newt*}
\end{eqnarray}
Following \cite{Kopeikin:1997,Kopeikin-book-2011,Turyshev-Toth:2017}, we define $b^\alpha\equiv  {\vec b}=[[{\vec k}\times{\vec x}_0]\times{\vec k}]+{\cal O}(G,{\omega_{\tt p}^2}/{\omega^2})$ to be the impact parameter of the unperturbed trajectory of the light ray. The vector ${\vec b}$ is directed from the origin of the coordinate system toward the point of the closest approach of the unperturbed path of light ray to that origin. We also introduce the parameter $\tau=\tau(t)$ as
{}
\begin{eqnarray}
\tau &=&({\vec k}\cdot {\vec x})=({\vec k}\cdot {\vec x}_{0})+c(t-t_0).
\label{eq:x-Newt*=}
\end{eqnarray}
Clearly, when the coordinate system is oriented along the inicdent direction of the light ray, then $\tau=({\vec k}\cdot {\vec x})\equiv  z$.

These quantities allow us to rewrite (\ref{eq:x-Newt*}) as
{}
\begin{eqnarray}
x^\alpha(\tau)&=&b^\alpha+k^\alpha \tau+{\cal O}(G,\frac{\omega_{\tt p}^2}{\omega^2}),
\qquad r(\tau)=\sqrt{b^2+\tau^2}+{\cal O}(G, \frac{\omega_{\tt p}^2}{\omega^2}).
\label{eq:b}
\end{eqnarray}
 The following relations hold:
{}
\begin{eqnarray}
r+\tau&=&\frac{b^2}{r-\tau}+{\cal O}(G,\frac{\omega_{\tt p}^2}{\omega^2}),\qquad r_0+\tau_0=\frac{b^2}{r_0-\tau_0}+{\cal O}(G,\frac{\omega_{\tt p}^2}{\omega^2}), \qquad {\rm and} \qquad
\frac{r+\tau}{r_0+\tau_0}=\frac{r_0-\tau_0}{r-\tau}+{\cal O}(G,\frac{\omega_{\tt p}^2}{\omega^2}).~~~~~~~~
\label{eq:rel}
\end{eqnarray}
They are useful for presenting the results of integration of the light ray equations in different forms.

Limiting our discussion to the monopole given by (\ref{eq:pot_w_1**}), we have $c^{-2}\partial^\alpha U=-(r_g/2r^2)\partial^\alpha r+{\cal O}(G^2,r^{-4})$. We recall that $\partial^\alpha r=\partial^\alpha \sqrt{-x_\epsilon x^\epsilon}=-x^\alpha/r.$ Then,  $c^{-2}\partial^\alpha U=(r_g/2r^3)x^\alpha +{\cal O}(r_g^2,r^{-4})$. In a similar manner, from (\ref{eq:n-eps}), for the plasma-related term we obtain: $\partial^\alpha n_e({\vec r})=\sum_k \alpha_k {\beta_k} \big({R_\odot}/{r}\big)^{\beta_k}(x^\alpha /r^2)$. As a result, (\ref{eq:X-eq2}) takes the form:
{}
\begin{eqnarray}
\frac{d^2x^\alpha}{dx^0{}^2}&=&
-r_g\frac{b^\alpha-k^\alpha \tau}{(b^2+\tau^2)^{3/2}}+
\frac{2\pi e^2}{m_e\omega^2}\sum_i \alpha_i {\beta_i} R_\odot^{\beta_i}\frac{b^\alpha-k^\alpha \tau}{(b^2+\tau^2)^{1+\frac{1}{2}\beta_i}}+{\cal O}(r_g^2,r_g\frac{\omega_{\tt p}^2}{\omega^2}).
\label{eq:X-eq2*}
\end{eqnarray}
Using (\ref{eq:x-Newt*=}), we make the substitution $d/d x^0=d/d\tau$, which leads to the following equation:
{}
\begin{eqnarray}
\frac{d^2x^\alpha}{d\tau^2}=-r_g\frac{b^\alpha-k^\alpha \tau}{(b^2+\tau^2)^{3/2}}+\frac{2\pi e^2}{m_e\omega^2}\sum_i \alpha_i {\beta_i} R_\odot^{\beta_i}\frac{b^\alpha-k^\alpha \tau}{(b^2+\tau^2)^{1+\frac{1}{2}\beta_i}}+{\cal O}(r_g^2,r_g\frac{\omega_{\tt p}^2}{\omega^2}).
\label{eq:X-eq2*/=}
\end{eqnarray}

We integrate (\ref{eq:X-eq2*/=}) from $-\infty$ to $\tau$ to get the following result:
{}
\begin{eqnarray}
\frac{dx^\alpha}{d\tau}&=&
k^\alpha-r_g\Big(\frac{k^\alpha}{\sqrt{b^2+\tau^2}}+\frac{b^\alpha}{b^2}\big(\frac{\tau}{\sqrt{b^2+\tau^2}}+1\big)\Big)+\nonumber\\
&&+\frac{2\pi e^2}{m_e\omega^2}\sum_i \alpha_i \Big(\frac{R_\odot}{b}\Big)^{\beta_i}\Big\{\frac{k^\alpha (b^2)^{\frac{1}{2}\beta_i}}{(b^2+\tau^2)^{\frac{1}{2}\beta_i}}+{\beta_i} \frac{b^\alpha}{b}\Big({}_2F_1\big[{\textstyle\frac{1}{2}},1+{\textstyle\frac{1}{2}}{\beta_k},{\textstyle\frac{3}{2}},-\frac{\tau^2}{b^2}\big]\frac{\tau}{b}\Big)\Big|_{-\infty}^\tau\Big\}+{\cal O}(r_g^2,r_g\frac{\omega_{\tt p}^2}{\omega^2}).
\label{eq:X-eq3*+}
\end{eqnarray}

Following \cite{Turyshev-Toth:2018-plasma}, we define the function $Q_{\beta_i}(\tau)$ as
{}
\begin{eqnarray}
Q_{\beta_i}(\tau)={}_2F_1\Big[{\textstyle\frac{1}{2}},{\textstyle\frac{1}{2}}{\beta_i},{\textstyle\frac{3}{2}},-\frac{\tau^2}{b^2}\Big]\frac{\tau}{b},
\label{eq:Q_b}
\end{eqnarray}
which is a smooth and finite function for all values of $\tau$ with the following relevant limits
{}
\begin{eqnarray}
\lim_{\tau\rightarrow0} Q_{\beta_i}\big(\tau\big)=0,
\qquad
\lim_{\tau\rightarrow\infty} Q_{\beta_i}\big(\tau\big)=Q_{\beta_i}^\star,
\qquad
\lim_{\beta_i\rightarrow\infty} Q^\star_{\beta_i}=0.
\label{eq:Q_b-app}
\end{eqnarray}
For $\beta_i$, typically present in solar corona models (i.e., given in  (\ref{eq:model})), the quantity $Q^\star_{\beta_i}$ has the following values:
{}
\begin{eqnarray}
Q^\star_{2}&=&\frac{\pi}{2}\approx 1.5708,\qquad Q^\star_{6}=\frac{3\pi}{16}\approx 0.5891,\qquad Q^\star_{16}=\frac{429\pi}{4096}\approx 0.3290, \nonumber\\
Q^\star_{4}&=&\frac{\pi}{4}\approx 0.7854,\qquad Q^\star_{8}=\frac{5\pi}{32}\approx 0.4909,\qquad Q^\star_{18}=\frac{6435\pi}{65536}\approx 0.3085.
\label{eq:sig*v}
\end{eqnarray}

Using the new function (\ref{eq:Q_b}), we can improve the form of (\ref{eq:X-eq3*+}) as:
{}
\begin{eqnarray}
\frac{dx^\alpha}{d\tau}&=&k^\alpha-r_g\Big(\frac{k^\alpha}{\sqrt{b^2+\tau^2}}+\frac{b^\alpha}{b^2}\big(\frac{\tau}{\sqrt{b^2+\tau^2}}+1\big)\Big)+\nonumber\\
&&+\frac{2\pi e^2}{m_e\omega^2}\sum_i \alpha_i \Big(\frac{R_\odot}{b}\Big)^{\beta_i}\Big\{\frac{k^\alpha (b^2)^{\frac{1}{2}\beta_i}}{(b^2+\tau^2)^{\frac{1}{2}\beta_i}}+{\beta_i} \frac{b^\alpha}{b}\Big(Q_{\beta_i+2}(\tau)+Q^\star_{\beta_i+2}\Big)\Big\}+{\cal O}\big(r_g^2,r_g\frac{\omega_{\tt p}^2}{\omega^2}\big).
\label{eq:X-eq3}
\end{eqnarray}

From  (\ref{eq:X-eq3}), and with the help of (\ref{eq:x-Newt*=})--(\ref{eq:b}),
we have the following expression for the wavevector $\kappa^\alpha$  from (\ref{eq:K-def}):
{}
\begin{eqnarray}
\kappa^\alpha=
\frac{dx^\alpha}{d\tau}&=&k^\alpha\big(1-\frac{r_g}{r}\big)-\frac{r_g}{b^2}b^\alpha\big(1+\frac{({\vec k}\cdot {\vec x})}{r}\big)+
\nonumber\\
&+&\,
\frac{2\pi e^2}{m_e\omega^2}\sum_i \alpha_i \Big(\frac{R_\odot}{b}\Big)^{\beta_i}\Big\{k^\alpha \Big(\frac{b}{r}\Big)^{\beta_i}+{\beta_i} \frac{b^\alpha}{b}\Big(Q_{\beta_i+2}\big(\sqrt{r^2-b^2}{\,}\big)+Q^\star_{\beta_i+2}\Big)\Big\}+
{\cal O}(r_g^2,r_g\frac{\omega_{\tt p}^2}{\omega^2}).
\label{eq:X-eq3*}
\end{eqnarray}

We may now integrate (\ref{eq:X-eq3}) from $\tau_0$ to $\tau$ to obtain
{}
\begin{eqnarray}
x^\alpha (\tau)&=&b^\alpha+k^\alpha \tau-r_g\int_{\tau_0}^\tau \Big(\frac{k^\alpha}{\sqrt{b^2+\tau'^2}}+\frac{b^\alpha}{b^2}\big(\frac{\tau'}{\sqrt{b^2+\tau'^2}}+1\big)\Big)d\tau'+\nonumber\\
&+&\frac{2\pi e^2}{m_e\omega^2}\sum_i \alpha_i \Big(\frac{R_\odot}{b}\Big)^{\beta_i}\int_{\tau_0}^\tau \Big\{\frac{k^\alpha (b^2)^{\frac{1}{2}\beta_i}}{(b^2+\tau'^2)^{\frac{1}{2}\beta_i}}+{\beta_i} \frac{b^\alpha}{b}\Big(Q_{\beta_i+2}(\tau')+Q^\star_{\beta_i+2}\Big)\Big\}d\tau'+
{\cal O}(r_g^2,r_g\frac{\omega_{\tt p}^2}{\omega^2}),
\label{eq:X-eq4}
\end{eqnarray}
which, to the order of  ${\cal O}(r_g^2,r_g{\omega_{\tt p}^2}/{\omega^2})$, results in
{}
\begin{eqnarray}
x^\alpha (\tau)&=&b^\alpha+k^\alpha \tau-r_g\Big(k^\alpha\ln\frac{\tau+\sqrt{b^2+\tau^2}}{\tau_0+\sqrt{b^2+\tau^2_0}}+\frac{b^\alpha}{b^2}\big(\sqrt{b^2+\tau^2}+\tau-\sqrt{b^2+\tau^2_0}-\tau_0\big)\Big)+\nonumber\\
&+&\frac{2\pi e^2 R_\odot}{m_e\omega^2}\sum_i \alpha_i \Big(\frac{R_\odot}{b}\Big)^{\beta_i-1}\Big\{k^\alpha \Big( Q_{\beta_i}(\tau)-Q_{\beta_i}(\tau_0)\Big)+{\beta_i} \frac{b^\alpha}{b^2}\Big(\int_{\tau_0}^\tau Q_{\beta_i+2}(\tau')d\tau'+(\tau-\tau_0)\,Q^\star_{\beta_i+2}\Big)\Big\},~~~
\label{eq:X-eq4*}
\end{eqnarray}
or, equivalently, substituting $\tau$ and $r$ from (\ref{eq:x-Newt*=})--(\ref{eq:b}), we have
{}
\begin{eqnarray}
x^\alpha (t)&=&x_0^\alpha+k^\alpha c(t-t_0)-r_g\Big(k^\alpha\,\ln\frac{r+({\vec k}\cdot{\vec x})}{r_0+({\vec k}\cdot{\vec x}_0)}+\frac{b^\alpha}{b^2}\big(r+({\vec k}\cdot{\vec x})-r_0-({\vec k}\cdot{\vec x}_0)\big)\Big)+\nonumber\\
&+&\frac{2\pi e^2R_\odot}{m_e\omega^2}\sum_i \alpha_i \Big(\frac{R_\odot}{b}\Big)^{\beta_i-1}\Big\{k^\alpha \Big(Q_{\beta_i}\big(({\vec k}\cdot{\vec x})\big)-Q_{\beta_i}\big(({\vec k}\cdot{\vec x}_0)\big)\Big)+\nonumber\\
&&\hskip 60pt
+\,{\beta_i} \frac{b^\alpha}{b^2}\Big(\int_{\tau_0}^\tau Q_{\beta_i+2}(\tau')d\tau'+\big({\vec k}\cdot({\vec x}-{\vec x}_0)\big)\,Q^\star_{\beta_i+2}\Big)\Big\}+ {\cal O}(r_g^2,r_g\frac{\omega_{\tt p}^2}{\omega^2}).
\label{eq:X-eq4**}
\end{eqnarray}

Therefore, the trajectory of a light ray in a static weak gravitational field with refractive medium (\ref{eq:eps}) and (\ref{eq:n-eps_n-ism}) is described by (\ref{eq:X-eq4*}), while the direction of its wavevector $\kappa^\alpha= {dx^\alpha}/{dx^0}$ is given by (\ref{eq:X-eq3*}).

For a radial light ray given by $k^\alpha =x^\alpha_0/r_0=n^\alpha_0$ and $b=0$, we integrate (\ref{eq:X-eq2*}) with $b=0$ to obtain:
{}
\begin{eqnarray}
\frac{dx^\alpha}{d\tau}&=&n^\alpha_0\Big\{1-\frac{r_g}{r}+\frac{2\pi e^2}{m_e\omega^2}\sum_i \alpha_i \Big(\frac{R_\odot}{r}\Big)^{\beta_i}\Big\} +{\cal O}(r_g^2,r_g\frac{\omega_{\tt p}^2}{\omega^2}),
\label{eq:X-eq3*_rad}\\
x^\alpha (t)&=&x_0^\alpha+n^\alpha_0 \Big(c(t-t_0)-r_g \ln\frac{r}{r_0}-\frac{2\pi e^2 R_\odot}{m_e\omega^2} \sum_i \frac{\alpha_i}{\beta_i-1} \Big\{\Big(\frac{R_\odot}{r}\Big)^{\beta_i-1}-\Big(\frac{R_\odot}{r_0}\Big)^{\beta_i-1}\Big\}\Big) +{\cal O}(r_g^2,r_g\frac{\omega_{\tt p}^2}{\omega^2}).~~~
\label{eq:X-eq4**_rad}
\end{eqnarray}

\subsection{Geometric optics approximation for the wave propagation in the vicinity of a massive body}
\label{sec:geom-optics}

In geometric optics, the phase $\varphi$ is a scalar function, a solution to the eikonal equation \cite{Fock-book:1959,Landau-Lifshitz:1988,Kopeikin:2009,Kopeikin-book-2011}:
\begin{equation}
g^{mn}\partial_m\varphi\partial_n\varphi=0.
\label{eq:eq_eik}
\end{equation}
We use this equation to determine the phase evolution in the presence of plasma and gravity. For this, we use the metric, $g_{mn}$, (\ref{eq:metric-gen*}) with plasma index of refraction, $n$, given by (\ref{eq:n-eps}).

Given the wavevector $K_m = \partial_m\varphi$, and its tangent $K^m = dx^m/d\lambda = g^{mn}\partial_n\varphi$ where $\lambda$ is an affine parameter, we note that (\ref{eq:eq_eik}) states that $K^m$ is null ($g_{mn}K^mK^n = 0$), thus
\begin{equation}
\frac{dK_m}{d\lambda} = \frac{1}{2}\partial_m g_{kl}K^kK^l.
\label{eq:eq_eik-K}
\end{equation}
Eq.~(\ref{eq:eq_eik}) can be solved by assuming an unperturbed solution that is a plane wave:
\begin{equation}
\varphi(t,{\vec x}) = \varphi_0+\int \underline{k}_m dx^m+\varphi_{\tt G} (t,{\vec x})+\varphi_{\tt p} (t,{\vec x})+{\cal O}(r_g^2,r_g\frac{\omega^2_{\tt p}}{\omega^2}),
\label{eq:eq_eik-phi}
\end{equation}
where $\varphi_0$ is an integration constant and, to Newtonian order, $\underline{k}^m = (k^0,k^\alpha)=k_0(1, {\vec k})$, where $k_0=\omega/c$, is a constant null vector of the unperturbed light ray trajectory, $\gamma_{mn}\underline{k}^m\underline{k}^n={\cal O}(r_g, {\omega^2_{\tt p}}/{\omega^2})$; also, $\varphi_G$ is the post-Newtonian perturbation of the eikonal, and $\varphi_{\tt p} (t,{\vec x})$ is the perturbation due to plasma. The wavevector $K^m(t,{\vec x})$ then also admits a series expansion in the form
\begin{equation}
K^m(t,{\vec x})=\frac{dx^m}{d\lambda}= g^{mn}\partial_n\varphi=\underline{k}^m+k_{\tt G}^m(t,{\vec x})+k_{\tt p}^m(t,{\vec x})+{\cal O}(r_g^2,r_g\frac{\omega^2_{\tt p}}{\omega^2}),
\label{eq:K}
\end{equation}
where $k^m_{\tt G}(t,{\vec x})=\gamma^{mn}\partial_n\varphi_{\tt G}(t,{\vec x})$ and $k^m_{\tt p}(t,{\vec x})=\gamma^{mn}\partial_n\varphi_{\tt p}(t,{\vec x})$ are the first order perturbations of the wavevector due to post-Newtonian gravity and plasma, correspondingly.

Substituting (\ref{eq:eq_eik-phi}) into (\ref{eq:eq_eik}) and defining $h^{mn}=g^{mn}-\gamma^{mn}$ with $g_{mn}$ from (\ref{eq:metric-gen*})--(\ref{eq:n-eps}), we obtain an ordinary differential equation to determine the perturbations $\varphi_{\tt G}$ and $\varphi_{\tt p}$:
\begin{equation}
\frac{d\varphi_{\tt G}}{d\lambda}+\frac{d\varphi_{\tt p}}{d\lambda}= -\frac{1}{2}h^{mn}\underline{k}_m\underline{k}_n = -\frac{k_0^2}{c^2}\Big(2U-\frac{2\pi e^2 n_e}{m_e\omega^2 }\Big) +{\cal O}(r_g^2,r_g\frac{\omega^2_{\tt p}}{\omega^2}),
\label{eq:eq_eik-phi-lamb}
\end{equation}
where ${d\varphi_{\tt G}}/{d\lambda}+{d\varphi_{\tt p}}/{d\lambda}= K_m\partial^m\varphi $. Similarly to (\ref{eq:x-Newt}),  to Newtonian order, we represent the light ray's trajectory  as
\begin{equation}
\{x^m\}=\Big(x^0=ct, ~~{\vec x}(t)={\vec x}_{\rm 0}+{\vec k} c(t-t_0)\Big)+{\cal O}(r_g,\frac{\omega^2_{\tt p}}{\omega^2}),
\label{eq:light-traj}
\end{equation}
and substituting a monopole potential characterized by the Schwarzschild radius $r_g$ for $U$ and $n_e$ from (\ref{eq:n-eps_n-ism}), we obtain
{}
\begin{eqnarray}
\frac{d\varphi_{\tt G}}{d\lambda}+\frac{d\varphi_{\tt p}}{d\lambda}&=&
- \frac{k_0^2r_g}{|{\vec x}_{\rm 0}+{\vec k} c(t-t_{\rm 0})|}+\frac{2\pi e^2 k_0^2}{m_e\omega^2}\sum_i\alpha_i R_\odot^{\beta_i}\frac{1}{|{\vec x}_{\rm 0}+{\vec k} c(t-t_{\rm 0})|^{\beta_i}}+{\cal O}(r_g^2,r_g\frac{\omega^2_{\tt p}}{\omega^2}).
\label{eq:eq_eik-phi-lamb-E}
\end{eqnarray}

The representation of the trajectory given by (\ref{eq:light-traj}) allows us to express the Newtonian part of the wavevector $K^m$, as given by  (\ref{eq:K}), as
$K^m= {dx^m}/{d\lambda} =k^0\big(1, {\vec k}\big)+{\cal O}(r_g,{\omega^2_{\tt p}}/{\omega^2})$, where $k^0$ is immediately derived to have the form  $k^0={cdt}/{d\lambda}+{\cal O}(r_g,{\omega^2_{\tt p}}/{\omega^2})$ and $|{\vec k}|=1$. Keeping in mind that $\underline{k}^m$ is constant and using (\ref{eq:x-Newt*=}), we establish an important relationship:
\begin{equation}
d\lambda= \frac{cdt}{k^0}+{\cal O}(r_g,\frac{\omega^2_{\tt p}}{\omega^2})=\frac{cdt}{k_0}+{\cal O}(r_g,\frac{\omega^2_{\tt p}}{\omega^2})=\frac{d\tau}{k_0}+{\cal O}(r_g,\frac{\omega^2_{\tt p}}{\omega^2}),
\label{eq:eq_eik-relat}
\end{equation}
which we use  together with (\ref{eq:b}) and (\ref{eq:rel}) to integrate (\ref{eq:eq_eik-phi-lamb-E}).

As a result, in the body's proper reference frame  \cite{Turyshev:2012nw,Turyshev-Toth:2013}, we obtain the following expression for the phase evolution of an EM wave that propagates on the background of a gravitating monopole and plasma to the order of ${\cal O}(r_g^2,r_g{\omega^2_{\tt p}}/{\omega^2})$
{}
\begin{eqnarray}
\varphi(t,{\vec x}) &=& \varphi_0+k_0\Big\{\tau-({\vec k}\cdot {\vec x})-r_g\ln\Big[\frac{\tau+\sqrt{b^2+\tau^2}}{\tau_0+\sqrt{b^2+\tau_0^2}}\Big]+
\frac{2\pi e^2R_\odot}{m_e\omega^2}\sum_i \alpha_i \Big(\frac{R_\odot}{b}\Big)^{\beta_i-1}\Big(Q_{\beta_i}(\tau)-Q_{\beta_i}(\tau_0)\Big)\Big\},~~~
\label{eq:phase_t}
\end{eqnarray}
or, equivalently,
{}
\begin{eqnarray}
\varphi(t,{\vec x}) &=& \varphi_0+k_0\Big\{c(t-t_0)-{\vec k}\cdot ({\vec x}-{\vec x}_0)-r_g\ln\Big[\frac{r+({\vec k}\cdot {\vec x})}{r_0+({\vec k}\cdot {\vec x}_0)}\Big]+\nonumber\\
&+&\frac{2\pi e^2R_\odot}{m_e\omega^2}\sum_i \alpha_i \Big(\frac{R_\odot}{b}\Big)^{\beta_i-1}\Big(Q_{\beta_i}\big(({\vec k}\cdot{\vec x})\big)- Q_{\beta_i}\big(({\vec k}\cdot{\vec x}_0)\big)\Big)\Big\}+
{\cal O}(r_g^2,r_g\frac{\omega_{\tt p}^2}{\omega^2}).
\label{eq:phase_t*}
\end{eqnarray}

For a radial light ray with $k^\alpha =x^\alpha_0/r_0=n^\alpha_0$ (similarly to (\ref{eq:X-eq4**_rad})), from (\ref{eq:eq_eik-phi-lamb-E})
accurate to ${\cal O}(r_g^2,r_g{\omega^2_{\tt p}}/{\omega^2})$ we have
\begin{eqnarray}
\varphi(t,{\vec x}) &=& \varphi_0+k_0\Big\{c(t-t_0)-(r-r_0)-r_g\ln\frac{r}{r_0}-\frac{2\pi e^2 R_\odot}{m_e\omega^2}\sum_i \frac{\alpha_i}{\beta_i-1} \Big[\Big(\frac{R_\odot}{r}\Big)^{\beta_i-1}-\Big(\frac{R_\odot}{r_0}\Big)^{\beta_i-1}\Big]\Big\}.
\label{eq:phase_t-rad}
\end{eqnarray}

It is worth pointing out that the results obtained here for the phase of an EM wave (\ref{eq:phase_t}) and (\ref{eq:phase_t-rad}) are consistent with those obtained in the preceding  section obtained for the geodesic trajectory of a light ray (\ref{eq:X-eq4**}) and (\ref{eq:X-eq4**_rad}).

\section{Solution for the radial equation in the WKB approximation}
\label{sec:rad_eq_wkb}

Here we focus on the equation for the radial function, $R$, given by (\ref{eq:R-bar-k*20}) with $\alpha= \ell(\ell+1)$:
{}
\begin{eqnarray}
\frac{d^2 R}{d r^2}+\Big(k^2(1+\frac{2r_g}{r})+\frac{r_g}{r^3}-\frac{\alpha}{r^2}
\Big)R&=&0.
\label{eq:R-bar-k*2a}
\end{eqnarray}

When the functional dependence of $V_{\tt sr}$ (\ref{eq:V-sr-m2]}) falls off faster than $r^{-2}$, this term represents an additional short range potential. No exact solution exists for such an equation, especially with the generic form of  $n_e$, (\ref{eq:n-eps_n-ism}), and, thus, $\omega^2_p$ in (\ref{eq:eps}). Nevertheless, following an approach presented in \cite{Herlt-Stephani:1976,Turyshev-Toth:2017}, we explore an approximate solution to (\ref{eq:R-bar-k*2}) using the method of stationary phase (i.e., the Wentzel--Kramers--Brillouin, or WKB approximation \cite{Friedrich-Trost:2004}). As we are interested in the case when $k$ is rather large (for optical wavelengths $k=2\pi/\lambda=6.28\cdot10^6\,{\rm m}^{-1}$), we are looking for an asymptotic solution as $k\rightarrow\infty$.   In fact, we are looking for a solution in the form
{}
\begin{eqnarray}
R=e^{ikS(r)}\Big[a_0(r)+k^{-1}a_1(r)+...+k^{-n}a_n(r)+...\Big].
\label{eq:R_bar}
\end{eqnarray}
Technically, however, it is more convenient to search for a solution to (\ref{eq:R-bar-k*2a}) in an exponential form:
{}
\begin{eqnarray}
R=\exp\Big[\int_{r_0}^r i \Big(k\alpha_{-1}(t)+\alpha_0(t)+k^{-1}\alpha_1(t)+...+k^{-n}\alpha_n(t)+...\Big)dt\Big].
\label{eq:R-exp-bar}
\end{eqnarray}
Defining $'= d/ d r$, with the help of a substitution of $R'/R=w$, for the function $w$ we obtain the following equation:
{}
\begin{eqnarray}
w'+w^2+k^2(1+\frac{2r_g}{r})+\frac{r_g}{r^3}-\frac{\alpha}{r^2}=0.
\label{eq:ricati_bar}
\end{eqnarray}
{}
Using this substitution, up to the terms $\propto k^{-5}$, we have
{}
\begin{eqnarray}
w=i\Big(k\alpha_{-1}(r)+\alpha_0(r)+k^{-1}\alpha_1(r)+k^{-2}\alpha_3(r)+k^{-3}\alpha_3(r)+k^{-4}\alpha_4(r)+k^{-5}\alpha_5(r)+...+
k^{-n}\alpha_n(r)+...\Big).~~
\label{eq:w_k_bar}
\end{eqnarray}
Substituting (\ref{eq:w_k_bar}) into (\ref{eq:ricati_bar}) we obtain
{}
\begin{eqnarray}
k^2\big[1+\frac{2r_g}{r}-\alpha^2_{-1}(r)\big]+k\big[i\alpha'_{-1}(r)-2\alpha_{-1}(r)\alpha_0(r)\big]&+&
\nonumber\\
+\,
i\alpha'_{0}(r)-\alpha^2_{0}(r)-2\alpha_{-1}(r)\alpha_{1}(r)+\frac{r_g}{r^3}-\frac{\alpha}{r^2}&+&\nonumber\\
+\,
k^{-1}\big[i\alpha'_{1}(r)-2\alpha_{-1}(r)\alpha_{2}(r)
-2\alpha_{0}(r)\alpha_{1}(r)\big] &+&
\nonumber\\
+\,
k^{-2}\big[i\alpha'_{2}(r)-\alpha^2_{1}(r)-2\alpha_{-1}(r)\alpha_{3}(r)
-2\alpha_{0}(r)\alpha_{2}(r)\big]&+&
\nonumber\\
+\,
k^{-3}\big[i\alpha'_{3}(r)-2\alpha_{-1}(r)\alpha_{4}(r)
-2\alpha_{0}(r)\alpha_{3}(r)-2\alpha_{1}(r)\alpha_{2}(r)\big]&+&
\nonumber\\
+\,
k^{-4}\big[i\alpha'_{4}(r)-\alpha^2_{2}(r)-2\alpha_{-1}(r)\alpha_{5}(r)
-2\alpha_{0}(r)\alpha_{4}(r)-2\alpha_{1}(r)\alpha_{3}(r)\big]&=&{\cal O}(k^{-5}, r_g^2).~~~~~~
\label{eq:series_bar}
\end{eqnarray}
Now, if we equate the terms with respect to the same powers of $k$, we get
{}
\begin{eqnarray}
\alpha^2_{-1}(r)=1+\frac{2r_g}{r}, \qquad i\alpha'_{-1}(r)-2\alpha_{-1}(r)\alpha_0(r)=0, \qquad i\alpha'_{0}(r)-\alpha^2_{0}(r)-2\alpha_{-1}(r)\alpha_{1}(r)+\frac{r_g}{r^3}-\frac{\alpha}{r^2}&=&0,\nonumber\\
 i\alpha'_{1}(r)-2\alpha_{-1}(r)\alpha_{2}(r)
-2\alpha_{0}(r)\alpha_{1}(r)=0, \qquad
i\alpha'_{2}(r)-\alpha^2_{1}(r)-2\alpha_{-1}(r)\alpha_{3}(r)
-2\alpha_{0}(r)\alpha_{2}(r)&=&0,\nonumber\\
i\alpha'_{3}(r)-2\alpha_{-1}(r)\alpha_{4}(r)
-2\alpha_{0}(r)\alpha_{3}(r)-2\alpha_{1}(r)\alpha_{2}(r)&=&0, \nonumber\\
i\alpha'_{4}(r)-\alpha^2_{2}(r)-2\alpha_{-1}(r)\alpha_{5}(r)
-2\alpha_{0}(r)\alpha_{4}(r)-2\alpha_{1}(r)\alpha_{3}(r)&=&0.
\label{eq:series2_bar}
\end{eqnarray}
These equations, to the order of ${\cal O}(k^{-5}, r_g^2)$, may be solved as
{}
\begin{eqnarray}
\alpha_{-1}(r)&=&\pm(1+\frac{r_g}{r}), \qquad
\alpha_0(r)=-i\frac{r_g}{2r^2}, \qquad
\alpha_{1}(r)=\mp\frac{\alpha}{2r^2}(1-\frac{r_g}{r}), \qquad
\alpha_{2}(r)=i\frac{\alpha}{2r^3}(1-\frac{3r_g}{r}),\nonumber\\
\alpha_{3}(r)&=&\pm\big[\frac{3\alpha}{4r^4}(1-\frac{16r_g}{3r})-\frac{\alpha^2}{8r^4}(1-\frac{3r_g}{r})\big], \qquad
\alpha_{4}(r)=i\big[-\frac{3\alpha}{2r^5}(1-\frac{95r_g}{12r})+\frac{\alpha^2}{2r^5}(1-\frac{5r_g}{r})\big], \nonumber\\
\alpha_{5}(r)&=&\pm\big[-\frac{15\alpha}{4r^6}(1-\frac{107r_g}{10r})+\frac{7\alpha^2}{4r^6}(1-\frac{101r_g}{14r})-\frac{\alpha^3}{16r^6}(1-\frac{5r_g}{r})\big], ...
\label{eq:series3_bar}
\end{eqnarray}
Note that the $\pm$ signs in these expressions are not independent; they all come from the solution for $\alpha_{-1}(r)$ in (\ref{eq:series3_bar}).

Substituting solutions (\ref{eq:series3_bar}) into (\ref{eq:R-exp-bar}) and keeping the integration bounds for brevity, we have
{}
\begin{eqnarray}
S_{-1}(r)&=&\int_{r_0}^r \alpha_{-1}(\tilde r)d\tilde r=\pm\int_{r_0}^r (1+\frac{r_g}{\tilde r}),d\tilde r=\pm \big(r+r_g\ln 2kr\big)\Big|^r_{r_0},
\label{eq:S-1}\\
S_{0}(r)&=&\int_{r_0}^r \alpha_{0}(\tilde r)d\tilde r=-i\int_{r_0}^r \frac{r_g}{2\tilde r^2}d\tilde r=i\frac{r_g}{2r}\Big|^r_{r_0},
\label{eq:S-0}\\
S_1(r)&=&\int_{r_0}^r \alpha_{1}(\tilde r)d\tilde r=\mp\frac{\alpha}{2}\int_{r_0}^r \frac{d\tilde r}{\tilde r^2}(1-\frac{r_g}{\tilde r})=\pm\frac{\alpha}{2r}\Big(1-\frac{r_g}{2r}\Big)\Big|^r_{r_0},
\label{eq:S1}\\
S_2(r)&=&\int_{r_0}^r\alpha_{2}(\tilde r)d\tilde r=i\frac{\alpha}{2}\int_{r_0}^r\frac{d\tilde r}{\tilde r^3}(1-\frac{3r_g}{\tilde r})=-i\frac{\alpha}{4r^2}\Big(1-\frac{2r_g}{r}\Big)\Big|^r_{r_0},
\label{eq:S2}\\
S_3(r)&=&\int_{r_0}^r\alpha_{3}(\tilde r)d\tilde r=\pm\int_{r_0}^r\Big(\frac{3\alpha}{4{\tilde r}^4}(1-\frac{16r_g}{3\tilde r})-\frac{\alpha^2}{8{\tilde r}^4}(1-\frac{3r_g}{\tilde r})\Big){d\tilde r}=
\mp\Big(\frac{\alpha}{4{r}^3}\big(1-\frac{4r_g}{r}\big)-
\frac{\alpha^2}{24{ r}^3}\big(1-\frac{9r_g}{4r}\big)\Big)\Big|_{r_0}^r,~~~~~~~~~
\label{eq:S3}\\
S_4(r)&=&\int_{r_0}^r\alpha_{4}(\tilde r)d\tilde r=i\int_{r_0}^r\Big(-\frac{3\alpha}{2{\tilde r}^5}(1-\frac{95r_g}{12\tilde r})+\frac{\alpha^2}{2{\tilde r}^5}(1-\frac{5r_g}{\tilde r})\Big){d\tilde r}=
i\Big(\frac{3\alpha}{8{r}^4}\big(1-\frac{19r_g}{3r}\big)-\frac{\alpha^2}{8{ r}^4}\big(1-\frac{4r_g}{r}\big)\Big)\Big|_{r_0}^r,
\label{eq:S4}\\
S_5(r)&=&\int_{r_0}^r\alpha_{5}(\tilde r)d\tilde r=\pm\int_{r_0}^r\Big(-\frac{15\alpha}{4{\tilde r}^6}(1-\frac{107r_g}{10\tilde r})+\frac{7\alpha^2}{4{\tilde r}^6}(1-\frac{101r_g}{14\tilde r})-\frac{\alpha^3}{16{\tilde r}^6}(1-\frac{5r_g}{\tilde r})\Big){d\tilde r}=\nonumber\\
&&\hskip 53 pt =\pm\Big(\frac{3\alpha}{4{r}^5}\big(1-\frac{214r_g}{15r}\big)-\frac{7\alpha^2}{20{ r}^5}\big(1-\frac{505r_g}{84r}\big)+\frac{\alpha^3}{80{ r}^5}\big(1-\frac{25r_g}{6r}\big)\Big)\Big|_{r_0}^r.
\label{eq:S5}
\end{eqnarray}

As we see,  for $i\geq 1$  functions $S_i$ from (\ref{eq:S1})--(\ref{eq:S5}) have factors of the type $(1-\beta \,r_g/r)$ in their structure, where $\beta$ is some constant. Clearly, outside the Sun, the ratio $r_g/r$ is very small; it reaches its maximum value at the solar radius and then it diminishes as $r_g/r=2.45\times 10^{-6}\times R_\odot/r$. As for any practical application $r\gg R_\odot$, this ratio may be neglected and factors $(1-\beta \,r_g/r)$ may be treated as being equal to 1 in all of such occurancies present in $S_i,  i\geq 1$.

We now obtain two approximate solutions for the partial radial function $R_\ell$, which is given, to ${\cal O}\big((kr)^{-6},r_g^2\big)$, as
{}
\begin{eqnarray}
R_\ell(r)&=& c_\ell \exp\Big\{i\big(kS_{-1}(r)+S_0(r)+k^{-1}S_1(r)+k^{-2}S_2(r)+k^{-3}S_3(r)+k^{-4}S_4(r)+k^{-5}S_5(r)\big)\Big\}+\nonumber\\
&&+\,
d_\ell \exp\Big\{-i\big(kS_{-1}(r)+S_0(r)+k^{-1}S_1(r)+k^{-2}S_2(r)+k^{-3}S_3(r)+k^{-4}S_4(r)+k^{-5}S_5(r)\big)\Big\},~~~~~
\label{eq:R_solWKB+=_bar+}
\end{eqnarray}
where $c_\ell$ and $d_\ell$ are arbitrary constants. Substituting  (\ref{eq:S-1})--(\ref{eq:S5}) in (\ref{eq:R_solWKB+=_bar+}), we obtain the following solution for $R_\ell$:
{}
\begin{eqnarray}
uR_\ell(r)&=&
\exp\Big[{\frac{\ell(\ell+1)}{4k^2r^2}}-{\frac{3\ell(\ell+1)}{8k^4r^4}}+{\frac{[\ell(\ell+1)]^2}{8k^4r^4}}\Big]\times
\nonumber\\
&&\hskip -36pt \times\,
\Big\{c_\ell \,\exp\Big[i\Big(k(r+r_g\ln2kr)+\frac{\ell(\ell+1)}{2kr}-\frac{\ell(\ell+1)}{4k^3r^3}+\frac{\big[\ell(\ell+1)\big]^2}{24k^3r^3}+
\frac{3\ell(\ell+1)}{4k^5{r}^5}-\frac{7[\ell(\ell+1)]^2}{20k^5{ r}^5}+\frac{[\ell(\ell+1)]^3}{80k^5{ r}^5}\Big)\Big]+\nonumber\\
&&\hskip -36pt +\,\,
d_\ell\,
\exp\Big[-i\Big(k(r+r_g\ln2kr)+\frac{\ell(\ell+1)}{2kr}-\frac{\ell(\ell+1)}{4k^3r^3}+\frac{\big[\ell(\ell+1)\big]^2}{24k^3r^3}+
\frac{3\ell(\ell+1)}{4k^5{r}^5}-\frac{7[\ell(\ell+1)]^2}{20k^5{ r}^5}+\frac{[\ell(\ell+1)]^3}{80k^5{ r}^5}\Big)\Big]\Big\}+
\nonumber\\&&\hskip -5pt +\,\,
{\cal O}\big((kr)^{-6},r_g^2\big),~~~
\label{eq:R_solWKB+=_bar}
\end{eqnarray}
where $c_\ell$ and $d_\ell$  now account for all the integration constants relevant to the point $r_0$ in (\ref{eq:S-1})--(\ref{eq:S5}).

As we discussed in \cite{Turyshev-Toth:2017}, omission of the ${r_g}/{r^3}$ term in (\ref{eq:R-bar-k*2a}) leads to appearance of  an ``uncompensated'' term ${r_g}/{4kr^2}=(1/8\pi)({r_g\lambda}/{r^2})$ in the exponent of (\ref{eq:R_solWKB+=_bar}). This term is extremely small; it decays fast as $r$ increases, and, thus, it may be neglected in the solution for the radial function.
A similar point was made in \cite{Matzner:1968}, suggesting that one can neglect the $r^{-3}$ terms in (\ref{eq:R-bar-k*2a}) (the same, of course, is true for (\ref{eq:R-bar-k*20})) and reduce the problem to the case of the equation for the radial function being the Schr\"odinger equation describing scattering in a Coulomb potential.

Expression (\ref{eq:R_solWKB+=_bar}) is used in Section~\ref{sec:radial-comp} where we apply the method of the stationary phase to develop expressions containing the scattering amplitude. As we saw previously (e.g., \cite{Turyshev-Toth:2017,Turyshev-Toth:2018}) the solution for the points of the stationary phase leads to a solution for $\ell$ of the form  $\ell\simeq kr \sin\theta$.  This observation allows us to somewhat simplify the expressions (\ref{eq:R_solWKB+=_bar}). Indeed, any term for which the exponent of $\ell$ in the numerator is less than the exponent of $(kr)$ in the denominator is extremely small compared to the other terms.  Indeed, the first and last terms in the amplitude are of order $\ell^2/(kr)^2$ and $\ell^4/(kr)^4$, correspondingly. However, the second term is of order $\ell^2/(kr)^4$, which is $1/(kr)^2$ times smaller than the first term and $1/\ell^2$ smaller than the third term. Thus, the second term may be neglected. On the same grounds we may neglect three terms in the phase of  expression (\ref{eq:R_solWKB+=_bar}).  In addition, similarly to \cite{Turyshev-Toth:2017}, we may further improve the asymptotic expression for $R_\ell$ from (\ref{eq:R_solWKB+=_bar}) by accounting for the Coulomb phase shifts, which can be done by simply redefining the constants $c_\ell$ and $d_\ell$ yet again \cite{Turyshev-Toth:2017} as
{}
\begin{eqnarray}
c_\ell\rightarrow c_\ell \exp\big[{i(\sigma_\ell-\frac{\pi \ell}{2})}\big],\qquad\qquad d_\ell\rightarrow d_\ell \exp\big[{-i(\sigma_\ell-\frac{\pi \ell}{2})}\big].
\label{eq:cd_ell}
\end{eqnarray}

As a result of the simplifications and rescaling of the constants discussed above, the expression for the asymptotic behavior of the partial radial function $R_\ell$ takes the following form:
\begin{eqnarray}
uR_\ell(r)&=&
\exp\Big[{\frac{\ell(\ell+1)}{4k^2r^2}}+{\frac{[\ell(\ell+1)]^2}{8k^4r^4}}\Big]
\times\nonumber\\
&&\hskip -50pt \times\,
\Big\{c_\ell \,\exp\Big[i\Big(k(r+r_g\ln 2kr)+\frac{\ell(\ell+1)}{2kr}+\frac{\big[\ell(\ell+1)\big]^2}{24k^3r^3}+\frac{[\ell(\ell+1)]^3}{80k^5r^5}+\sigma_\ell-\frac{\pi \ell}{2}\Big)\Big]+\nonumber\\
&&\hskip -40pt
+\,
d_\ell\,
\exp\Big[-i\Big(k(r+r_g\ln 2kr)+\frac{\ell(\ell+1)}{2kr}+\frac{\big[\ell(\ell+1)\big]^2}{24k^3r^3}+\frac{[\ell(\ell+1)]^3}{80k^5r^5}+\sigma_\ell-\frac{\pi \ell}{2}\Big)\Big]\Big\}
+{\cal O}\big((kr)^{-6}, r_g^2\big).~~~~~
\label{eq:R_solWKB+=_bar-imp}
\end{eqnarray}

In \cite{Turyshev-Toth:2017} the asymptotic behavior of the Coulomb function was obtained for very larger distances from the turning point for $r\gg r_{\tt t}$; the solution (\ref{eq:R_solWKB+=_bar-imp}) improves it further by extending the argument of these functions to shorter distances, closer to the turning point (as was done in \cite{Turyshev-Toth:2018-plasma} for Riccati--Bessel functions in the flat space-time.)

\end{document}